\documentclass[onecolumn,english]{IEEEtran}
\usepackage[T1]{fontenc}
\usepackage[latin9]{inputenc}
\usepackage{color}
\usepackage{amsmath}
\usepackage{amsthm}
\usepackage{amssymb}
\usepackage{graphicx}

\makeatletter
\theoremstyle{plain}
\newtheorem{thm}{\protect\theoremname}
\theoremstyle{definition}
\newtheorem{defn}[thm]{\protect\definitionname}
\theoremstyle{plain}
\newtheorem{prop}[thm]{\protect\propositionname}
\theoremstyle{remark}
\newtheorem{rem}[thm]{\protect\remarkname}
\theoremstyle{plain}
\newtheorem{lem}[thm]{\protect\lemmaname}
\theoremstyle{plain}
\newtheorem{cor}[thm]{\protect\corollaryname}

\makeatother

\usepackage{babel}
\providecommand{\corollaryname}{Corollary}
\providecommand{\definitionname}{Definition}
\providecommand{\lemmaname}{Lemma}
\providecommand{\propositionname}{Proposition}
\providecommand{\remarkname}{Remark}
\providecommand{\theoremname}{Theorem}

\begin{document}
\title{New Second-Order Achievability Bounds for Coding with Side Information via Type Deviation Convergence}
\author{Xiang Li and Cheuk Ting Li\\
Department of Information Engineering, The Chinese University of Hong Kong, Hong Kong, China\\
Email: lx024@ie.cuhk.edu.hk, ctli@ie.cuhk.edu.hk}
\maketitle
\begin{abstract}
We propose a framework for second-order achievability, called type deviation convergence, that is generally applicable to settings in network information theory, and is especially suitable for lossy source coding and channel coding with cost. We give a second-order achievability bound for lossy source coding with side information at the decoder (Wyner-Ziv problem) that improves upon all known bounds (e.g., Watanabe-Kuzuoka-Tan, Yassaee-Aref-Gohari and Li-Anantharam).  We also give second-order achievability bounds for  lossy compression where side information may be absent (Heegard-Berger problem) and channels with noncausal state information at the encoder and cost constraint (Gelfand-Pinsker problem with cost) that improve upon previous bounds.
\end{abstract}


\section{Introduction}

In information theory, first-order analysis concerns the characterization of the limit $\mathsf{R}^{*}$ of the optimal coding rate $\mathsf{R}^{*}(n,\epsilon)$ (e.g., the maximal message rate for channel coding, or the minimal compression rate for lossy source coding) as a function of the blocklength $n$ and the error probability $\epsilon$, as $n\to\infty$; whereas second-order analysis concerns a more refined characterization of how $\mathsf{R}^{*}(n,\epsilon)$ approaches $\mathsf{R}^{*}$. Most second-order results are in the form
\[
\mathsf{R}^{*}(n,\epsilon)=\mathsf{R}^{*}+\mathsf{W}/\sqrt{n}+o(1/\sqrt{n}),
\]
where $\mathsf{W}$ is the coefficient of the second-order term that may depend on $\epsilon$. For example, in channel coding, we have $\mathsf{W}=-\sqrt{\mathsf{V}}\mathcal{Q}^{-1}(\epsilon)$, where $\mathsf{V}=\mathrm{Var}[\iota(X;Y)]$ is the channel dispersion, and $\mathcal{Q}^{-1}$ is the inverse of the Q-function \cite{strassen1962asymptotic,hayashi2009information,polyanskiy2010channel}. In lossy source coding, we have $\mathsf{W}=\sqrt{\mathsf{V}}\mathcal{Q}^{-1}(\epsilon)$, where $\mathsf{V}$ is the dispersion of lossy source coding \cite{kontoyiannis2000pointwise,kostina2012fixed,ingber2011dispersion} (see Section \ref{sec:sc}).

Second-order analysis has also been performed on more complex settings where side information is present. An example is the Wyner-Ziv problem \cite{wyner1976ratedistort,wyner1978rate}, where the encoder compresses a source $X^{n}$ into a description $M$ so that the decoder who observes $M$ and a side information $Y^{n}$ (correlated with $X^{n}$, unknown to the encoder) can output a lossy reconstruction of $X^{n}$. Second-order acheivability results have been given in \cite{verdu2012nonasymp,yassaee2013oneshot,watanabe2015nonasymp,li2021unified,liu2024one}. For the lossless case, the Wyner-Ziv problem reduces to a special case of the Slepian-Wolf problem \cite{slepianwolf1973a}, where the dispersion has been characterized \cite{tan2013dispersions}. Nevertheless, the optimal dispersion for the general lossy Wyner-Ziv problem remains an open problem. The Wyner-Ziv problem can be generalized to the scenario where the side information may be absent \cite{heegard1985rate,kaspi1994rate}, where a second-order result was given in \cite{yassaee2013oneshot}. Also see \cite{draper2004side,kochman2010excess,phan2024importance,gkagkos2024structural,wei2025non,watanabe2025tight} for related results.

Another example is the Gelfand-Pinsker problem \cite{gelfand1980coding,heegard1983capacity,costa1983writing} about a channel coding setting where the channel depends on a state sequence known noncausally to the encoder. The current best second-order results were given in \cite{scarlett2015dispersions} (for discrete and Gaussian channels) and \cite{li2021unified} (for general channels). A second-order result for Gelfand-Pinsker problem with a cost constraint has been derived in \cite{watanabe2015nonasymp}.

The method of types \cite{csiszar1998method,csiszar2011information}, which concerns the type (or the empirical distribution) of sequences in the coding setting, is a common tool in second-order analysis. For example, it has been used in \cite{ingber2011dispersion} for the second-order result for lossy source coding, and in \cite{tan2013dispersions} for the Slepian-Wolf problem, multiple-access channels and asymmetric broadcast channels. See \cite{tan2014asymptotic,zhou2023finite} and references therein for discussions on more settings in network information theory. Constant-composition codes, where each codeword has the same type, have been applied to prove second-order achievability results in multiple access channels \cite{huang2012finite,scarlett2014second} and the Gelfand-Pinsker problem \cite{scarlett2015dispersions}.  However, to the best of the authors' knowledge, the method of types had not been successful for lossy source coding problems with side information (e.g., Wyner-Ziv), likely due to the complexity of the analysis.

Another useful technique is the Poisson matching lemma \cite{li2021unified} and the Poisson functional representation \cite{sfrl_trans}, which have been applied to derive various finite-blocklength and refined asymptoical results about coding with side information \cite{phan2024importance,liu2025one,liu2024hiding,wei2025non,liu2025nonasymptotic,watanabe2025tight}. Nevertheless, \cite{li2021unified} mostly focused on i.i.d. codebooks, and it was unclear whether the Poisson matching lemma would still be useful for code constructions with a more precise control over the type. In particular, the second-order achievability bound for the Wyner-Ziv problem in \cite{li2021unified}, which was already one of the best bounds at that time, fails to subsume the dispersion of lossy source coding without side information \cite{kontoyiannis2000pointwise,kostina2012fixed,ingber2011dispersion} as a special case.

The purpose of this paper is to propose a new definition of ``well-behaved'' random sequences, called \emph{type deviation convergent sequences}, where the type deviates from its limit by an amount that has a convergent distribution. This includes i.i.d. sources (where the deviation is asymptotically Gaussian by the central limit theorem), constant-composition codes (the deviation is zero), as well as sequences with a non-Gaussian deviation. We give a small, essential toolbox of properties of type deviation convergent sequences that, together with the Poisson matching lemma, can be applied to a wide range of problems to give simple second-order achievability proofs that recover and sometimes improve upon the state of the art. We list some advantages of our approach:
\begin{itemize}
\item \textbf{Improving upon best known bounds.} Our approach yields second-order achievability results that improves upon the state of the art for the following problems:
\begin{itemize}
\item For the Wyner-Ziv problem, our approach gives a dispersion term that significantly improves upon \cite{verdu2012nonasymp,watanabe2015nonasymp,yassaee2013oneshot,li2021unified,liu2025one}, and subsumes the achievability of the dispersion of lossy source coding without side information \cite{kontoyiannis2000pointwise,kostina2012fixed,ingber2011dispersion} as a special case\footnote{To the best of the authors' knowledge, this is the first second-order achievability result for the Wyner-Ziv problem that subsumes the achievability of the dispersion of lossy source coding as a special case. Although the finite-blocklength bound in \cite{watanabe2015nonasymp} subsumes the dispersion of lossy source coding, the second-order bound stated in \cite{watanabe2015nonasymp} does not subsume the dispersion of lossy source coding.} (Section \ref{sec:wz}). This improvement is partly due to the use of codes with non-Gaussian deviations, which differ from the usual Gaussian deviation from the central limit theorem.  
\item For the indirect Wyner-Ziv problem \cite{yamamoto1980source,rebollo2005generalization}, our bound improves upon \cite{wei2025non} (Section \ref{sec:noisy_wz}). 
\item For lossy compression where side information may be absent (Heegard-Berger) \cite{heegard1985rate,kaspi1994rate}, our bound improves upon \cite{yassaee2013oneshot} (Section \ref{sec:hbk}).
\item For Gelfand-Pinsker coding with cost constraint, our bound improves upon \cite{watanabe2015nonasymp} and \cite{li2021unified} (after a straightforward generalization to include cost) (Section \ref{sec:gp}).
\end{itemize}
\item \textbf{Recovering best known bounds.} For other settings, our approach can recover state-of-the-art second-order results. This includes channel coding (with or without cost constraint) \cite{feinstein1954new,hayashi2009information,polyanskiy2010channel,kostina2015channels}, lossy source coding \cite{kontoyiannis2000pointwise,kostina2012lossy,ingber2011dispersion}, indirect or noisy lossy source coding \cite{kostina2016nonasymptotic,yang2024jointdatasemanticslossy}, Gelfand-Pinsker coding \cite{scarlett2015dispersions,li2021unified} and broadcast channels \cite{yassaee2013binning}.
\item \textbf{Simplicity.} The main novelty of our approach is to couple the source and channel input/output sequences for different blocklengths $n$ together in the same probability space. This allows the type of a sequence to be approximated by a single random vector that does not keep changing as $n$ increases. As a result, the proofs are shorter than conventional proofs using the method of types. See Remark \ref{rem:why_coupling}.\footnote{While the proof of the second-order result for Wyner-Ziv (Theorem \ref{thm:wz}) is short by itself, it is no longer short if we also count the proofs of the basic properties of type deviation convergent sequences in Sections \ref{sec:tdc}, \ref{sec:gcc}. This should not be considered as an argument against the simplicity of our approach, since ``factoring out'' the complicated parts of the proofs to reusable basic properties can simplify subsequent proofs of new theorems. This is similar to typicality and asymptotic equipartition property, which are considered to be valuable tools for simplifying proofs in information theory, even though the basic properties of typicality often have complicated proofs.}
\item \textbf{A unified workflow.} The proofs in this paper all follow the same unified workflow using the same sequence of tools, which can accomodate i.i.d. sequences and constant-composition codes under the same framework. This eliminates the need of devising ad-hoc arguments for each coding settings, and significantly reduces the difficulty of deriving second-order achievability results for new settings. This is in the same spirit as previous unified workflows for refined asymptotical analysis in network information theory such as \cite{verdu2012nonasymp,yassaee2013oneshot,yassaee2014achievability,yassaee2013binning,li2021unified,liu2025one}, though our approach yields better second-order bounds than previous unified approaches for settings with distortion or cost constraints.
\end{itemize}
\medskip{}

Since the purpose of this paper is to present a new technique for achievability results, we will not discuss converse results which require completely different techniques, and hence are out of the scope of this paper. For refined asymptotical converse bounds for the Wyner-Ziv problem, see \cite{oohama2018exponential,watanabe2025tight}.

\medskip{}

\section{Preliminaries}

\subsection{Notations}

Throughout this paper, we assume all coding settings concern discrete sources and channels with finite alphabets. Entropy is in bits, and logarithms are to the base $2$. We use upper-case serif letters (e.g., $X,U$) for random variables, bold letters (e.g., $\mathbf{X},\mathbf{U}$) for random processes, and san-serif letters (e.g., $\mathsf{R},\mathsf{D}$) for non-random parameters. Write $[n]:=\{1,\ldots,n\}$. For a discrete random variable $X\in\mathcal{X}$, write $P_{X}:\mathcal{X}\to\mathbb{R}$ for its probability mass function. For random variables $X,Y$, write $\iota_{X}(x):=-\log P_{X}(x)$, $\iota_{Y|X}(x):=-\log P_{Y|X}(y|x)$, $\iota_{X;Y}(x;y):=\iota_{Y}(y)-\iota_{Y|X}(y|x)$ (we sometimes omit the subscripts and simply write $\iota(x;y)$). For a random sequence $X_{1},\ldots,X_{n}$, we say that it has an exchangeable distribution if $P_{X_{1},\ldots,X_{n}}(x_{1},\ldots,x_{n})=P_{X_{1},\ldots,X_{n}}(x_{\pi(1)},\ldots,x_{\pi(n)})$ for any permutation $\pi$ over $[n]$. For a random vector $X$, write $\mathrm{Var}[X]$ for its covariance matrix. Write the Q-function as $\mathcal{Q}(x)=\mathbb{P}(X\ge x)$ where $X\sim\mathrm{N}(0,1)$, and its inverse as $\mathcal{Q}^{-1}(t)$.

For a finite set $\mathcal{X}$, write $\mathbb{R}^{\mathcal{X}}$ for the space of real vectors with entries indexed by $\mathcal{X}$, or equivalently, the space of all functions $\mathcal{X}\to\mathbb{R}$ (we use $\mathbb{R}^{\mathcal{X}}$ and $\mathcal{X}\to\mathbb{R}$ interchangeably). For function $f:\mathcal{X}\to\mathbb{R}$, write $\Vert f\Vert:=\sqrt{\sum_{x\in\mathcal{X}}(f(x))^{2}}$, $\Vert f\Vert_{\infty}:=\max_{x\in\mathcal{X}}|f(x)|$. For $f,g\in\mathbb{R}^{\mathcal{X}}$, we say that $f$ is dominated by $g$, denoted as $f\ll g$, if $g(x)=0$ implies $f(x)=0$ for $x\in\mathcal{X}$.

For a vector $x^{n}\in\mathcal{X}^{n}$, write $\hat{P}_{n}(x^{n})\in\mathbb{R}^{\mathcal{X}}$ to be its type or empirical distribution, i.e., $\hat{P}_{n}(x^{n})(x')=|\{i\in[n]:x_{i}=x'\}|/n$. Note that when $n=1$, $\hat{P}_{1}(x)$ is the one-hot encoding of $x\in\mathcal{X}$. Its range $\hat{P}_{n}(\mathcal{X}):=\{\hat{P}_{n}(x^{n}):\,x^{n}\in\mathcal{X}^{n}\}\subseteq\mathbb{R}^{\mathcal{X}}$ is the set of all probability mass functions over $\mathcal{X}$ where each entry is a multiple of $1/n$. We sometimes omit the subscript and simply write $\hat{P}$ if $n$ is clear in the context.

For sets $\mathcal{A}_{1},\mathcal{A}_{2}\subseteq\mathbb{R}^{\mathcal{X}}$ and $c\in\mathbb{R}$, write the Minkowski sum as $\mathcal{A}_{1}+\mathcal{A}_{2}:=\{a_{1}+a_{2}:\,a_{1}\in\mathcal{A}_{1},a_{2}\in\mathcal{A}_{2}\}$, and $c\mathcal{A}_{1}:=\{ca_{1}:\,a_{1}\in\mathcal{A}_{1}\}$. Write $\mathcal{B}^{\mathcal{X}}:=\{x\in\mathbb{R}^{\mathcal{X}}:\,\Vert x\Vert<1\}$ for the open $\ell_{2}$ ball. For two probability distributions $P,Q$ over a vector space $\mathbb{R}^{\mathcal{X}}$, their \emph{L\'{e}vy-Prokhorov distance} is \cite{dudley1968distances}
\begin{align}
d_{\Pi}(P,Q) & :=\inf\Big\{\epsilon:\,P(\mathcal{A})\le Q(\mathcal{A}+\epsilon\mathcal{B}^{\mathcal{X}})+\epsilon,\nonumber \\
 & \quad\quad\quad\quad\forall\;\text{closed}\;\mathcal{A}\subseteq\mathbb{R}^{\mathcal{X}}\Big\}.\label{eq:lp_dist}
\end{align}
For two random vectors $X,Y\in\mathbb{R}^{\mathcal{X}}$, their \emph{Ky Fan distance} \cite{dudley2018real} is defined as 
\begin{equation}
d_{\mathrm{KF}}(X,Y):=\inf\{\epsilon:\,\mathbb{P}(\Vert X-Y\Vert>\epsilon)\le\epsilon\}.\label{eq:kyfan}
\end{equation}
These two metrics are related via the \emph{Strassen-Dudley theorem} \cite{dudley2018real}, \cite[Theorem 6.9]{billingsley2013convergence}, which states that $d_{\Pi}(P,Q)=\inf_{X\sim P,Y\sim Q}d_{\mathrm{KF}}(X,Y)$ (the infimum is over couplings $(X,Y)$ of $P$ and $Q$). 

\medskip{}

\subsection{Distributions and Perturbations}

We now define some notations that are useful for studying discrete distributions and their perturbations. Write $\Delta(\mathcal{X}):=\{p\in\mathbb{R}^{\mathcal{X}}:\,\min_{x}p(x)\ge0,\,\sum_{x}p(x)=1\}$ for the set of probability mass functions over $\mathcal{X}$. Write $\Delta(\mathcal{Y}|\mathcal{X}):=\{p(\cdot|\cdot)\in\mathbb{R}^{\mathcal{X}\times\mathcal{Y}}:\,p(\cdot|x)\in\Delta(\mathcal{Y}),\forall x\}$ for the set of conditional probability mass functions from $\mathcal{X}$ to $\mathcal{Y}$, where $p(\cdot|x)$ denotes the function $y\mapsto p(y|x)$. Note that if we write $p(\cdot|\cdot)\in\mathbb{R}^{\mathcal{X}\times\mathcal{Y}}$, then the function $p:\mathbb{R}^{\mathcal{X}\times\mathcal{Y}}\to\mathbb{R}$ is being used as $p(y|x)$ (instead of $p(x,y)$) to conform with the usual notation of conditional distributions (we use $\mathbb{R}^{\mathcal{X}\times\mathcal{Y}}$ instead of $\mathbb{R}^{\mathcal{Y}\times\mathcal{X}}$, despite $y$ being written before $x$ in $p(y|x)$, because $x$ is usually generated before $y$). For $p\in\Delta(\mathcal{X})$, write
\[
\mathrm{Tan}(p):=\big\{ v\in\mathbb{R}^{\mathcal{X}}:\,v\ll p,\,\sum_{x}v(x)=0\big\}.
\]
Note that $\mathrm{Tan}(p)$ is the tangent space of $\Delta(\mathcal{X})$ at $p$, in the sense that there exists $\epsilon>0$ such that $p+v\in\Delta(\mathcal{X})$ for every $v\in\mathrm{Tan}(p)$ with $\Vert v\Vert<\epsilon$. Hence, we can understand $\mathrm{Tan}(p)$ as the ``space of small perturbations'' of the distribution $p$. Write $\mathrm{Tan}(\mathcal{X}):=\{v\in\mathbb{R}^{\mathcal{X}}:\,\sum_{x}v(x)=0\}$. Similarly, for $p\in\Delta(\mathcal{Y}|\mathcal{X})$, write $\mathrm{Tan}(p):=\{v\in\mathbb{R}^{\mathcal{X}\times\mathcal{Y}}:\,v\ll p,\,\sum_{y}v(y|x)=0,\forall x\}$.

For a function $f:\mathcal{X}\to[0,\infty)$, write $\sqrt{f}:\mathcal{X}\to[0,\infty)$ for the function $x\mapsto\sqrt{f(x)}$. For functions $f:\mathcal{X}\to\mathbb{R}$ and $g:\mathcal{Y}\to\mathbb{R}$, denote their product as $f\times g:\mathcal{X}\times\mathcal{Y}\to\mathbb{R}$, $(f\times g)(x,y):=f(x)g(y)$. For functions $f:\mathcal{X}\to\mathbb{R}$ and $g:\mathcal{X}\times\mathcal{Y}\to\mathbb{R}$, denote their \emph{semidirect product} as $f\circ g:\mathcal{X}\times\mathcal{Y}\to\mathbb{R}$, 
\[
(f\circ g)(x,y):=f(x)g(x,y).
\]
For example, for random variables $(X,Y)\sim P_{X,Y}$, we have $P_{X}\circ P_{Y|X}=P_{X,Y}$. If some parts of the domain of $f$ is not present in the domain of $g$, then the domain of $g$ is suitably extended to include those parts. For example, if we have $f:\mathcal{X}\times\mathcal{Z}\to\mathbb{R}$ instead where the part $\mathcal{Z}$ is not present in the domain of $g:\mathcal{X}\times\mathcal{Y}\to\mathbb{R}$, then we extend $g$ to $g:\mathcal{X}\times\mathcal{Z}\times\mathcal{Y}\to\mathbb{R}$, $g(x,z,y)=g(x,y)$, so $(f\circ g)(x,z,y)=f(x,z)g(x,y)$. Note that $P_{X,Z}\circ P_{Y|X}=P_{X,Z,Y}$ if $Z\leftrightarrow X\leftrightarrow Y$ forms a Markov chain. For functions $f,g:\mathcal{X}\to\mathbb{R}$ write
\[
\left\langle f,g\right\rangle :=\sum_{x\in\mathcal{X}}f(x)g(x).
\]
If the domain of $f$ is larger than the domain of $g$, then the domain of $g$ is suitably extended to match the domain of $f$ (e.g., if we have $f:\mathcal{X}\times\mathcal{Y}\to\mathbb{R}$, then we extend $g$ to $g(x,y)=g(x)$ so $\left\langle f,g\right\rangle =\sum_{x,y}f(x,y)g(x)$). Note that if $(X,Y)\sim P_{X,Y}$, then $\left\langle P_{X,Y},g\right\rangle =\mathbb{E}[g(X)]$, so $\left\langle f,g\right\rangle $ is a generalization of expectation where $f$ may not be a probability mass function.\footnote{Similar notations have appeared, for example, in \cite{liu2018dispersion}.}

To demonstrate the use of these notations, note that the derivative of entropy $H(P_{X})$ along the direction $V\in\mathrm{Tan}(P_{X})$ is
\begin{equation}
\frac{\mathrm{d}H(P_{X}+tV)}{\mathrm{d}t}\Big|_{t=0}=\left\langle V,\iota_{X}\right\rangle ,\label{eq:H_derivative}
\end{equation}
where $\iota_{X}:\mathcal{X}\to\mathbb{R}$, $\iota_{X}(x)=-\log P_{X}(x)$ is the self-information. As a result, the derivative of $I(X;Y)$ when $(X,Y)\sim P_{X,Y}+tV$ along the direction $V\in\mathrm{Tan}(P_{X,Y})$ is $\left\langle V,\iota_{X;Y}\right\rangle $.

\medskip{}

\subsection{Gaussian-Multinomial Distribution}

The main technique in this paper is to approximate the distribution of the type $\hat{P}(X^{(n)})$ of a random sequence $X^{(n)}\in\mathcal{X}^{n}$ by a Gaussian distribution. We first consider a simple example where $X^{(n)}\sim P_{X}^{n}$ is an i.i.d. sequence. Since $\hat{P}(X^{(n)})=n^{-1}\sum_{i=1}^{n}\hat{P}_{1}(X_{i}^{(n)})$, where $\hat{P}_{1}(X_{i}^{(n)})\in\mathbb{R}^{\mathcal{X}}$ is the one-hot encoding of $X_{i}^{(n)}\in\mathcal{X}$, we know that $n\hat{P}(X^{(n)})$ follows a multinomial distribution, and $\sqrt{n}(\hat{P}(X^{(n)})-P_{X})$ is approximately Gaussian with mean $0$ and covariance matrix $\mathrm{Var}[\hat{P}_{1}(X)]$ where $X\sim P_{X}$, by the central limit theorem. This distribution is defined as follows. It is a special case of the Gaussian-multinomial distribution in \cite{chen1999forecasting}, and hence we adopt this name.

\medskip{}

\begin{defn}
[Gaussian-multinomial distribution]\label{def:nm}For a probability mass function $P_{X}\in\Delta(\mathcal{X})$, denote 
\[
\mathrm{NM}(P_{X}):=\mathrm{N}(\mathbf{0},\Sigma)
\]
to be the multivariate Gaussian distribution over $\mathrm{Tan}(P_{X})\subseteq\mathbb{R}^{\mathcal{X}}$ with covariance matrix $\Sigma(x,x)=P_{X}(x)(1-P_{X}(x))$, $\Sigma(x,x')=-P_{X}(x)P_{X}(x')$ for $x\neq x'$. Equivalently, $\Sigma=\mathrm{Var}[\hat{P}_{1}(X)]$ where $X\sim P_{X}$, and $\hat{P}_{1}(X)\in\mathbb{R}^{\mathcal{X}}$ is the one-hot encoding of $X$.
\end{defn}
\medskip{}

We now define a conditional version of this distribution, which consists of independent Gaussian-multinomial vectors stacked together to form a matrix.

\medskip{}

\begin{defn}
[Conditional Gaussian-multinomial distribution]For a conditional probability mass function $P_{Y|X}\in\Delta(\mathcal{Y}|\mathcal{X})$, denote $\mathrm{NM}(P_{Y|X})$ to be the distribution of $(V_{x})_{x\in\mathcal{X}}\in\mathrm{Tan}(P_{Y|X})\subseteq\mathbb{R}^{\mathcal{X}\times\mathcal{Y}}$ (i.e., stacking $V_{x}$ for $x\in\mathcal{X}$ together), where $V_{x}\in\mathbb{R}^{\mathcal{Y}}$, $V_{x}\sim\mathrm{NM}(P_{Y|X}(\cdot|x))$ are independent across $x\in\mathcal{X}$.
\end{defn}
\medskip{}

We now state some simple facts about $\mathrm{NM}(P_{X})$ and $\mathrm{NM}(P_{Y|X})$. The proofs are immediate and hence omitted.

\smallskip{}

\begin{prop}
\label{prop:nm_prop}Let $(X,Y)\sim P_{X,Y}=P_{X}\circ P_{Y|X}$. Let $G_{X}\sim\mathrm{NM}(P_{X})$ and $G_{Y|X}\sim\mathrm{NM}(P_{Y|X})$ be independent. We have the following:
\begin{itemize}
\item For $f:\mathcal{X}\to\mathbb{R}$,
\begin{equation}
\left\langle G_{X},f\right\rangle \sim\mathrm{N}(0,\,\mathrm{Var}[f(X)]).\label{eq:NM_var}
\end{equation}
\item 
\[
G_{X}\circ P_{Y|X}+\sqrt{P_{X}}\circ G_{Y|X}\sim\mathrm{NM}(P_{X,Y}).
\]
\item For $f:\mathcal{X}\times\mathcal{Y}\to\mathbb{R}$,
\[
\left\langle G_{X}\circ P_{Y|X},f\right\rangle \sim\mathrm{N}\left(0,\,\mathrm{Var}[\mathbb{E}[f(X,Y)|X]]\right),
\]
\[
\langle\sqrt{P_{X}}\circ G_{Y|X},f\rangle\sim\mathrm{N}\left(0,\,\mathbb{E}[\mathrm{Var}[f(X,Y)|X]]\right).
\]
\end{itemize}
\end{prop}
Note that $\mathrm{Var}[\mathbb{E}[f(X,Y)|X]]+\mathbb{E}[\mathrm{Var}[f(X,Y)|X]]=\mathrm{Var}[f(X,Y)]$ by the law of total variance. Hence, $G_{X}\circ P_{Y|X}+\sqrt{P_{X}}\circ G_{Y|X}$ decomposes the randomness in $\mathrm{NM}(P_{X,Y})$ into two parts: the part $G_{X}\circ P_{Y|X}$ that comes from the randomness in $X$, and the part $\sqrt{P_{X}}\circ G_{Y|X}$ from the randomness in $Y$.

\medskip{}

\section{Type Deviation Convergence\label{sec:tdc}}

We now introduce a framework, called \emph{type deviation convergence}, which is a collection of notations and results that simplifies second-order analyses. We consider families of random sequences in the form $\mathbf{X}=(X^{(n)})_{n\in\mathbb{N}}$, where $X^{(n)}\in\mathcal{X}^{n}$ is a random sequence and $\mathcal{X}$ is finite. Such an $\mathbf{X}$ is called a \emph{general source} \cite{han2006reliability}. Note that we do not require $X^{(n)}$ to be the prefix of $X^{(n+1)}$, i.e., $X_{i}^{(n)}$ (the $i$-th entry of $X^{(n)}$) may not equal $X_{i}^{(n+1)}$. This is the reason we use the notation $X^{(n)}$ instead of $X^{n}$. 

We focus on general sources with a type $\hat{P}(X^{(n)})$ that can be approximated by a distribution $P_{X}$ in the sense that $\hat{P}(X^{(n)})-P_{X}$ is $O(n^{-1/2})$, and $\sqrt{n}(\hat{P}(X^{(n)})-P_{X})$ converges to a subgaussian random vector. This is captured by the following definition.

\medskip{}

\begin{defn}
[Type deviation convergence]\label{def:typegaussian}For a general source $\mathbf{X}=(X^{(n)})_{n\in\mathbb{N}}$, we say that it is\emph{ type deviation convergent} if $X^{(n)}$ has an exchangeable distribution, and there exists a probability vector $P_{X}\in\Delta(\mathcal{X})$ and a subgaussian\footnote{A random vector $G\in\mathbb{R}^{\mathcal{X}}$ is subgaussian if there exists $\eta>0$ such that $\mathbb{P}(\Vert G\Vert\ge t)\le2\exp(-t^{2}/\eta^{2})$ for all $t\ge0$.} random vector $G_{\mathbf{X}}\in\mathrm{Tan}(P_{X})$ such that $\hat{P}(X^{(n)})\ll P_{X}$ almost surely, and $G_{\mathbf{X}}^{(n)}:=\sqrt{n}(\hat{P}(X^{(n)})-P_{X})$ approaches $G_{\mathbf{X}}$ in the order $O(n^{-1/2})$ with respect to the Ky Fan distance (\ref{eq:kyfan}), that is, there exists $c>0$ such that
\[
\mathbb{P}\left(\Vert G_{\mathbf{X}}^{(n)}-G_{\mathbf{X}}\Vert>\frac{c}{\sqrt{n}}\right)<\frac{c}{\sqrt{n}},
\]
for all $n$. If this is satisfied, we say that $\mathbf{X}$ is type deviation convergent with \emph{center} $P_{X}$ and \emph{asymptotic deviation} $G_{\mathbf{X}}$. 

For two general sources $\mathbf{X},\mathbf{Y}$, we say that $(\mathbf{X},\mathbf{Y})$ is (jointly) type deviation convergent if $((X_{i}^{(n)},Y_{i}^{(n)})_{i\in[n]})_{n\in\mathbb{N}}$ is type deviation convergent (which implies that $\mathbf{X}$ and $\mathbf{Y}$ are both type deviation convergent). 
\end{defn}
\medskip{}

Note that we do not require $G_{\mathbf{X}}$ to be Gaussian or have zero mean. Definition \ref{def:typegaussian} allows many interesting statistics of  $\mathbf{X}$ to be summarized by the first-order term $P_{X}$ and the second-order term $G_{\mathbf{X}}$. We remark that the $X^{(n)}$'s for different $n$'s are assumed to be coupled together (i.e., defined in the same probability space), so we can talk about the limit $G_{\mathbf{X}}$. This is a main novelty in our approach which significantly simplifies the proofs, compared to previous achievability proofs in information theory where the source sequences for different blocklengths are effectively in separate probability spaces.\footnote{In a conventional proof of the source coding theorem, the source $X^{(n)}\in\mathcal{X}^{n}$ is either assumed to be in different probability spaces for different $n$'s, or assumed to be the first $n$ symbols of the same infinite sequences $X_{1},X_{2},\ldots$ (and hence $X^{(n)}$'s are ``coupled'' together). It makes no difference which of these two definitions is used, and hence the $X^{(n)}$'s are ``effectively in separate probability spaces''. In our approach, the coupling of $X^{(n)}$ is important, and the aforementioned infinite sequence ``coupling'' will not work.}   Refer to Remark \ref{rem:why_coupling} for further discussions on why a coupling is desirable.

\medskip{}

\subsection{Examples}

The ``prototypical example'' of a type deviation convergent source is the i.i.d. memoryless source in which the entries of $X^{(n)}$ are i.i.d. following $P_{X}$, where $G_{\mathbf{X}}\sim\mathrm{NM}(P_{X})$ is a Gaussian-multinomial random variable.  Nevertheless, the memoryless source is not technically type deviation convergent since the $X^{(n)}$'s for different $n$'s are not defined in the same probability space.\footnote{If we put them in the same probability space simply by taking $X^{(n)}=(X_{1},\ldots,X_{n})$ where $X_{i}\stackrel{\mathrm{iid}}{\sim}P_{X}$, then it is not type deviation convergent since $G_{\mathbf{X}}^{(n)}=\sqrt{n}(\hat{P}(X^{(n)})-P_{X})$ does not converge by Donsker's theorem.} Therefore, it is usually more reasonable to ask whether there exists a coupling of the $X^{(n)}$'s that is type deviation convergent. We say that $\tilde{\mathbf{X}}=(\tilde{X}^{(n)})_{n}$ is a coupling of $\mathbf{X}=(X^{(n)})_{n}$ if the $\tilde{X}^{(n)}$'s for $n\in\mathbb{N}$ are defined in the same probability space, and $\tilde{X}^{(n)}$ has the same marginal distribution as $X^{(n)}$. The following proposition gives the condition for a type deviation convergent coupling to exist. 

\medskip{}

\begin{prop}
\label{prop:lp_coupling}For a general source $\mathbf{X}=(X^{(n)})_{n}$, there exists a type deviation convergent coupling if and only if $X^{(n)}$ has an exchangeable distribution, and there exists $P_{X}\in\Delta(\mathcal{X})$ and a subgaussian random vector $G_{\mathbf{X}}\in\mathrm{Tan}(P_{X})$ such that $\hat{P}(X^{(n)})\ll P_{X}$ almost surely, and the distribution of $G_{\mathbf{X}}^{(n)}=\sqrt{n}(\hat{P}(X^{(n)})-P_{X})$ approaches the distribution of $G_{\mathbf{X}}$ in the order $O(n^{-1/2})$ with respect to the L\'{e}vy-Prokhorov distance (\ref{eq:lp_dist}), i.e., there exists $c>0$ such that
\[
d_{\Pi}(G_{\mathbf{X}}^{(n)},G_{\mathbf{X}})\le cn^{-1/2}.
\]
\end{prop}
\medskip{}

\begin{IEEEproof}
For the ``only if'' part, if $\tilde{\mathbf{X}}=(\tilde{X}^{(n)})_{n}$ is a type deviation convergent coupling, then for any closed $\mathcal{A}\subseteq\mathrm{Tan}(P_{X})$, we have
\begin{align*}
\mathbb{P}(G_{\tilde{\mathbf{X}}}\in\mathcal{A}) & \le\mathbb{P}(G_{\tilde{\mathbf{X}}}^{(n)}\in\mathcal{A}+cn^{-1/2}\mathcal{B})+cn^{-1/2},
\end{align*}
where $\mathcal{B}$ is the open unit ball in $\mathrm{Tan}(P_{X})$, and hence $d_{\Pi}(G_{\mathbf{X}}^{(n)},G_{\tilde{\mathbf{X}}})=d_{\Pi}(G_{\tilde{\mathbf{X}}}^{(n)},G_{\tilde{\mathbf{X}}})\le cn^{-1/2}$. For the ``if'' part, by the Strassen-Dudley theorem \cite[Theorem 6.9]{billingsley2013convergence}, there exists a coupling of $G_{\mathbf{X}}$ and $G_{\mathbf{X}}^{(n)}$ with a Ky Fan distance (\ref{eq:kyfan}) $d_{\mathrm{KF}}(G_{\mathbf{X}}^{(n)},G_{\mathbf{X}})\le2cn^{-1/2}$. Hence, $(X^{(n)})_{n}$ can be coupled with $G_{\mathbf{X}}$ such that $d_{\mathrm{KF}}(G_{\mathbf{X}}^{(n)},G_{\mathbf{X}})\le2cn^{-1/2}$ holds for all $n$.
\end{IEEEproof}

\medskip{}

We can now show that a type deviation convergent coupling exists for an i.i.d. source $\mathbf{X}$. 

\medskip{}

\begin{prop}
\label{prop:gauss_iid}If $X^{(n)}\sim P_{X}^{n}$ (the $n$-fold i.i.d. distribution), then $\mathbf{X}$ has a type deviation convergent coupling with center $P_{X}$ and asymptotic deviation being a Gaussian-multinomial random variable $G_{\mathbf{X}}\sim\mathrm{NM}(P_{X})$ (Definition \ref{def:nm}).
\end{prop}
\medskip{}

\begin{IEEEproof}
Recall that $G_{\mathbf{X}}^{(n)}=\sqrt{n}(\hat{P}(X^{(n)})-P_{X})$. We have $\hat{P}(X^{n})=n^{-1}\sum_{i=1}^{n}\hat{P}_{1}(X_{i})$, where $\hat{P}_{1}(X_{i})$ are i.i.d. with $\mathbb{E}[\hat{P}_{1}(X_{i})]=P_{X}$. By the result in \cite{jurinskii1975smoothing} on the $O(n^{-1/2})$ convergence in the central limit theorem with respect to $d_{\Pi}$, there exists a constant $c$ such that $d_{\Pi}(G_{\mathbf{X}}^{(n)},G_{\mathbf{X}})\le cn^{-1/2}$. The result follows from Proposition \ref{prop:lp_coupling}.
\end{IEEEproof}
\medskip{}

More generally, if we pass a type deviation convergent source through a memoryless channel, then the source and the output can be coupled to be jointly type deviation convergent. The proof is in Appendix \ref{subsec:pf_gauss_memoryless}.

\medskip{}

\begin{prop}
[Type deviation convergence of memoryless channels]\label{prop:gauss_memoryless}Assume $\mathbf{X}$ is type deviation convergent with center $P_{X}$ and asymptotic deviation $G_{\mathbf{X}}$, and $P_{Y|X}$ is a conditional distribution from $\mathcal{X}$ to $\mathcal{Y}$. Then there exists a source $\mathbf{Y}$ such that $Y^{(n)}|X^{(n)}\sim P_{Y|X}^{n}$ (i.e., the conditional marginal distribution of $Y^{(n)}$ given $X^{(n)}$ is as if $Y^{(n)}$ is the output when $X^{(n)}$ is passed through the memoryless channel $P_{Y|X}$), and $(\mathbf{X},\mathbf{Y})$ is type deviation convergent with center $P_{X}\circ P_{Y|X}$ and asymptotic deviation 
\[
G_{\mathbf{X},\mathbf{Y}}=G_{\mathbf{X}}\circ P_{Y|X}+\sqrt{P_{X}}\circ G_{\mathbf{Y}|\mathbf{X}},
\]
where  $G_{\mathbf{Y}|\mathbf{X}}\sim\mathrm{NM}(P_{Y|X})$ is independent of $G_{\mathbf{X}}$.
\end{prop}

\medskip{}

\subsection{Average Properties}

To see why the interesting statistics of a type deviation convergent source $\mathbf{X}$ are summarized by $P_{X}$ and $G_{\mathbf{X}}$, first we note that for any function $f:\mathcal{X}\to\mathbb{R}$, we have
\begin{align}
\frac{1}{n}\sum_{i=1}^{n}f(X_{i}^{(n)}) & =\left\langle \hat{P}(X^{(n)}),f\right\rangle \nonumber \\
 & =\left\langle P_{X},f\right\rangle +\frac{1}{\sqrt{n}}\left\langle G_{\mathbf{X}},\iota_{X}\right\rangle +O\Big(\frac{1}{n}\Big)\label{eq:f_avg}
\end{align}
with probability $1-O(n^{-1/2})$,\footnote{``$X=Y+O(1/n)$ with probability $1-O(n^{-1/2})$'' means that there exists $c>0$ such that $\mathbb{P}(|X-Y|\le c/n)\ge1-cn^{-1/2}$.} which follows directly from the definition. Next, we show that the self-information $\iota_{X^{(n)}}(X^{(n)})=-\log P_{X^{(n)}}(X^{(n)})$ can be approximated as
\begin{equation}
\iota_{X^{(n)}}(X^{(n)})=nH(X)+\sqrt{n}\left\langle G_{\mathbf{X}},\iota_{X}\right\rangle +O(\log n)\label{eq:i_avg}
\end{equation}
with probability $1-O(n^{-1/2})$. The proof is in Appendix \ref{subsec:pf_self_info}.

\medskip{}

\begin{prop}
\label{prop:self_info}Assume $\mathbf{X}$ is type deviation convergent with center $P_{X}$ and asymptotic deviation $G_{\mathbf{X}}$. Then
\[
\mathbb{P}\left(\left|nH(X)+\sqrt{n}\left\langle G_{\mathbf{X}},\iota_{X}\right\rangle -\iota_{X^{(n)}}(X^{(n)})\right|>c\log n\right)\le\frac{c}{\sqrt{n}},
\]
for some constant $c$ that only depends on the distribution of $\mathbf{X}$, where $H(X)$ and $\iota_{X}$ are computed using $P_{X}$. 
\end{prop}
\medskip{}

As a result, if $(\mathbf{X},\mathbf{Y})$ is type deviation convergent with center $P_{X,Y}$ and asymptotic deviation $G_{\mathbf{X},\mathbf{Y}}$, applying Proposition \ref{prop:self_info} on $\mathbf{X}$, $\mathbf{Y}$ and $(\mathbf{X},\mathbf{Y})$, we have
\begin{align}
 & \iota_{X^{(n)};Y^{(n)}}(X^{(n)};Y^{(n)})\nonumber \\
 & =nI(X;Y)+\sqrt{n}\left\langle G_{\mathbf{X},\mathbf{Y}},\iota_{X;Y}\right\rangle +O(\log n)\label{eq:info_density}
\end{align}
with probability $1-O(n^{-1/2})$. As we will see, for many settings, sample averages $n^{-1}\sum_{i=1}^{n}f(X_{i}^{(n)})$ (e.g., cost and distortion constraints) and combinations of $\iota_{X^{(n)}}(X^{(n)})$ are all the statistics we need to prove coding theorems.

\medskip{}

\begin{rem}
\label{rem:why_coupling}The use of coupling in Definition \ref{def:typegaussian} is a main novelty of our technique. Alternatively, one might simply take Proposition \ref{prop:lp_coupling} as the definition of type deviation convergent sources, which will not require $X^{(n)}$'s to be coupled together. This alternative definition is more similar to conventional proofs based on the method of types. However, it comes with a significant downside that we can no longer have simple statements like (\ref{eq:f_avg}) and (\ref{eq:i_avg}), since it is no longer reasonable to talk about convergence of random variables as $n\to\infty$ (we can only talk about convergence of distributions). We argue that this is a main source of complexity in previous proofs, and might have been a reason why the method of types has not been applied to more complicated settings, e.g., Heegard-Berger and the broadcast channel, where our approach can be applied with relative ease (Theorems \ref{thm:hb} and \ref{thm:bc}). The proofs in this paper (e.g., Theorems \ref{thm:sc} and \ref{thm:wz}) are comparatively simpler (and also stronger).
\end{rem}
\smallskip{}

\section{General Constant-Composition Codes\label{sec:gcc}}

Constant-composition code is a common technique for proving second-order achievability results \cite{huang2012finite,scarlett2014second,scarlett2015dispersions}. In this paper, we consider a general definition of constant-composition codes, where the conditional type of the output $U^{(n)}$ can depend on the type of the source $X^{(n)}$ in a general manner through a ``deviation function'' $\zeta:\mathrm{Tan}(P_{X})\to\mathrm{Tan}(P_{U|X})$. Intuitively, when $\mathbf{X}$ is type deviation convergent, we want the conditional type of $U^{(n)}$ given $X^{(n)}$ to be approximately $\zeta(G_{\mathbf{X}})$, which can depend on $G_{\mathbf{X}}$. Note that constructions where the type of $U^{(n)}$ depend of the type of $X^{(n)}$ have appeared in previous proofs (e.g., \cite{liu2018dispersion}), though the level of generality in Definition \ref{def:acc}, which allows $\zeta$ to be nonlinear, appears to be novel.

\medskip{}

\begin{defn}
[General constant-composition (GCC) channel]\label{def:acc}Given a joint distribution $P_{X,U}=P_{X}\circ P_{U|X}$ and a Lipschitz continuous function $\zeta:\mathrm{Tan}(P_{X})\to\mathrm{Tan}(P_{U|X})$, we say that a family of conditional distributions $(P_{U^{(n)}|X^{(n)}})_{n}$ (from $\mathcal{X}^{n}$ to $\mathcal{U}^{n}$) for $n\in\mathbb{N}$ is \emph{general constant-composition} with center $P_{X}\circ P_{U|X}$ and deviation function $\zeta$ if $P_{U^{(n)}|X^{(n)}}$ is jointly exchangeable (i.e., $P_{U^{(n)}|X^{(n)}}(u^{n}|x^{n})=P_{U^{(n)}|X^{(n)}}((u_{\pi(i)})_{i\in[n]}|(x_{\pi(i)})_{i\in[n]})$ for every permutation $\pi:[n]\to[n]$), and there exists $c>0$ such that for every $n\ge c$ and $x^{n}\in\mathcal{X}^{n}$ with $\hat{P}(x^{n})\ll P_{X}$ and $\Vert\hat{P}(x^{n})-P_{X}\Vert\le1/c$, and for every $u^{n}$ with $P_{U^{(n)}|X^{(n)}}(u^{n}|x^{n})>0$, we must have $\hat{P}(x^{n},u^{n})\ll P_{X,U}$ and 
\[
\big\Vert G_{x^{n}}^{(n)}\circ P_{U|X}+P_{X}\circ\zeta(G_{x^{n}}^{(n)})-G_{x^{n},u^{n}}^{(n)}\big\Vert\le cn^{-1/2},
\]
where
\begin{align*}
G_{x^{n}}^{(n)} & :=\sqrt{n}(\hat{P}(x^{n})-P_{X}),\\
G_{x^{n},u^{n}}^{(n)} & :=\sqrt{n}(\hat{P}(x^{n},u^{n})-P_{X,U}).
\end{align*}
\end{defn}
\medskip{}

We first prove the existence of a GCC channel. The proof is in Appendix \ref{subsec:pf_acc_exists}.

\smallskip{}

\begin{prop}
\label{prop:acc_exists}For every $P_{X,U}$ and Lipschitz continuous function $\zeta:\mathrm{Tan}(P_{X})\to\mathrm{Tan}(P_{U|X})$, there exists a GCC channel $(P_{U^{(n)}|X^{(n)}})_{n}$. 
\end{prop}
\medskip{}

Next, we show that passing a type deviation convergent source as the input to a GCC channel results in an output that is jointly type deviation convergent with the input.

\medskip{}

\begin{prop}
[Type deviation convergence of GCC channels]\label{prop:acc_product}Assume $\mathbf{X}$ is type deviation convergent with center $P_{X}$ and asymptotic deviation $G_{\mathbf{X}}$, and $(P_{U^{(n)}|X^{(n)}})_{n}$ is a GCC channel with center $P_{X,U}=P_{X}\circ P_{U|X}$ and deviation function $\zeta$. Let $\mathbf{U}=(U^{(n)})_{n}$ be random sequences such that $U^{(n)}$ follows $P_{U^{(n)}|X^{(n)}}$ given $X^{(n)}$. Then $(\mathbf{X},\mathbf{U})$ is type deviation convergent with center $P_{X,U}$ and asymptotic deviation
\[
G_{\mathbf{X},\mathbf{U}}=G_{\mathbf{X}}\circ P_{U|X}+P_{X}\circ\zeta(G_{\mathbf{X}}).
\]
\end{prop}
\medskip{}

\begin{IEEEproof}
Recall that $G_{\mathbf{X}}^{(n)}=\sqrt{n}(\hat{P}(X^{(n)})-P_{X})$. Let $c$ be the maximum of the constant in Definition \ref{def:typegaussian} for $\mathbf{X}$, and the constant in Definition \ref{def:acc} for $(P_{U^{(n)}|X^{(n)}})_{n}$. Since $\mathbf{X}$ is type deviation convergent, we have $\Vert\hat{P}(X^{(n)})-P_{X}\Vert\le1/c$ with probability $1-O(n^{-1/2})$. Consider 
\begin{align*}
G_{\mathbf{X},\mathbf{U}}^{(n)} & :=\sqrt{n}(\hat{P}(X^{(n)},U^{(n)})-P_{X,U}),\\
\hat{G}_{\mathbf{X},\mathbf{U}}^{(n)} & :=G_{\mathbf{X}}^{(n)}\circ P_{U|X}+P_{X}\circ\zeta(G_{\mathbf{X}}^{(n)}).
\end{align*}
By Definition \ref{def:acc}, $\Vert G_{\mathbf{X},\mathbf{U}}^{(n)}-\hat{G}_{\mathbf{X},\mathbf{U}}^{(n)}\Vert\le cn^{-1/2}$ with probability $1-O(n^{-1/2})$. We also have $\Vert\hat{G}_{\mathbf{X},\mathbf{U}}^{(n)}-G_{\mathbf{X},\mathbf{U}}\Vert=O(n^{-1/2})$ with probability $1-O(n^{-1/2})$ since $\zeta$ is Lipschitz and $\Vert G_{\mathbf{X}}^{(n)}-G_{\mathbf{X}}\Vert\le cn^{-1/2}$ with probability $1-O(n^{-1/2})$. The result follows. 
\end{IEEEproof}
\smallskip{}

We remark that if $G_{\mathbf{X}}$ is Gaussian and $\zeta$ is an affine function, then $G_{\mathbf{X},\mathbf{U}}$ is also Gaussian. Nevertheless, we do not require $\zeta$ to be affine, so $G_{\mathbf{X},\mathbf{U}}$ may not be Gaussian even if $G_{\mathbf{X}}$ is Gaussian. Non-affine $\zeta$ is useful for the Wyner-Ziv problem in Theorem \ref{thm:wz}.

\smallskip{}

\section{Second-Order Lossy Source Coding\label{sec:sc}}

We will now utilize type deviation convergence to prove coding theorems. Before we proceed to the theorems, we state the Poisson matching lemma introduced in \cite{li2021unified}, which is another basic tool for constructing coding schemes via exponential random variables that we will use throughout this paper.\footnote{\cite{li2021unified} proved the Poisson matching lemma for the general case where $X,U,Y$ may not be discrete, which involve Poisson processes. Here, we focus on the discrete case, which has a simpler statement and proof in \cite{li2021unified}.}

\smallskip{}

\begin{lem}
[Poisson matching lemma on $X,U,Y$ \cite{li2021unified}]\label{lem:pml}Consider a joint distribution $P_{X,U,Y}=P_{X}\circ P_{U|X}\circ P_{Y|X,U}$. Let $T_{u}\sim\mathrm{Exp}(1)$, i.i.d. across $u\in\mathcal{U}$. Assume $X\sim P_{X}$, $U=\mathrm{argmin}_{u}T_{u}/P_{U|X}(u|X)$, $Y|(X,U)\sim P_{Y|X,U}$, and $\hat{U}=\mathrm{argmin}_{u}T_{u}/P_{U|Y}(u|Y)$. Then $(X,U,Y)\sim P_{X,U,Y}$, and 
\begin{equation}
\mathbb{P}\big(U\neq\hat{U}\,\big|\,X,U,Y\big)\le2^{\iota_{U;X}(U;X)-\iota_{U;Y}(U;Y)}\label{eq:ppl}
\end{equation}
holds almost surely. As a result, for any error event $E$ that depends only on $(X,U,Y)$ and any $\gamma\in\mathbb{R}$,
\begin{align*}
 & \mathbb{P}(E\;\mathrm{or}\;U\neq\hat{U})\\
 & \le\mathbb{P}\left(E\;\mathrm{or}\;\iota_{U;X}(U;X)-\iota_{U;Y}(U;Y)>-\gamma\right)+2^{-\gamma}.
\end{align*}
\end{lem}
\smallskip{}

We first demonstrate the use of type deviation convergence and the Poisson matching lemma, by recovering the second-order result on lossy source coding for a discrete memoryless source \cite{kontoyiannis2000pointwise,ingber2011dispersion,kostina2012lossy}. In this setting, the encoder encodes a discrete memoryless source $X^{(n)}\sim P_{X}^{n}$ into a message $M\in[\lceil2^{n\mathsf{R}}\rceil]$ with rate $\mathsf{R}>0$. The decoder observes $M$ and recovers $\hat{Y}^{(n)}\in\mathcal{Y}^{n}$. The goal is to minimize the probability of excess distortion
\[
P_{e}:=\mathbb{P}\big(d(X^{(n)},\hat{Y}^{(n)})>\mathsf{D}\big),
\]
where $d:\mathcal{X}\times\mathcal{Y}\to\mathbb{R}$ is a distortion function, $d(X^{(n)},\hat{Y}^{(n)}):=n^{-1}\sum_{i=1}^{n}d(X_{i}^{(n)},\hat{Y}_{i}^{(n)})$, and $\mathsf{D}\in\mathbb{R}$ is the allowed distortion level. The optimal asymptotic rate needed to have $P_{e}\to0$ is given by the rate-distortion function $\mathsf{R}(\mathsf{D}):=\min_{P_{Y|X}:\,\mathbb{E}[d(X,Y)]\le\mathsf{D}}I(X;Y)$ \cite{shannon1959coding}. We now recover the second-order result in \cite{kontoyiannis2000pointwise,kostina2012lossy,ingber2011dispersion}, given in terms of the $d$\emph{-tilted information }\cite{kontoyiannis2000pointwise,kostina2012lossy}
\begin{align}
\jmath_{X,\mathsf{D}}(x) & :=-\log\mathbb{E}\left[2^{-\lambda(d(x,Y^{*})-\mathsf{D})}\right]\nonumber \\
 & \stackrel{(a)}{=}\iota_{X;Y}(x;y)+\lambda(d(x,y)-\mathsf{D})\label{eq:d_tilted}\\
 & \stackrel{(b)}{=}\mathbb{E}\big[\iota_{X;Y}(X;Y)+\lambda(d(X,Y)-\mathsf{D})\,\big|\,X=x\big],\nonumber 
\end{align}
where $\lambda:=-\mathrm{d}\mathsf{R}(\mathsf{D})/\mathrm{d}\mathsf{D}$, we assume $P_{Y|X}$ is the distribution attaining $\mathsf{R}(\mathsf{D})$, and $Y^{*}\sim P_{Y}$ (the marginal distribution of $Y$ in $P_{X}\circ P_{Y|X}$), (a) holds for $P_{Y}$-almost all $y$'s \cite{csiszar1974extremum}, and (b) is assuming $(X,Y)\sim P_{X}\circ P_{Y|X}$.

\medskip{}

\begin{thm}
\label{thm:sc}For discrete lossy source coding, if $\mathsf{D}>\min_{P_{Y|X}}\mathbb{E}[d(X,Y)]$, then for any $0<\epsilon<1$ and any large enough $n$, there is a scheme achieving a probability of excess distortion $P_{e}\le\epsilon$, with rate
\[
\mathsf{R}=\mathsf{R}(\mathsf{D})+\sqrt{\frac{\mathsf{V}}{n}}\mathcal{Q}^{-1}(\epsilon)+O\left(\frac{\log n}{n}\right),
\]
where $\mathsf{V}:=\mathrm{Var}[\jmath_{X,\mathsf{D}}(X)]$,  and the constant in $O((\log n)/n)$ depends on $P_{X}$, $d$, $\mathsf{D}$ and $\epsilon$.
\end{thm}
\medskip{}

\begin{figure*}
\begin{centering}
\includegraphics[scale=0.9]{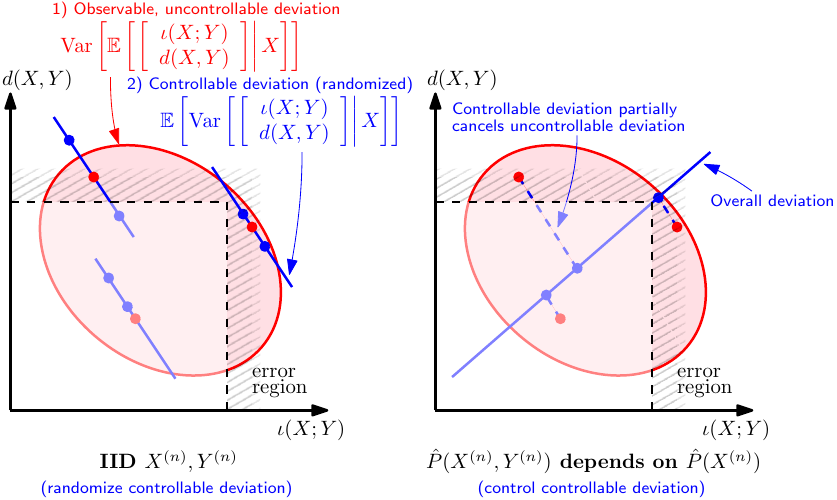}
\par\end{centering}
\caption{\label{fig:sc}Left: Illustration for lossy source coding with i.i.d. $X^{(n)},Y^{(n)}$. The three red dots are drawn from a Gaussian distribution with covariance matrix given by the first term in (\ref{eq:sc_gauss}) (red ellipse is a contour of the Gaussian distribution), and the blue dots are the red dots plus a Gaussian vector with covariance matrix given by the second term in (\ref{eq:sc_gauss}). Right: The optimal scheme where we control the deviation of the type of $Y^{(n)}$ according to the type of $X^{(n)}$, moving the red dots to the blue dots along the blue line.}
\end{figure*}

We give an intuitive explanation of Theorem \ref{thm:sc}. Loosely speaking, there are two error events in lossy source coding: excess distortion $d(X^{(n)},Y^{(n)})>\mathsf{D}$, and excess information $n^{-1}\iota(X^{(n)};Y^{(n)})>\mathsf{R}$ (the $n\mathsf{R}$-bit message is insufficient to store $Y^{(n)}$). In case if $(X^{(n)},Y^{(n)})\stackrel{\mathrm{iid}}{\sim}P_{X}\circ P_{Y|X}$, then by the central limit theorem, the vector $[n^{-1}\iota(X^{(n)};Y^{(n)}),d(X^{(n)},Y^{(n)})]^{\top}$ will be approximately Gaussian with covariance matrix
\begin{align}
 & \frac{1}{n}\mathrm{Var}\left[\left[\begin{array}{c}
\iota\\
d
\end{array}\right]\right]\nonumber \\
 & =\frac{1}{n}\mathrm{Var}\left[\mathbb{E}\left[\left.\left[\begin{array}{c}
\iota\\
d
\end{array}\right]\right|X\right]\right]+\frac{1}{n}\mathbb{E}\left[\mathrm{Var}\left[\left.\left[\begin{array}{c}
\iota\\
d
\end{array}\right]\right|X\right]\right],\label{eq:sc_gauss}
\end{align}
due to the law of total covariance, where $\iota:=\iota(X;Y)$, $d:=d(X,Y)$. The first term and the second term above are the contribution of the randomness in $X^{(n)}$ (an observable but uncontrollable component of the deviation) and $Y^{(n)}$ (a controllable component of the deviation), respectively. Error occurs if one of the two coordinates of the random vector is too large; see Figure \ref{fig:sc} (left). The second term is rank-one since $\iota(X;Y)+\lambda d(X,Y)$ depends only on $X$ by \ref{eq:d_tilted}. The intuitive reason is that if changing $Y$ can affect $\iota+\lambda d$, then we can perturb $P_{Y|X}$ to reduce $I(X;Y)+\lambda\mathbb{E}[d(X,Y)]$, contradicting the optimality of $P_{Y|X}$.

To reduce the error probability, note that the encoder has full control of $Y^{(n)}$, and there is no reason to randomize $Y^{(n)}$ and increase the error probability. The second term in (\ref{eq:sc_gauss}) can be eliminated via a constant composition code for $Y^{(n)}$. Moreover, we can control the deviation of the type of $Y^{(n)}$ according to the type of $X^{(n)}$ in order to combine the two error events into one. Since $\iota(X;Y)+\lambda d(X,Y)$ depends only on $X$, controlling $Y$ can only allow trading off $\iota(X;Y)$ and $d(X,Y)$ along a diagonal line with slope $-1/\lambda$. We can move the point $[\iota,d]^{\top}$ along this diagonal line to push it out of the error region as much as possible, for example, by moving it to the blue line in Figure \ref{fig:sc} (right). 

\smallskip{}

We now prove Theorem \ref{thm:sc} using type deviation convergence.
\begin{IEEEproof}
The proof is divided into five steps.

\textbf{1) Code construction.} Consider the $P_{Y|X}$ that attains $\mathsf{R}(\mathsf{D})$. Consider a GCC channel $(P_{Y^{(n)}|X^{(n)}})_{n}$ with center $P_{X}\circ P_{Y|X}$ and Lipschitz deviation function $\zeta_{Y|X}:\mathrm{Tan}(P_{X})\to\mathrm{Tan}(P_{Y|X})$ to be specified later. Let $T_{m,y^{n}}\sim\mathrm{Exp}(1)$, i.i.d. across $m\in[\lfloor2^{n\mathsf{R}}\rfloor]$, $y^{n}\in\mathcal{Y}^{n}$, which serves as a random codebook available to the encoder and the decoder. Consider the joint distribution
\[
(M,X^{(n)},Y^{(n)})\sim\mathrm{Unif}([\lceil2^{n\mathsf{R}}\rceil])\times(P_{X}^{n}\circ P_{Y^{(n)}|X^{(n)}}),
\]
i.e., $M\sim\mathrm{Unif}([\lceil2^{n\mathsf{R}}\rceil])$ is independent of $(X^{(n)},Y^{(n)})\sim P_{X}^{n}\circ P_{Y^{(n)}|X^{(n)}}$. The encoder observes $X^{(n)}$, finds 
\[
(M,Y^{(n)})=\mathrm{argmin}_{m,y^{n}}T_{m,y^{n}}/P_{M,Y^{(n)}|X^{(n)}}(m,y^{n}|X^{(n)}),
\]
and sends $M$. Note that $P_{M,Y^{(n)}|X^{(n)}}(m,y^{n}|X^{(n)})=P_{M}(m)P_{Y^{(n)}|X^{(n)}}(y^{n}|X^{(n)})$. The decoder observes $M$ and computes 
\[
(\hat{M},\hat{Y}^{(n)})=\mathrm{argmin}_{m,y^{n}}T_{m,y^{n}}/P_{M,Y^{(n)}|M}(m,y^{n}|M).
\]
Note that $P_{M,Y^{(n)}|M}(m,y^{n}|M)=\mathbf{1}\{m=M\}P_{Y^{(n)}}(y^{n})$.

\textbf{2) Computing the asymptotic deviation of $(\mathbf{X},\mathbf{Y})$.}  By Proposition (\ref{prop:gauss_iid}), $\mathbf{X}$ can be coupled to be type deviation convergent with center $P_{X}$ and asymptotic deviation $G_{\mathbf{X}}\sim\mathrm{NM}(P_{X})$. By Proposition \ref{prop:acc_product}, $(\mathbf{X},\mathbf{Y})$ is type deviation convergent with center $P_{X}\circ P_{Y|X}$ and asymptotic deviation
\begin{equation}
G_{\mathbf{X},\mathbf{Y}}=G_{\mathbf{X}}\circ P_{Y|X}+P_{X}\circ\zeta_{Y|X}(G_{\mathbf{X}}).\label{eq:sc_GXY}
\end{equation}

\textbf{3) Error bound via Poisson matching lemma and average properties.} We slightly relax the error condition and consider the probability that $d(X^{(n)},\hat{Y}^{(n)})=\langle\hat{P}(X^{(n)},\hat{Y}^{(n)}),\,d\rangle>\mathsf{D}+\delta/n$, where $\delta>0$ will be specified later.  We have
\begin{align}
P_{e} & =\mathbb{P}\left(\left\langle \hat{P}(X^{(n)},\hat{Y}^{(n)}),\,d\right\rangle >\mathsf{D}+\delta/n\right)\nonumber \\
 & \le\mathbb{P}\left(\left\langle \hat{P}(X^{(n)},Y^{(n)}),\,d\right\rangle >\mathsf{D}+\delta/n\;\mathrm{or}\;Y^{(n)}\neq\hat{Y}^{(n)}\right)\nonumber \\
 & \stackrel{(a)}{\le}\mathbb{P}\bigg(\left\langle \hat{P}(X^{(n)},Y^{(n)}),\,d\right\rangle >\mathsf{D}+\delta/n\;\mathrm{or}\nonumber \\
 & \;\;\;\;\;\iota(M,Y^{(n)};X^{(n)})-\iota(M,Y^{(n)};M)>-\frac{\log n}{2}\bigg)+O(n^{-1/2})\nonumber \\
 & \stackrel{(b)}{\le}\mathbb{P}\bigg(\left\langle P_{X,Y},d\right\rangle +n^{-1/2}\left\langle G_{\mathbf{X},\mathbf{Y}},d\right\rangle +O(n^{-1})>\mathsf{D}+\delta/n\;\mathrm{or}\;\nonumber \\
 & \quad\left\langle P_{X,Y},\iota_{X;Y}\right\rangle +n^{-1/2}\left\langle G_{\mathbf{X},\mathbf{Y}},\iota_{X;Y}\right\rangle -\mathsf{R}>-O\Big(\frac{\log n}{n}\Big)\bigg)+O(n^{-1/2})\nonumber \\
 & \stackrel{(c)}{\le}\mathbb{P}\left(\left\langle G_{\mathbf{X},\mathbf{Y}},\,d\right\rangle >0\;\mathrm{or}\;\left\langle G_{\mathbf{X},\mathbf{Y}},\iota_{X;Y}\right\rangle >\mathsf{W}\right)+O(n^{-1/2}),\label{eq:sc_bd_On}
\end{align}
where the constants in $O(\cdots)$ depend only on the distribution of $(\mathbf{X},\mathbf{Y})$, (a) is by the Poisson matching lemma (Lemma \ref{lem:pml}), (b) is by (\ref{eq:f_avg}) and (\ref{eq:i_avg}), and (c) is by $\left\langle P_{X,Y},d\right\rangle =\mathsf{D}$, $\left\langle P_{X,Y},\iota_{X;Y}\right\rangle =I(X;Y)=\mathsf{R}(\mathsf{D})$, taking $\delta$ to be the constant in the $O(n^{-1})$ term, and taking
\[
\mathsf{R}=\mathsf{R}(\mathsf{D})+\mathsf{W}/\sqrt{n}+O((\log n)/n),
\]
where $\mathsf{W}$ will be specified later. 

\textbf{4) Simplification via Gaussian vector manipulation.} The problem is now reduced to bounding the probability that the Gaussian vector $G_{\mathbf{X},\mathbf{Y}}$ violates any of the constraints $\left\langle G_{\mathbf{X},\mathbf{Y}},\,d\right\rangle \le0$ (the distortion constraint) and $\left\langle G_{\mathbf{X},\mathbf{Y}},\iota_{X;Y}\right\rangle \le\mathsf{W}$ (the decodability constraint). Although we can leave the answer as an optimization problem over all choices of $\zeta_{Y|X}$, we can further simplify the answer via some simple linear algebra manipulations.

By the optimality of $P_{Y|X}$, it also minimizes $I(X;Y)+\lambda\mathbb{E}[d(X,Y)]$. If $(X,Y)\sim P_{X}\circ(P_{Y|X}+tV_{Y|X})$ (where $t\in\mathbb{R}$) for $V_{Y|X}\in\mathrm{Tan}(P_{Y|X})$, by (\ref{eq:H_derivative}),
\begin{align}
 & \frac{\mathrm{d}}{\mathrm{d}t}\left(I(X;Y)+\lambda\mathbb{E}[d(X,Y)]\right)\Big|_{t=0}\nonumber \\
 & =\left\langle P_{X}\circ V_{Y|X},\,\iota_{X;Y}+\lambda d\right\rangle .\label{eq:sc_dV}
\end{align}
Hence, the above must be $0$ for every $V_{Y|X}\in\mathrm{Tan}(P_{Y|X})$. Letting 
\begin{align*}
J & :=\left\langle G_{\mathbf{X}}\circ P_{Y|X},\iota_{X;Y}\right\rangle ,\\
D & :=\left\langle G_{\mathbf{X}}\circ P_{Y|X},\,d\right\rangle ,\\
K & :=\left\langle P_{X}\circ\zeta_{Y|X}(G_{\mathbf{X}}),\,d\right\rangle ,
\end{align*}
we have $\left\langle P_{X}\circ\zeta_{Y|X}(G_{\mathbf{X}}),\,\iota_{X;Y}\right\rangle =-\lambda K$ by (\ref{eq:sc_dV}). By (\ref{eq:sc_GXY}),
\begin{align}
 & \mathbb{P}\left(\left\langle G_{\mathbf{X},\mathbf{Y}},\,d\right\rangle >0\;\mathrm{or}\;\left\langle G_{\mathbf{X},\mathbf{Y}},\iota_{X;Y}\right\rangle >\mathsf{W}\right)\nonumber \\
 & =\mathbb{P}\left(D+K>0\;\mathrm{or}\;J-\lambda K>\mathsf{W}\right)\nonumber \\
 & \ge\mathbb{P}\left(\lambda(D+K)+J-\lambda K>\mathsf{W}\right)\label{eq:sc_ineq}\\
 & =\mathbb{P}\left(J+\lambda D>\mathsf{W}\right),
\end{align}
where the inequality in (\ref{eq:sc_ineq}) is an equality when $D+K$ and $J-\lambda K-\mathsf{W}$ are nonnegative constant multiples of the same random variable, for example, when $K=-D$ or $K=(J-\mathsf{W})/\lambda$ (since $\zeta_{Y|X}$ is arbitrary, we can choose $\zeta_{Y|X}$ to make $K$ any Lipschitz function of $(J,D)$).\footnote{The only exception is when $\left\langle P_{X}\circ V_{Y|X},\,d\right\rangle =0$ for every $V_{Y|X}\in\mathrm{Tan}(P_{Y|X})$, which forces $K=0$. Note that (\ref{eq:d_tilted}) implies that $P_{Y|X}(y|x)>0$ whenever $P_{X}(x)>0$, $P_{Y}(y)>0$ (we only consider finite $d(x,y)$ here). Hence, if $\left\langle P_{X}\circ V_{Y|X},\,d\right\rangle =0$ for every $V_{Y|X}\in\mathrm{Tan}(P_{Y|X})$, this implies that $d(x,y)$ is only a function of $x$. This means that no communication is required, and the theorem is clearly true.} Taking
\begin{align*}
\mathsf{V} & =\mathrm{Var}\left[J+\lambda D\right]=\mathrm{Var}\left[\jmath_{X,\mathsf{D}}(X)\right],
\end{align*}
\[
\mathsf{W}=\sqrt{\mathsf{V}}\mathcal{Q}^{-1}\left(\epsilon-\frac{\gamma}{\sqrt{n}}\right)=\sqrt{\mathsf{V}}\mathcal{Q}^{-1}(\epsilon)-O(n^{-1/2}),
\]
where $\gamma$ is chosen be the constant in the $O(n^{-1/2})$ term in (\ref{eq:sc_bd_On}), we have $P_{e}\le\epsilon$ by (\ref{eq:sc_bd_On}). 

\textbf{5) Minor technical steps.}  Although the codebook is random, we can fix a particular codebook $(T_{m,y^{n}})_{m,y^{n}}$ that minimizes $P_{e}$. To complete the proof, we need to strengthen $\left\langle \hat{P}(X^{(n)},\hat{Y}^{(n)}),\,d\right\rangle \le\mathsf{D}+\delta/n$ to $\left\langle \hat{P}(X^{(n)},\hat{Y}^{(n)}),\,d\right\rangle \le\mathsf{D}$. We can do so by concatenating $Y^{(n)}$ with $k$ symbols $Y_{n+1},\ldots,Y_{n+k}$, where $Y_{n+i}=\mathrm{argmin}_{y}d(X_{n+i},y)$. Recall that $\mathsf{D}>\mathsf{D}_{\min}:=\min_{P_{Y|X}}\mathbb{E}[d(X,Y)]=\mathbb{E}[d(X_{n+i},Y_{n+i})]$. By Cram\'{e}r's theorem, $\mathbb{P}(k^{-1}\sum_{i=1}^{k}d(X_{n+i},Y_{n+i})>(\mathsf{D}+\mathsf{D}_{\min})/2)$ decays exponentially with $k$, and hence we can make this probability $O(n^{-1/2})$ by taking $k=O(\log n)$. As long as $k^{-1}\sum_{i=1}^{k}d(X_{n+i},Y_{n+i})\le(\mathsf{D}+\mathsf{D}_{\min})/2$, the average cost is reduced by $O(k/n)=O(\log n/n)$, which is more than enough to cancel out the $\delta/n$ term. Transmitting $Y_{n+1},\ldots,Y_{n+k}$ requires $O(k)=O(\log n)$ bits, incurring only a $O((\log n)/n)$ increase in $\mathsf{R}$. 

\end{IEEEproof}
\medskip{}

This result can be generalized to the indirect or noisy lossy source coding \cite{dobrushin1962information} where only a noisy version of the source is observed at the encoder. We will show in Section \ref{sec:noisy_wz} that our approach can recover the optimal second order result \cite{kostina2016nonasymptotic}.

\medskip{}

\section{Second-Order Wyner-Ziv Coding\label{sec:wz}}

\subsection{Second-Order Achievability for Wyner-Ziv\label{subsec:wz_statement}}

We now prove a new second-order achievability result for lossy source coding with side information at the decoder, also known as the Wyner-Ziv problem \cite{wyner1976ratedistort,wyner1978rate}, which improves upon existing results \cite{verdu2012nonasymp,yassaee2013oneshot,watanabe2015nonasymp,li2021unified,liu2024one}. In this setting, there is a $2$-discrete memoryless source $(X^{(n)},Y^{(n)})\sim P_{X,Y}^{n}$. The encoder encodes $X^{(n)}$ into a message $M\in[\lceil2^{n\mathsf{R}}\rceil]$ with rate $\mathsf{R}>0$. The decoder observes $M$ and the side information $Y^{(n)}$, and outputs $\hat{Z}^{(n)}\in\mathcal{Z}^{n}$. The goal is to minimize the probability of excess distortion
\[
P_{e}:=\mathbb{P}\big(d(X^{(n)},\hat{Z}^{(n)})>\mathsf{D}\big),
\]
where $d:\mathcal{X}\times\mathcal{Z}\to\mathbb{R}$ is a distortion function, $d(X^{(n)},\hat{Z}^{(n)}):=n^{-1}\sum_{i=1}^{n}d(X_{i}^{(n)},\hat{Z}_{i}^{(n)})$, and $\mathsf{D}\in\mathbb{R}$ is the allowed distortion level. The optimal asymptotic rate needed to have $P_{e}\to0$ is given by Wyner-Ziv theorem \cite{wyner1976ratedistort,wyner1978rate}
\begin{equation}
\mathsf{R}(\mathsf{D}):=\min_{P_{U|X},z:\,\mathbb{E}[d(X,Z)]\le\mathsf{D}}\left(I(U;X)-I(U;Y)\right),\label{eq:wz_rate}
\end{equation}
where the minimum is over $P_{U|X}$ and functions $z:\mathcal{U}\times\mathcal{Y}\to\mathcal{Z}$, subject to the constraint that $\mathbb{E}[d(X,Z)]\le\mathsf{D}$ where $(X,Y,U)\sim P_{X,Y}\circ P_{U|X}$ and $Z=z(U,Y)$. We now prove a second-order result which improves upon existing results \cite{verdu2012nonasymp,watanabe2015nonasymp,yassaee2013oneshot,li2021unified,liu2025one} (see Section \ref{subsec:wz_compare} for a comparison).

\medskip{}

\begin{thm}
\label{thm:wz}For discrete Wyner-Ziv coding, assume these two conditions are satisfied: 1) $\mathsf{D}$ is a value such that $\lambda:=-\mathrm{d}\mathsf{R}(\mathsf{D})/\mathrm{d}\mathsf{D}>0$ is finite at $\mathsf{D}$; and 2) letting $(P_{U|X},z)$ be a minimizer in $\mathsf{R}(\mathsf{D})$, $(X,Y,U)\sim P_{X,Y}\circ P_{U|X}$ and $Z=z(U,Y)$, they satisfy $\mathbb{E}[\mathrm{Var}[\mathbb{E}[d(X,Z)|X,U]|X]]>0$. Fix any $\mathsf{W}>0$. For any large enough $n$, there is a scheme with rate
\begin{equation}
\mathsf{R}=\mathsf{R}(\mathsf{D})+\frac{\mathsf{W}}{\sqrt{n}}+O\left(\frac{\log n}{n}\right),\label{eq:wz_R}
\end{equation}
achieving a probability of excess distortion 
\[
P_{e}\le\mathbb{E}\left[P_{e}^{*}(\mathsf{W}-A_{\mathbf{X}})\right],
\]
where $A_{\mathbf{X}}\in\mathbb{R}$ is a zero mean Gaussian random variable with variance
\[
\mathrm{Var}\left[\mathbb{E}\left[\left.\iota_{U;X}(U;X)-\iota_{U;Y}(U;Y)+\lambda d(X,Z)\,\right|\,X\right]\right],
\]
and $P_{e}^{*}:\mathbb{R}\to\mathbb{R}$ is defined as
\begin{equation}
P_{e}^{*}(\alpha):=\min_{t\in\mathbb{R}}\mathbb{P}\left(D_{\mathbf{Y}}>t\;\mathrm{or}\;J_{\mathbf{Y}}>\alpha-\lambda t\right),\label{eq:wz_Pe_a}
\end{equation}
with $[J_{\mathbf{Y}},D_{\mathbf{Y}}]\in\mathbb{R}^{2}$ being a zero mean Gaussian vector with covariance matrix 
\begin{equation}
\mathbb{E}\left[\mathrm{Var}\left[\left.\left[\begin{array}{c}
-\iota_{U;Y}(U;Y)\\
d(X,Z)
\end{array}\right]\,\right|\,X,U\right]\right].\label{eq:wz_noise_var}
\end{equation}
\end{thm}
\medskip{}

To obtain a simpler but looser bound, applying the union bound on (\ref{eq:wz_Pe_a}) gives the following corollary. This corollary subsumes the second-order result for lossy source coding in Theorem \ref{thm:sc}.

\medskip{}

\begin{cor}
\label{cor:wz_cor-1}For discrete Wyner-Ziv coding, under the same assumptions as Theorem \ref{thm:wz}, for any fixed $0<\epsilon<1$ and any large enough $n$, there is a scheme achieving a probability of excess distortion $P_{e}\le\epsilon$, with rate
\begin{equation}
\mathsf{R}=\mathsf{R}(\mathsf{D})+\sqrt{\frac{\mathsf{V}_{\mathrm{GCC}}(\mathsf{D})}{n}}\mathcal{Q}^{-1}(\epsilon/2)+O\left(\frac{\log n}{n}\right),\label{eq:wz_cor_R-1}
\end{equation}
where 
\begin{align*}
 & \mathsf{V}_{\mathrm{GCC}}(\mathsf{D})\\
 & :=\mathrm{Var}\left[\mathbb{E}\left[\left.\iota(U;X)-\iota(U;Y)+\lambda d(X,Z)\,\right|\,X\right]\right]\\
 & \;\;+\left(\sqrt{\mathbb{E}\left[\mathrm{Var}\left[\left.\iota(U;Y)\right|X,U\right]\right]}+\lambda\sqrt{\mathbb{E}\left[\mathrm{Var}\left[\left.d(X,Z)\right|X,U\right]\right]}\right)^{2}.
\end{align*}
The $\mathcal{Q}^{-1}(\epsilon/2)$ term can be improved to $\mathcal{Q}^{-1}(\epsilon)$ if $d(X,Z)$ is a function of $(X,U)$, e.g., if $U=Z$.
\end{cor}
\medskip{}

Let $\mathsf{R}(\mathsf{D},n,\epsilon)$ be the infimum of achievable rates when the blocklength is $n$ and the probability of excess distortion is upper-bounded by $\epsilon$, and
\begin{equation}
\mathsf{V}^{*}(\mathsf{D}):=\lim_{\epsilon\to0}\underset{n\to\infty}{\mathrm{limsup}}\frac{n\left(\mathsf{R}(\mathsf{D},n,\epsilon)-\mathsf{R}(\mathsf{D})\right)^{2}}{-2\ln\epsilon}\label{eq:wz_op_disp}
\end{equation}
be the operational dispersion (defined similarly as \cite{polyanskiy2009dispersion,tan2013dispersions}). Corollary \ref{cor:wz_cor-1} gives an upper bound 
\[
\mathsf{V}^{*}(\mathsf{D})\le\mathsf{V}_{\mathrm{GCC}}(\mathsf{D}).
\]

\begin{figure*}
\begin{centering}
\includegraphics[scale=0.9]{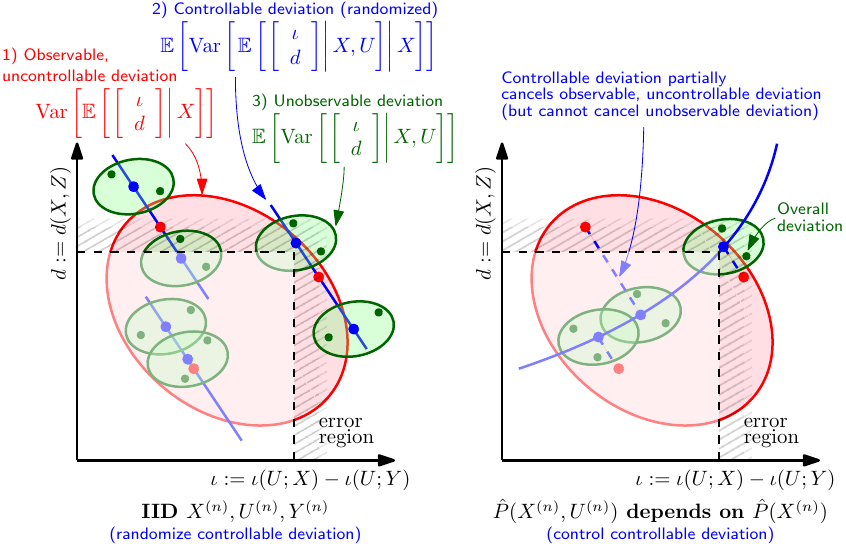}
\par\end{centering}
\caption{\label{fig:wz}Left: Illustration for Wyner-Ziv coding with i.i.d. $X^{(n)},U^{(n)}$. The red dots, blue dots and green dots are samples of $\mathbb{E}[[\iota,d]^{\top}|X]$, $\mathbb{E}[[\iota,d]^{\top}|X,U]$ and $[\iota,d]^{\top}$, respectively (the 3 stages in Section \ref{subsec:wz_statement}). Right: The optimal scheme where we control the deviation of the type of $U^{(n)}$ according to the type of $X^{(n)}$, moving the red dots to the blue dots along the blue curve.}
\end{figure*}

We give an intuitive explanation of Theorem \ref{thm:wz}. Loosely speaking, there are two error events: excess distortion $d(X^{(n)},Z^{(n)})>\mathsf{D}$, and excess information $n^{-1}(\iota(U^{(n)};X^{(n)})-\iota(U^{(n)};Y^{(n)}))>\mathsf{R}$ (the $n\mathsf{R}$-bit message is insufficient to store $U^{(n)}$). In case if $(X^{(n)},U^{(n)})\stackrel{\mathrm{iid}}{\sim}P_{X}\circ P_{U|X}$, then by the central limit theorem, the vector $[n^{-1}(\iota(U^{(n)};X^{(n)})-\iota(U^{(n)};Y^{(n)})),d(X^{(n)},Z^{(n)})]^{\top}$ will be approximately Gaussian with covariance matrix
\begin{align}
 & \frac{1}{n}\mathrm{Var}\left[\left[\begin{array}{c}
\iota\\
d
\end{array}\right]\right]\nonumber \\
 & =\frac{1}{n}\mathrm{Var}\left[\mathbb{E}\left[\left.\!\left[\begin{array}{c}
\!\iota\!\\
\!d\!
\end{array}\right]\right|X\right]\right]+\frac{1}{n}\mathbb{E}\left[\mathrm{Var}\left[\left.\mathbb{E}\left[\left.\!\left[\begin{array}{c}
\!\iota\!\\
\!d\!
\end{array}\right]\right|X,U\right]\right|X\right]\right]\nonumber \\
 & \quad+\frac{1}{n}\mathbb{E}\left[\mathrm{Var}\left[\left.\left[\begin{array}{c}
\iota\\
d
\end{array}\right]\right|X,U\right]\right],\label{eq:wz_gauss}
\end{align}
due to the law of total covariance, where $\iota:=\iota(U;X)-\iota(U;Y)$, $d:=d(X,Z)$. The first, second and third term above are the contribution of the randomness in $X^{(n)}$, $U^{(n)}$ and $Y^{(n)}$, respectively. Hence, we can think of the generation $[\iota,d]^{\top}$ as a 3-stage process: 1) generate $\mathbb{E}[[\iota,d]^{\top}|X]$ with covariance matrix given by the first term in (\ref{eq:wz_gauss}) (the observable but uncontrollable component of the deviation, contributed by $X^{(n)}$); 2) obtain $\mathbb{E}[[\iota,d]^{\top}|X,U]$ by adding a noise to $\mathbb{E}[[\iota,d]^{\top}|X]$, with covariance matrix given by the second term (the controllable deviation by $U^{(n)}$); and 3) obtain $[\iota,d]^{\top}$ by adding a noise with covariance matrix given by the third term (the deviation unobservable by the encoder, contributed by $Y^{(n)}$, which appears in (\ref{eq:wz_noise_var})). Error occurs if one of the two components of the random vector is too large; see Figure \ref{fig:wz} (left). The second term is rank-one since $\mathbb{E}[\iota+\lambda d|X,U]$ depends only on $X$ by the first-order optimality of $P_{U|X}$.

To reduce the error probability, note that the encoder has full control of $U^{(n)}$, and there is no reason to randomize $U^{(n)}$ and increase the error probability. The second term in (\ref{eq:wz_gauss}) can be eliminated. Moreover, we can control the deviation of the type of $U^{(n)}$ according to the type of $X^{(n)}$. Controlling $U$ can only allow trading off $\mathbb{E}[\iota|X,U]$ and $\mathbb{E}[d|X,U]$ along a diagonal line with slope $-1/\lambda$. We move the point $\mathbb{E}[[\iota,d]^{\top}|X]$ along this diagonal line according to (\ref{eq:wz_Pe_a}) in order to minimize the probability that the randomness of $Y^{(n)}$ (noise in the third stage with covariance (\ref{eq:wz_noise_var})) would push that point into the error region. For example, we can moving the point to the blue curve in Figure \ref{fig:wz} (right). Also, the blue curve not being a straight line implies that $[\iota,d]^{\top}$ has a non-Gaussian deviation. This is generally necessary to attain the minimum in (\ref{eq:wz_Pe_a}).

\smallskip{}

We now prove Theorem \ref{thm:wz} and Corollary \ref{cor:wz_cor-1}.
\begin{IEEEproof}
The proof is divided into four steps.

\textbf{1) Code construction.} Consider the $P_{U|X}$, function $z:\mathcal{U}\times\mathcal{Y}\to\mathcal{Z}$ that attains $\mathsf{R}(\mathsf{D})$. Consider a GCC channel $(P_{U^{(n)}|X^{(n)}})_{n}$ with center $P_{X}P_{U|X}$ and Lipschitz deviation function $\zeta_{U|X}:\mathrm{Tan}(P_{X})\to\mathrm{Tan}(P_{U|X})$ to be specified later. Consider the joint distribution
\[
(M,X^{(n)},U^{(n)},Y^{(n)})\sim\mathrm{Unif}([\lceil2^{n\mathsf{R}}\rceil])\times(P_{X}^{n}\circ P_{U^{(n)}|X^{(n)}}\circ P_{Y|X}^{n}).
\]
Let $T_{m,u^{n}}\sim\mathrm{Exp}(1)$, i.i.d. across $m\in[\lfloor2^{n\mathsf{R}}\rfloor]$, $u^{n}\in\mathcal{U}^{n}$, which serves as a random codebook available to the encoder and the decoder. The encoder observes $X^{(n)}$, find 
\[
(M,U^{(n)})=\mathrm{argmin}_{m,u^{n}}T_{m,u^{n}}/P_{M,U^{(n)}|X^{(n)}}(m,u^{n}|X^{(n)}),
\]
and sends $M$. The decoder observes $M,Y^{(n)}$ and computes
\[
(\hat{M},\hat{U}^{(n)})=\mathrm{argmin}_{m,u^{n}}T_{m,u^{n}}/P_{M,U^{(n)}|M,Y^{(n)}}(m,u^{n}|M,Y^{(n)}).
\]

\textbf{2) Computing the asymptotic deviation of $(\mathbf{X},\mathbf{U},\mathbf{Y})$.} By Proposition (\ref{prop:gauss_iid}), $\mathbf{X}$ can be coupled to be type deviation convergent with center $P_{X}$ and asymptotic deviation $G_{\mathbf{X}}\sim\mathrm{NM}(P_{X})$. By Proposition \ref{prop:acc_product}, $(\mathbf{X},\mathbf{U})$ is type deviation convergent with center $P_{X}P_{U|X}$ and asymptotic deviation
\begin{equation}
G_{\mathbf{X},\mathbf{U}}=G_{\mathbf{X}}\circ P_{U|X}+P_{X}\circ\zeta_{U|X}(G_{\mathbf{X}}).\label{eq:sc_GXY-1-1}
\end{equation}
By Proposition \ref{prop:gauss_memoryless}, $(\mathbf{X},\mathbf{U},\mathbf{Y})$ can be coupled to be type deviation convergent with center $P_{X}P_{U|X}P_{Y|X}$ and asymptotic deviation
\begin{align}
G_{\mathbf{X},\mathbf{U},\mathbf{Y}} & =(G_{\mathbf{X}}\circ P_{U|X}+P_{X}\circ\zeta_{U|X}(G_{\mathbf{X}}))\circ P_{Y|X}\nonumber \\
 & \quad+\sqrt{P_{X}\circ P_{U|X}}\circ G_{\mathbf{Y}|\mathbf{X},\mathbf{U}},\label{eq:wz_GXUY}
\end{align}
where $G_{\mathbf{Y}|\mathbf{X},\mathbf{U}}\sim\mathrm{NM}(P_{Y|X,U})$ independent of $G_{\mathbf{X}}$. 

\textbf{3) Error bound.} We slightly relax the error condition and consider the probability that $d(X^{(n)},\hat{Z}^{(n)})=\langle\hat{P}(X^{(n)},\hat{U}^{(n)},Y^{(n)}),\,d\rangle>\mathsf{D}+\delta/n$, where we let $d(x,u,y)=d(x,z(u,y))$, and $\delta>0$ will be specified later. Using the Poisson matching lemma in a similar manner as in the proof of Theorem \ref{thm:sc},
\begin{align}
P_{e} & =\mathbb{P}\left(\langle\hat{P}(X^{(n)},\hat{U}^{(n)},Y^{(n)}),\,d\rangle>\mathsf{D}+\delta/n\right)\nonumber \\
 & \le\mathbb{P}\left(\left\langle G_{\mathbf{X},\mathbf{U},\mathbf{Y}},\,d\right\rangle >0\;\mathrm{or}\;\left\langle G_{\mathbf{X},\mathbf{U},\mathbf{Y}},\iota_{U;X}-\iota_{U;Y}\right\rangle >\mathsf{W}\right)+O(n^{-1/2}),\label{eq:wz_bd_On}
\end{align}
with a suitable $\delta$ (see Theorem \ref{thm:sc}), by taking
\[
\mathsf{R}=\mathsf{R}(\mathsf{D})+\mathsf{W}/\sqrt{n}+O((\log n)/n).
\]

\textbf{4) Simplification via Gaussian vector manipulation.} The problem is now reduced to bounding the probability that the Gaussian vector $G_{\mathbf{X},\mathbf{U},\mathbf{Y}}$ violates any of the constraints in (\ref{eq:wz_bd_On}). Although we can leave the answer as an optimization problem over all choices of $\zeta_{Y|X}$, we can further simplify the answer. Let $\lambda:=-\mathrm{d}\mathsf{R}(\mathsf{D})/\mathrm{d}\mathsf{D}$. We now consider the terms in (\ref{eq:wz_GXUY}). Let
\begin{align*}
J_{\mathbf{X}} & :=\langle G_{\mathbf{X}}\circ P_{U|X}\circ P_{Y|X},\,\iota_{U;X}-\iota_{U;Y}\rangle,\\
D_{\mathbf{X}} & :=\langle G_{\mathbf{X}}\circ P_{U|X}\circ P_{Y|X},\,d\rangle,\\
A_{\mathbf{X}} & :=J_{\mathbf{X}}+\lambda D_{\mathbf{X}},\\
J_{\mathbf{U}} & :=\langle P_{X}\circ\zeta_{U|X}(G_{\mathbf{X}})\circ P_{Y|X},\,\iota_{U;X}-\iota_{U;Y}\rangle,\\
D_{\mathbf{U}} & :=\langle P_{X}\circ\zeta_{U|X}(G_{\mathbf{X}})\circ P_{Y|X},\,d\rangle,\\
J_{\mathbf{Y}} & :=\langle\sqrt{P_{X}\circ P_{U|X}}\circ G_{\mathbf{Y}|\mathbf{X},\mathbf{U}},\,\iota_{U;X}-\iota_{U;Y}\rangle,\\
D_{\mathbf{Y}} & :=\langle\sqrt{P_{X}\circ P_{U|X}}\circ G_{\mathbf{Y}|\mathbf{X},\mathbf{U}},\,d\rangle.
\end{align*}
Note that $J_{\mathbf{Y}}$, $D_{\mathbf{Y}}$, $A_{\mathbf{X}}$ follow the distribution stated in the theorem by Proposition \ref{prop:nm_prop}, and $(J_{\mathbf{Y}},D_{\mathbf{Y}})$ is independent of $A_{\mathbf{X}}$. By the optimality of $P_{U|X}$, using the same arguments as Theorem \ref{thm:sc}, for any $V_{U|X}\in\mathrm{Tan}(P_{U|X})$,
\begin{equation}
\left\langle P_{X}\circ V_{U|X}\circ P_{Y|X},\,\iota_{U;X}-\iota_{U;Y}+\lambda d\right\rangle =0.\label{eq:VUX_sum}
\end{equation}
Since $\zeta_{U|X}(G_{\mathbf{X}})\in\mathrm{Tan}(P_{U|X})$, we have $J_{\mathbf{U}}=-\lambda D_{\mathbf{U}}$. Hence, the probability in (\ref{eq:wz_bd_On}) is
\begin{align}
 & \mathbb{P}\left(\left\langle G_{\mathbf{X},\mathbf{U},\mathbf{Y}},\,d\right\rangle >0\;\mathrm{or}\;\left\langle G_{\mathbf{X},\mathbf{U},\mathbf{Y}},\iota_{U;X}-\iota_{U;Y}\right\rangle >\mathsf{W}\right)\nonumber \\
 & =\mathbb{E}\left[\mathbb{P}\left(D_{\mathbf{Y}}+D_{\mathbf{X}}+D_{\mathbf{U}}>0\;\mathrm{or}\;J_{\mathbf{Y}}+J_{\mathbf{X}}+J_{\mathbf{U}}>\mathsf{W}\big|A_{\mathbf{X}}\right)\right]\nonumber \\
 & \ge\mathbb{E}\left[P_{e}^{*}(\mathsf{W}-A_{\mathbf{X}})\right],\label{eq:wz_Pe_lb}
\end{align}
where
\[
P_{e}^{*}(\alpha):=\min_{t\in\mathbb{R}}\mathbb{P}\left(D_{\mathbf{Y}}-t>0\;\mathrm{or}\;J_{\mathbf{Y}}-\alpha+\lambda t>0\right).
\]
The inequality (\ref{eq:wz_Pe_lb}) is shown by considering $t=-D_{\mathbf{X}}-D_{\mathbf{U}}$, which makes $J_{\mathbf{Y}}+J_{\mathbf{X}}+J_{\mathbf{U}}=J_{\mathbf{Y}}+A_{\mathbf{X}}+\lambda t$. We now discuss how to make (\ref{eq:wz_Pe_lb}) hold with equality. Letting $G_{U|X}\sim\mathrm{NM}(P_{U|X})$, by Proposition \ref{prop:nm_prop}, $\langle P_{X}\circ G_{U|X}\circ P_{Y|X},\,d\rangle$ is zero-mean Gaussian with variance $\mathbb{E}[\mathrm{Var}[\mathbb{E}[d(X,Z)|X,U]|X]]>0$ by the assumption of Theorem \ref{thm:wz}. ,  Hence, there exists a fixed $V_{U|X}\in\mathrm{Tan}(P_{U|X})$ with $\langle P_{X}\circ V_{U|X}\circ P_{Y|X},\,d\rangle=1$. We take
\[
\zeta_{U|X}(G_{\mathbf{X}})=\left(-\psi(\mathsf{W}-A_{\mathbf{X}})-D_{\mathbf{X}}\right)\cdot V_{U|X},
\]
where $\psi:\mathbb{R}\to\mathbb{R}$ is a Lipschitz function to be specified later. We have $-D_{\mathbf{X}}-D_{\mathbf{U}}=\psi(\mathsf{W}-A_{\mathbf{X}})$. Hence, (\ref{eq:wz_Pe_lb}) hold with equality with 
\[
\psi(\alpha)=\underset{t\in\mathbb{R}}{\mathrm{argmin}}\mathbb{P}\left(D_{\mathbf{Y}}-t>0\;\mathrm{or}\;J_{\mathbf{Y}}-\alpha+\lambda t>0\right).
\]
 The technical proof that $\psi$ is Lipschitz is in Appendix \ref{subsec:pf_Lipschitz}. Note that $\psi$ is not affine, and hence $G_{\mathbf{X},\mathbf{U},\mathbf{Y}}$ is non-Gaussian. The remainder of the proof is similar to Theorem \ref{thm:sc}, and is omitted. 

For Corollary \ref{cor:wz_cor-1}, by union bound, $P_{e}^{*}(\alpha,t)\le\mathbb{P}(D_{\mathbf{Y}}>t)+\mathbb{P}(J_{\mathbf{Y}}>\alpha-\lambda t)$. Choosing $t=\tilde{\psi}(\alpha):=\alpha\sigma_{D_{\mathbf{Y}}}/(\sigma_{J_{\mathbf{Y}}}+\lambda\sigma_{D_{\mathbf{Y}}})$ (where $\sigma_{D_{\mathbf{Y}}}:=\sqrt{\mathrm{Var}[D_{\mathbf{Y}}]}$) makes these two probabilities equal, giving $P_{e}^{*}(\alpha,t)\le2\mathcal{Q}(\alpha/(\sigma_{J_{\mathbf{Y}}}+\lambda\sigma_{D_{\mathbf{Y}}}))$. The bound in the corollary follows immediately.  In case if $\sigma_{D_{\mathbf{Y}}}=0$ (e.g., if $d(X,Z)$ is a function of $(X,U)$), we have $t=0$ and $\mathbb{P}(D_{\mathbf{Y}}>t)=0$, so we have $P_{e}^{*}(\alpha,t)\le\mathcal{Q}(\alpha/(\sigma_{J_{\mathbf{Y}}}+\lambda\sigma_{D_{\mathbf{Y}}}))$ instead. 

\end{IEEEproof}

\medskip{}

\begin{rem}
We conjecture that the second-order term in Theorem \ref{thm:wz} is optimal for Wyner-Ziv coding. The informal reason is that the first-order optimality of (\ref{eq:wz_rate}) suggests that a scheme achieving the rate $\mathsf{R}(\mathsf{D})$ should be close to the random coding scheme with auxiliary $U$. The proof of Theorem \ref{thm:wz} is optimizing over all schemes with type deviation convergent auxiliaries where the deviation from the random coding scheme is small, that is, we have searched over all possible small perturbations of the first-order optimal scheme.

Nevertheless, proving the optimality requires a second-order converse, which is generally difficult for coding theorems involving auxiliary random variables (e.g., see \cite{watanabe2016second,zhou2023finite}). Considering that Theorem \ref{thm:wz} is in a different form compared to any previous second-order result, its converse will likely require new techniques. Since the focus of this paper is the type deviation convergence framework which is purely an achievability technique, converse proofs are out of the scope of this paper. We leave the converse for future research.
\end{rem}
\medskip{}

\subsection{Comparison with Existing Bounds\label{subsec:wz_compare}}

Theorem \ref{thm:wz} and Corollary \ref{cor:wz_cor-1} improve upon the following existing achievability bounds (we assume that the same assumptions in Theorem \ref{thm:wz} hold). 
\begin{itemize}
\item Yassaee-Aref-Gohari \cite{yassaee2013oneshot} (which improves upon \cite{verdu2012nonasymp}): Achieves a rate $\mathsf{R}=\mathsf{R}(\mathsf{D})+\mathsf{W}/\sqrt{n}+O((\log n)/n)$ with probability of excess distortion
\begin{equation}
P_{e}\le\min_{t,\tau\in\mathbb{R}}\mathbb{P}(\bar{J}_{\mathbf{X}}>\mathsf{W}-\lambda t-\tau\;\mathrm{or}\;\bar{J}_{\mathbf{Y}}>\tau\;\mathrm{or}\;\bar{D}>t),\label{eq:wz_Pe_yassaee}
\end{equation}
where $[\bar{J}_{\mathbf{X}},\bar{J}_{\mathbf{Y}},\bar{D}]$ is a zero-mean Gaussian vector with the same covariance matrix as $[\iota(U;X),-\iota(U;Y),d(X,Z)]$. The bounds in \cite{verdu2012nonasymp,yassaee2013oneshot} imply the following upper bound on the operational dispersion (\ref{eq:wz_op_disp}): $\mathsf{V}^{*}(\mathsf{D})\le\mathsf{V}_{\mathrm{VYAG}}(\mathsf{D})$ where
\begin{equation}
\mathsf{V}_{\mathrm{VYAG}}(\mathsf{D}):=\left(\sqrt{\mathrm{Var}\iota(U;X)}+\sqrt{\mathrm{Var}\iota(U;Y)}+\lambda\sqrt{\mathrm{Var}d(X,Z)}\right)^{2}.\label{eq:wz_yassaee}
\end{equation}
Theorem \ref{thm:wz} improves upon (\ref{eq:wz_Pe_yassaee}), and Corollary \ref{cor:wz_cor-1} improves upon (\ref{eq:wz_yassaee}), which follow from the law of total variance. See Appendix \ref{subsec:pf_compare}. 
\end{itemize}
\smallskip{}

\begin{itemize}
\item Watanabe-Kuzuoka-Tan \cite{watanabe2015nonasymp}: Achieves $\mathsf{R}=\mathsf{R}(\mathsf{D})+\mathsf{W}/\sqrt{n}+O((\log n)/n)$ with 
\begin{equation}
P_{e}\le\min_{t,\tau\in\mathbb{R}}\mathbb{P}(\tilde{J}_{\mathbf{X}}>\mathsf{W}-\lambda t-\tau\;\mathrm{or}\;\tilde{J}_{\mathbf{Y}}>\tau\;\mathrm{or}\;\tilde{D}>t),\label{eq:wz_Pe_wkt}
\end{equation}
where $[\tilde{J}_{\mathbf{X}},\tilde{J}_{\mathbf{Y}},\tilde{D}]$ is a zero-mean Gaussian vector with covariance matrix 
\[
\mathbb{E}\big[\mathrm{Var}[[\iota(\tilde{U};X|T),-\iota(\tilde{U};Y|T),d(X,Z)]\,|\,T]\big],
\]
where $(X,Y,T,\tilde{U},Z)\sim P_{X,Y}\circ P_{T}\circ P_{\tilde{U}|Y,T}\circ P_{Z|Y,T,\tilde{U}}$ with $I(\tilde{U};X|T)-I(\tilde{U};Y|T)=\mathsf{R}(\mathsf{D})$ and $\mathbb{E}[d(X,Z)]=\mathsf{D}$. This implies $\mathsf{V}^{*}(\mathsf{D})\le\mathsf{V}_{\mathrm{WKT}}(\mathsf{D})$ where
\begin{align}
\mathsf{V}_{\mathrm{WKT}}(\mathsf{D}):= & \Big(\sqrt{\mathbb{E}\big[\mathrm{Var}[[\iota(\tilde{U};X|T)|T]\big]}+\sqrt{\mathbb{E}\big[\mathrm{Var}[[\iota(\tilde{U};Y|T)|T]\big]}\nonumber \\
 & \;\;\;\;+\lambda\sqrt{\mathbb{E}\big[\mathrm{Var}[[d(X,Z)|T]\big]}\Big)^{2}.\label{eq:wz_wkt}
\end{align}
This bound is tighter than (\ref{eq:wz_Pe_yassaee}), (\ref{eq:wz_yassaee}) due to the inclusion of a time-sharing random variable $T$.  Nevertheless, such a $T$ is unnecessary in Theorem \ref{thm:wz} since $T$ can be absorbed into $U$ there. We can show that Theorem \ref{thm:wz} and Corollary \ref{cor:wz_cor-1} improve upon (\ref{eq:wz_Pe_wkt}) and (\ref{eq:wz_wkt}). See Appendix \ref{subsec:pf_compare}.
\end{itemize}
\smallskip{}

\begin{itemize}
\item Li-Anantharam \cite{li2021unified} (also see \cite{liu2025one}): Achieves $\mathsf{R}=\mathsf{R}(\mathsf{D})+\mathsf{W}/\sqrt{n}+O((\log n)/n)$ with 
\begin{equation}
P_{e}\le\min_{t\in\mathbb{R}}\mathbb{P}(\bar{J}_{\mathbf{X}}+\bar{J}_{\mathbf{Y}}>\mathsf{W}-\lambda t\;\mathrm{or}\;\bar{D}>t),\label{eq:wz_Pe_prev_poisson}
\end{equation}
with the same variables defined in (\ref{eq:wz_yassaee}). This implies $\mathsf{V}^{*}(\mathsf{D})\le\mathsf{V}_{\mathrm{LA}}(\mathsf{D})$ where
\begin{equation}
\mathsf{V}_{\mathrm{LA}}(\mathsf{D}):=\left(\sqrt{\mathrm{Var}[\iota(U;X)-\iota(U;Y)]}+\lambda\sqrt{\mathrm{Var}d(X,Z)}\right)^{2}.\label{eq:wz_prev_poisson}
\end{equation}
This bound is tighter than (\ref{eq:wz_Pe_yassaee}), (\ref{eq:wz_yassaee}). Theorem \ref{thm:wz} and Corollary \ref{cor:wz_cor-1} improve upon (\ref{eq:wz_Pe_prev_poisson}) and (\ref{eq:wz_prev_poisson}). See Appendix \ref{subsec:pf_compare}.
\end{itemize}
\smallskip{}

In Section \ref{subsec:wz_binary}, we will further compare these bounds using a numerical example.

\smallskip{}

\subsection{Binary-Hamming Wyner-Ziv Coding\label{subsec:wz_binary}}

\begin{figure*}
\begin{centering}
\includegraphics[scale=0.52]{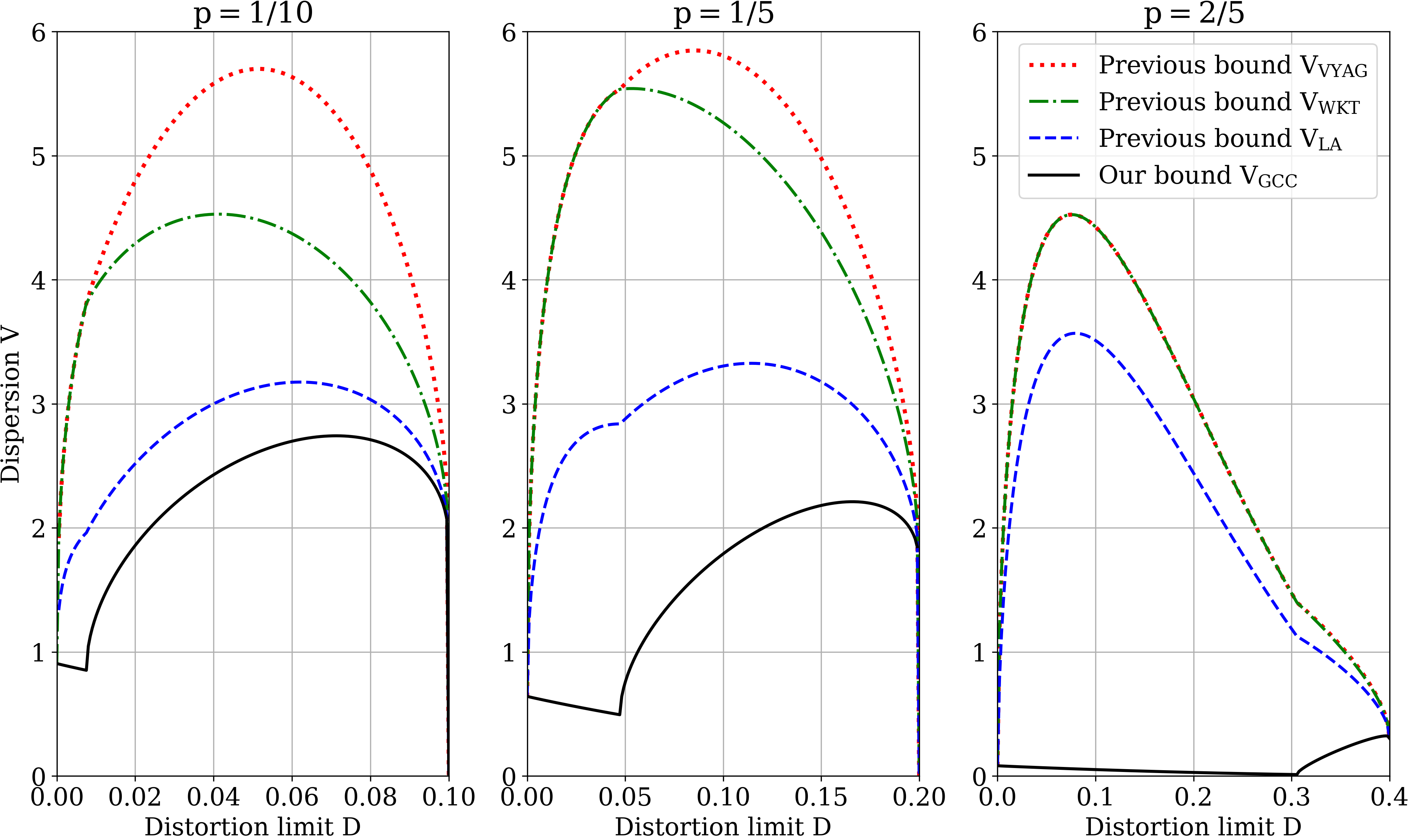}
\par\end{centering}
\caption{\label{fig:wz_bin}Our upper bound $\mathsf{V}_{\mathrm{GCC}}$ on $\mathsf{V}^{*}(\mathsf{D})$ for binary-Hamming Wyner-Ziv, and previous upper bounds $\mathsf{V}_{\mathrm{VYAG}}$, $\mathsf{V}_{\mathrm{WKT}}$, $\mathsf{V}_{\mathrm{LA}}$ for $p\in\{1/10,1/5,2/5\}$ and $\mathsf{D}\in[0,p]$.}
\end{figure*}

For the binary-Hamming Wyner-Ziv problem where $X\sim\mathrm{Bern}(1/2)$, $P_{Y|X}$ is a binary symmetric channel with crossover probability $p$, and $d(x,z)=\mathbf{1}\{x\neq z\}$ for $z\in\{0,1\}$, the optimal rate \cite{wyner1976ratedistort,barron2003duality} is the lower convex envelope of
\[
f(\mathsf{D})=\begin{cases}
H_{\mathrm{b}}(p*\mathsf{D})-H_{\mathrm{b}}(\mathsf{D}) & \text{if}\;0\le\mathsf{D}<p,\\
0 & \text{if}\;\mathsf{D}\ge p,
\end{cases}
\]
where $H_{\mathrm{b}}$ is the binary entropy function, and $a*b:=a(1-b)+b(1-a)$. To obtain the dispersion bound $\mathsf{V}_{\mathrm{GCC}}$, we apply Corollary \ref{cor:wz_cor-1} on $U\in\{0,1,2\}$ with $P_{U|X}$ given by the conditional probability matrix $\left[\begin{array}{ccc}
\gamma(1-\beta) & \gamma\beta & 1-\gamma\\
\gamma\beta & \gamma(1-\beta) & 1-\gamma
\end{array}\right]$, and $z(u,y)=u$ if $u\in\{0,1\}$, or $z(u,y)=y$ if $u=2$, where $\beta,\gamma$ are chosen to minimize $I(U;X)-I(U;Y)$ subject to $\mathbb{E}[d(X,Z)]\le\mathsf{D}$. Figure \ref{fig:wz_bin} plots our upper bound $\mathsf{V}_{\mathrm{GCC}}$, and previous upper bounds $\mathsf{V}_{\mathrm{VYAG}}$, $\mathsf{V}_{\mathrm{WKT}}$, $\mathsf{V}_{\mathrm{LA}}$ (see Section \ref{subsec:wz_compare}), for $p\in\{1/10,1/5,2/5\}$ and $\mathsf{D}\in[0,p]$. We can see that our bound significantly improves upon previous bounds, especially for $p$ close to $1/2$.

\textcolor{red}{}

\medskip{}

\section{Indirect Wyner-Ziv Coding\label{sec:noisy_wz}}

The Wyner-Ziv setting can be generalized to a noisy setting where only a noisy version of the source is available to the encoder \cite{yamamoto1980source,rebollo2005generalization,wei2025non}. In this setting, there is a $3$-discrete memoryless source $(F^{(n)},X^{(n)},Y^{(n)})\sim P_{F,X,Y}^{n}$, where $F^{(n)}$ is the source. The encoder encodes the observation $X^{(n)}$ into a message $M\in[\lceil2^{n\mathsf{R}}\rceil]$ with rate $\mathsf{R}>0$. The decoder observes $M$ and the side information $Y^{(n)}$, and recovers $\hat{Z}^{(n)}\in\mathcal{Z}^{n}$. The goal is to minimize the probability of excess distortion
\[
P_{e}:=\mathbb{P}\Big(n^{-1}\sum_{i=1}^{n}d(F_{i}^{(n)},\hat{Z}_{i}^{(n)})>\mathsf{D}\Big),
\]
where $d:\mathcal{F}\times\mathcal{Z}\to\mathbb{R}$ is a distortion function, and $\mathsf{D}\in\mathbb{R}$ is the allowed distortion level. The optimal asymptotic rate needed to have $P_{e}\to0$ is given by \cite{yamamoto1980source}
\begin{equation}
\mathsf{R}(\mathsf{D}):=\min_{P_{U|X},z:\,\mathbb{E}[d(F,Z)]\le\mathsf{D}}\left(I(U;X)-I(U;Y)\right),\label{eq:wz_rate-1}
\end{equation}
where the minimum is over $P_{U|X}$ and functions $z:\mathcal{U}\times\mathcal{Y}\to\mathcal{Z}$, subject to the constraint that $\mathbb{E}[d(F,Z)]\le\mathsf{D}$ where $(F,X,Y,U)\sim P_{F,X,Y}P_{U|X}$ and $Z=z(U,Y)$. We now extend Theorem \ref{thm:wz} to this indirect setting. 

\medskip{}

\begin{thm}
\label{thm:wz_noisy}For discrete indirect Wyner-Ziv coding, assume these two conditions are satisfied: 1) $\mathsf{D}$ is a value such that $\lambda:=-\mathrm{d}\mathsf{R}(\mathsf{D})/\mathrm{d}\mathsf{D}>0$ is finite at $\mathsf{D}$; and 2) letting $(P_{U|X},z)$ be a minimizer in $\mathsf{R}(\mathsf{D})$, $(F,X,Y,U)\sim P_{F,X,Y}P_{U|X}$ and $Z=z(U,Y)$, they satisfy $\mathbb{E}[\mathrm{Var}[\mathbb{E}[d(F,Z)|X,U]|X]]>0$. Fix any $\mathsf{W}>0$. For any large enough $n$, there is a scheme with rate $\mathsf{R}=\mathsf{R}(\mathsf{D})+\mathsf{W}/\sqrt{n}+O((\log n)/n)$, achieving a probability of excess distortion 
\[
P_{e}\le\mathbb{E}\left[P_{e}^{*}(\mathsf{W}-A_{\mathbf{X}})\right],
\]
where $A_{\mathbf{X}}\in\mathbb{R}$ is a zero mean Gaussian random variable with variance $\mathrm{Var}[\mathbb{E}[\iota(U;X)-\iota(U;Y)+\lambda d(F,Z)\,|\,X]]$, $P_{e}^{*}:\mathbb{R}\to\mathbb{R}$, 
\begin{equation}
P_{e}^{*}(\alpha):=\min_{t\in\mathbb{R}}\mathbb{P}\left(D_{\mathbf{Y}}>t\;\mathrm{or}\;J_{\mathbf{Y}}>\alpha-\lambda t\right),\label{eq:wz_noisy_Pe_a}
\end{equation}
with $[J_{\mathbf{Y}},D_{\mathbf{Y}}]\in\mathbb{R}^{2}$ being a zero mean Gaussian vector with covariance matrix 
\[
\mathbb{E}\left[\mathrm{Var}\left[\left.\left[\begin{array}{c}
-\iota(U;Y)\\
d(F,Z)
\end{array}\right]\,\right|\,X,U\right]\right].
\]
\end{thm}
\medskip{}

\begin{IEEEproof}
 We use the same coding scheme as in Theorem \ref{thm:wz}. We can show that $(\mathbf{X},\mathbf{U},\mathbf{Y},\mathbf{F})$ can be coupled to be type deviation convergent with center $P_{X}P_{U|X}P_{Y,F|X}$ and asymptotic deviation
\begin{align}
G_{\mathbf{X},\mathbf{U},\mathbf{Y},\mathbf{F}} & =(G_{\mathbf{X}}\circ P_{U|X}+P_{X}\circ\zeta_{U|X}(G_{\mathbf{X}}))\circ P_{Y|X}\nonumber \\
 & \quad+\sqrt{P_{X}\circ P_{U|X}}\circ G_{\mathbf{Y},\mathbf{F}|\mathbf{X},\mathbf{U}}.\label{eq:wz_GXUY-1}
\end{align}
Instead of (\ref{eq:wz_bd_On}), we have
\begin{align*}
P_{e} & \le\mathbb{P}\Big(\left\langle G_{\mathbf{X},\mathbf{U},\mathbf{Y},\mathbf{F}},\,d\right\rangle >0\;\mathrm{or}\;\\
 & \quad\;\left\langle G_{\mathbf{X},\mathbf{U},\mathbf{Y},\mathbf{F}},\iota_{U;X}-\iota_{U;Y}\right\rangle >\mathsf{W}\Big)+O(n^{-1/2}),
\end{align*}
where $d(x,u,y,f)=d(f,z(u,y))$. The remaining steps are the same as Theorem \ref{thm:wz}, and are omitted.
\end{IEEEproof}
\medskip{}

We now show that Theorem \ref{thm:wz_noisy} recovers the second order result for indirect or noisy lossy source coding \cite{dobrushin1962information} given in \cite{kostina2016nonasymptotic} by taking $Y=\emptyset$ and $U=Z$. We have $J_{\mathbf{Y}}=0$, and hence we take $t=\alpha/\lambda$ in (\ref{eq:wz_noisy_Pe_a}) to give
\begin{align*}
P_{e} & \le\mathbb{P}\left(D_{\mathbf{Y}}>\frac{\mathsf{W}-A_{\mathbf{X}}}{\lambda}\right)=\mathbb{P}\left(A_{\mathbf{X}}+\lambda D_{\mathbf{Y}}>\mathsf{W}\right),
\end{align*}
where $A_{\mathbf{X}}\sim\mathrm{N}(0,\mathrm{Var}[\mathbb{E}[\iota(Z;X)+\lambda d(F,Z)\,|\,X]])$ independent of $D_{\mathbf{Y}}\sim\mathrm{N}(0,\mathbb{E}[\mathrm{Var}[d(F,Z)\,|\,X,Z]])$. Hence, we have $P_{e}\le\epsilon$ for $\mathsf{W}=\sqrt{\mathsf{V}}\mathcal{Q}^{-1}(\epsilon)$, where
\begin{align*}
\mathsf{V} & =\mathrm{Var}[\mathbb{E}[\iota(Z;X)+\lambda d(F,Z)\,|\,X]]\\
 & \quad+\lambda^{2}\mathbb{E}[\mathrm{Var}[d(F,Z)\,|\,X,Z]]\\
 & \stackrel{(a)}{=}\mathrm{Var}[\iota(Z;X)+\lambda d(F,Z)]\\
 & \quad-\mathbb{E}[\mathrm{Var}[\mathbb{E}[\iota(Z;X)+\lambda d(F,Z)\,|\,X,Z]|X]]\\
 & \stackrel{(b)}{=}\mathrm{Var}[\iota(Z;X)+\lambda d(F,Z)],
\end{align*}
which coincides with the dispersion in \cite{kostina2016nonasymptotic}, where (a) is by the law of total variance, and (b) is because $\iota(Z;X)+\lambda\mathbb{E}[d(F,Z)\,|\,X,Z]$ depends only on $X$ since it is the tilted information for the surrogate distortion function $\mathbb{E}[d(F,Z)\,|\,X,Z]$ \cite{kostina2016nonasymptotic}.

We can extend Theorem \ref{thm:wz_noisy} to the situation where there are multiple distortion functions $d_{1},\ldots,d_{k}:\mathcal{F}\times\mathcal{Z}\to\mathbb{R}$, and the distortion constraint is violated if $n^{-1}\sum_{j=1}^{n}d_{i}(F_{j}^{(n)},\hat{Z}_{j}^{(n)})>\mathsf{D}_{i}$ for any $i$. In this case, the rate-distortion function is 
\[
\mathsf{R}(\vec{\mathsf{D}}):=\min_{P_{U|X},z:\,\mathbb{E}[\vec{d}(F,Z)]\le\vec{\mathsf{D}}}\left(I(U;X)-I(U;Y)\right),
\]
where we write $\vec{\mathsf{D}}=(\mathsf{D}_{i})_{i=1,\ldots,k}\in\mathbb{R}^{k}$, $\vec{d}(f,z)=(d_{i}(f,z))_{i=1,\ldots,k}$. The following theorem is a straightforward extension of Theorem \ref{thm:wz_noisy}. The proof is omitted.

\smallskip{}

\begin{thm}
\label{thm:wz_noisy_multi}For discrete indirect Wyner-Ziv coding with multiple distortion functions, assume that 1) $\vec{\lambda}:=\nabla\mathsf{R}(\vec{\mathsf{D}})\in\mathbb{R}^{k}$ has positive finite entries; and 2)  letting $(P_{U|X},z)$ be a minimizer in $\mathsf{R}(\vec{\mathsf{D}})$, $(F,X,Y,U)\sim P_{F,X,Y}P_{U|X}$ and $Z=z(U,Y)$, they satisfy that $\mathbb{E}[\mathrm{Var}[\mathbb{E}[\vec{d}(F,Z)|X,U]|X]]\in\mathbb{R}^{k\times k}$ is a full-rank matrix. Fix any $\mathsf{W}>0$. For any large enough $n$, there is a scheme with rate $\mathsf{R}=\mathsf{R}(\vec{\mathsf{D}})+\mathsf{W}/\sqrt{n}+O((\log n)/n)$, achieving a probability of excess distortion $P_{e}\le\mathbb{E}[P_{e}^{*}(\mathsf{W}-A_{\mathbf{X}})],$ where $A_{\mathbf{X}}\in\mathbb{R}$ is a zero mean Gaussian random variable with variance, 
\[
\mathrm{Var}\Big[\mathbb{E}\Big[\iota(U;X)-\iota(U;Y)+\langle\vec{\lambda},\vec{d}(F,Z)\rangle\,\Big|\,X\Big]\Big],
\]
and $P_{e}^{*}:\mathbb{R}\to\mathbb{R}$ is defined as 
\[
P_{e}^{*}(\alpha):=\min_{\vec{t}\in\mathcal{D}}\mathbb{P}\Big(J_{\mathbf{Y}}>\alpha-\langle\vec{\lambda},\vec{t}\rangle\;\mathrm{or}\;\mathrm{not}\;\vec{D}_{\mathbf{Y}}\le\vec{t}\Big),
\]
with $J_{\mathbf{Y}}\in\mathbb{R}$, $\vec{D}_{\mathbf{Y}}\in\mathbb{R}^{k}$ such that $[J_{\mathbf{Y}},\vec{D}_{\mathbf{Y}}]\in\mathbb{R}^{k+1}$ is a zero mean Gaussian vector with covariance matrix 
\[
\mathbb{E}\left[\mathrm{Var}\left[\left.\left[\begin{array}{c}
-\iota(U;Y)\\
\vec{d}(F,Z)
\end{array}\right]\,\right|\,X,U\right]\right].
\]
\end{thm}
\smallskip{}

Theorem \ref{thm:wz_noisy_multi} improves upon the second-order result in \cite{wei2025non} where there are two distortion functions,\footnote{In \cite{wei2025non}, the decoder outputs $\hat{F}^{n},\hat{X}^{n}$, and the two distortion constraints are $d_{1}(F^{n},\hat{F}^{n})\le\mathsf{D}_{1}$, $d_{2}(X^{n},\hat{X}^{n})\le\mathsf{D}_{2}$. This is covered by the setting in Theorem \ref{thm:wz_noisy_multi} since we can take $F\leftarrow(F,X)$ and $Z\leftarrow(\hat{F},\hat{X})$, so both distortion functions can be defined as functions of $(F,Z)$.} in a manner similar to how Theorem \ref{thm:wz} improves upon \cite{li2021unified} (see Section \ref{subsec:wz_compare}). Theorem \ref{thm:wz_noisy_multi} also recovers the dispersion of joint data and semantics lossy compression setting in \cite{yang2024jointdatasemanticslossy}.

\smallskip{}

\section{Lossy Compression where Side Information may be Absent\label{sec:hbk}}

We consider a generalization of the Wyner-Ziv problem where the side information $Y^{(n)}$ may be absent, known as the Heegard-Berger problem \cite{heegard1985rate}, and also studied by Kaspi \cite{kaspi1994rate}. In this setting, there is a $2$-discrete memoryless source $(X^{(n)},Y^{(n)})\sim P_{X,Y}^{n}$. The encoder encodes $X^{(n)}$ into a message $M\in[\lceil2^{n\mathsf{R}}\rceil]$ with rate $\mathsf{R}>0$. There are two decoders, where Decoder 1 observes $M$ and recovers $\hat{Z}_{1}^{(n)}\in\mathcal{Z}_{1}^{n}$, and Decoder 2 observes $M$ and the side information $Y^{(n)}$, and recovers $\hat{Z}_{2}^{(n)}\in\mathcal{Z}_{2}^{n}$. The goal is to minimize the probability of excess distortion
\[
P_{e}:=\mathbb{P}\Big(d_{1}(X^{(n)},\hat{Z}_{1}^{(n)})>\mathsf{D}_{1}\;\mathrm{or}\;d_{2}(X^{(n)},\hat{Z}_{2}^{(n)})>\mathsf{D}_{2}\Big),
\]
where $d_{1}:\mathcal{X}\times\mathcal{Z}_{1}\to\mathbb{R}$, $d_{2}:\mathcal{X}\times\mathcal{Z}_{2}\to\mathbb{R}$ are distortion functions, $d_{1}(x^{n},z^{n})=n^{-1}\sum_{i}d_{1}(x_{i},z_{i})$, and $\mathsf{D}_{1},\mathsf{D}_{2}\in\mathbb{R}$. The optimal asymptotic rate needed to have $P_{e}\to0$ is given by \cite{heegard1985rate,kaspi1994rate}
\begin{equation}
\mathsf{R}(\mathsf{D}_{1},\mathsf{D}_{2}):=\min_{P_{U_{1},U_{2}|X},z_{1},z_{2}}\left(I(U_{1};X)+I(U_{2};X|U_{1},Y)\right),\label{eq:wz_rate-2}
\end{equation}
where the minimum is over $P_{U_{1},U_{2}|X}$ and functions $z_{1}:\mathcal{U}_{1}\to\mathcal{Z}_{1}$, $z_{2}:\mathcal{U}_{1}\times\mathcal{U}_{2}\times\mathcal{Y}\to\mathcal{Z}_{2}$, subject to the constraint that $\mathbb{E}[d_{i}(X,Z_{i})]\le\mathsf{D}_{i}$ for $i=1,2$ where $(X,Y,U_{1},U_{2})\sim P_{X,Y}\circ P_{U_{1},U_{2}|X}$, $Z_{1}=z_{1}(U_{1})$ and $Z_{2}=z_{2}(U_{1},U_{2},Y)$. We now prove a new second-order achievability result which improves upon \cite{yassaee2013oneshot}. We utilize a new technique which we call \emph{dependent rate splitting}.

\smallskip{}

\begin{thm}
\label{thm:hb}For discrete Heegard-Berger coding, assume these two conditions are satisfied: 1) $\lambda_{i}:=-\partial\mathsf{R}(\mathsf{D}_{1},\mathsf{D}_{2})/\partial\mathsf{D}_{i}>0$ is finite for $i=1,2$; 2) letting $(P_{U_{1},U_{2}|X},z_{1},z_{2})$ be a minimizer in $\mathsf{R}(\mathsf{D}_{1},\mathsf{D}_{2})$, $(X,Y,U_{1},U_{2})\sim P_{X,Y}P_{U_{1},U_{2}|X}$, $Z_{1}=z_{1}(U_{1})$ and $Z_{2}=z_{2}(U_{1},U_{2},Y)$, they satisfy that $I(U_{1};X),I(U_{2};X|U_{1},Y)>0$, and
\begin{equation}
\mathbb{E}\big[\mathrm{Var}\big[\mathbb{E}[[d_{1}(X,Z_{1}),d_{2}(X,Z_{2})]^{\top}\,|\,X,U_{1},U_{2}]\big|X\big]\big]\label{eq:hb_fullrank}
\end{equation}
is a $2\times2$ full-rank matrix. Fix any $\mathsf{W}>0$. For any large enough $n$, there is a scheme with rate
\[
\mathsf{R}=\mathsf{R}(\mathsf{D})+\mathsf{W}/\sqrt{n}+O((\log n)/n),
\]
achieving a probability of excess distortion $P_{e}\le\mathbb{E}[P_{e}^{*}(\mathsf{W}-A_{\mathbf{X}})]$, where $A_{\mathbf{X}}\in\mathbb{R}$ is a zero mean Gaussian random variable with variance
\[
\mathrm{Var}\bigg[\mathbb{E}\Big[\iota(U_{1};X)+\iota(U_{2};X|U_{1},Y)+\sum_{i=1}^{2}\lambda_{i}d_{i}(X,Z_{i})\,\Big|\,X\Big]\bigg],
\]
and $P_{e}^{*}:\mathbb{R}\to\mathbb{R}$ is defined as 
\begin{align*}
P_{e}^{*}(\alpha) & :=\min_{(t_{1},t_{2})\in\mathbb{R}^{2}}\mathbb{P}\Big(J_{\mathbf{Y}}>\alpha-\lambda_{1}t_{1}-\lambda_{2}t_{2}\;\\
 & \qquad\mathrm{or}\;D_{1,\mathbf{Y}}>t_{1}\;\mathrm{or}\;D_{2,\mathbf{Y}}>t_{2}\Big),
\end{align*}
with $[J_{\mathbf{Y}},D_{1,\mathbf{Y}},D_{2,\mathbf{Y}}]\in\mathbb{R}^{3}$ being a zero mean Gaussian vector with covariance matrix 
\[
\mathbb{E}\left[\mathrm{Var}\left[\left.\left[\begin{array}{c}
\iota(U_{1};X)+\iota(U_{2};X|U_{1},Y)\\
d_{1}(X,Z_{1})\\
d_{2}(X,Z_{2})
\end{array}\right]\,\right|\,X,U_{1},U_{2}\right]\right].
\]
\end{thm}
\smallskip{}

\begin{IEEEproof}
The proof is divided into five steps.

\textbf{1) Rate splitting.} Conventionally, the coding scheme for the Heegard-Berger problem (and many other problems in network information theory) is constructed via rate splitting, where we allocate a rate $I(U_{1};X)$ for a message $M_{1}$ for Decoder 1, and a rate $I(U_{2};X|U_{1},Y)$ for a message $M_{2}$ for Decoder 2, and put $M_{1},M_{2}$ together to form the message $M$. The lengths of $M_{1}$ and $M_{2}$ are fixed in previous proofs (e.g., \cite{yassaee2013oneshot}). Here, we use a new technique which we call \emph{dependent rate splitting}, where the lengths can depend on the type of $X^{(n)}$. Suppose $M_{1}$ has a length $\ell_{1}=\lceil n\mathsf{R}_{1}\rceil$ where
\begin{equation}
\mathsf{R}_{1}=I(U_{1};X)+\frac{\mathsf{W}_{1}+\kappa(G_{\mathbf{X}}^{(n)})}{\sqrt{n}}+O\left(\frac{\log n}{n}\right),\label{eq:hb_R1}
\end{equation}
where $G_{\mathbf{X}}^{(n)}:=\sqrt{n}(\hat{P}(X^{(n)})-P_{X})$, and $\kappa:\mathrm{Tan}(P_{X})\to\mathbb{R}$ is a Lipschitz function that will be determined later; and $M_{2}$ has a length $\ell_{2}=\lceil n\mathsf{R}_{2}\rceil$ where
\begin{equation}
\mathsf{R}_{2}=I(U_{2};X|U_{1},Y)+\frac{\mathsf{W}_{2}-\kappa(G_{\mathbf{X}}^{(n)})}{\sqrt{n}}+O\left(\frac{\log n}{n}\right).\label{eq:hb_R2}
\end{equation}
Then we have $\mathsf{R}_{1}+\mathsf{R}_{2}=\mathsf{R}(\mathsf{D}_{1},\mathsf{D}_{2})+(\mathsf{W}_{1}+\mathsf{W}_{2})/\sqrt{n}+O((\log n)/n)$. Also, since $I(U_{1};X),I(U_{2};X|U_{1},Y)>0$, the probability that $\mathsf{R}_{1}<0$ or $\mathsf{R}_{2}<0$ is at most $O(n^{-1/2})$ if $X^{(n)}$ is type deviation convergent, and can be absorbed into $\epsilon$.

\textbf{2) Code construction.} Consider a GCC channel $(P_{U_{1}^{(n)},U_{2}^{(n)}|X^{(n)}})_{n}$ with center $P_{X}P_{U_{1},U_{2}|X}$ and Lipschitz deviation function $\zeta_{U_{1},U_{2}|X}:\mathrm{Tan}(P_{X})\to\mathrm{Tan}(P_{U_{1},U_{2}|X})$ to be specified later. Consider the joint distribution
\[
(X^{(n)},U_{1}^{(n)},U_{2}^{(n)},Y^{(n)})\sim P_{X}^{n}\circ P_{U_{1}^{(n)},U_{2}^{(n)}|X^{(n)}}\circ P_{Y|X}^{n}.
\]
We also let $M_{1}^{(\ell_{1})}\sim\mathrm{Unif}([2^{\ell_{1}}])$, $M_{2}^{(\ell_{2})}\sim\mathrm{Unif}([2^{\ell_{2}}])$ to be independent over all $\ell_{1},\ell_{2}\in\mathbb{Z}_{\ge0}$. Since there are two components of the encoding, we require two independent exponential codebooks $(T_{m_{1},u_{1}^{n}}^{(\ell_{1})})_{m_{1}\in[2^{\ell_{1}}],u_{1}^{n}\in\mathcal{U}_{1}^{n}}$ and $(T_{m_{2},u_{2}^{n}}^{(\ell_{2})})_{m_{2}\in[2^{\ell_{2}}],u_{2}^{n}\in\mathcal{U}_{2}^{n}}$. The encoder observes $X^{(n)}$, computes $\ell_{1}=\lceil n\mathsf{R}_{1}\rceil$ and $\ell_{2}=\lceil n\mathsf{R}_{2}\rceil$ according to (\ref{eq:hb_R1}) and (\ref{eq:hb_R2}), finds
\[
(M_{1},U_{1}^{(n)})=\mathrm{argmin}T_{m_{1},u_{1}^{n}}^{(\ell_{1})}/P_{M_{1}^{(\ell_{1})},U_{1}^{(n)}|X^{(n)}}(m_{1},u_{1}^{n}|X^{(n)}),
\]
\[
(M_{2},U_{2}^{(n)})=\mathrm{argmin}T_{m_{2},u_{2}^{n}}^{(\ell_{2})}/P_{M_{2}^{(\ell_{2})},U_{2}^{(n)}|X^{(n)},U_{1}^{(n)}}(m_{2},u_{2}^{n}|X^{(n)},U_{1}^{(n)}),
\]
and sends $\ell_{1},\ell_{2}$ (which only takes $O(\log n)$ bits; they are required so the decoders can split $M$ into $M_{1},M_{2}$) and $M_{1},M_{2}$. Decoder 1 observes $\ell_{1},\ell_{2},M_{1},M_{2}$ and computes
\[
(\hat{M}_{1},\hat{U}_{1}^{(n)})=\mathrm{argmin}T_{m_{1},u_{1}^{n}}^{(\ell_{1})}/P_{M_{1},U_{1}^{(n)}|M_{1}}(m_{1},u_{1}^{n}|M_{1}).
\]
Decoder 2 observes $\ell_{1},\ell_{2},M_{1},M_{2},Y^{(n)}$, computes the same $(\hat{M}_{1},\hat{U}_{1}^{(n)})$, and computes
\[
(\hat{M}_{2},\hat{U}_{2}^{(n)})=\mathrm{argmin}T_{m_{2},u_{2}^{n}}^{(\ell_{2})}/P_{M_{2},U_{2}^{(n)}|M_{1},U_{1}^{(n)},Y^{(n)}}(m_{2},u_{2}^{n}|M_{2},\hat{U}_{1}^{(n)},Y^{(n)}).
\]

\textbf{3) Computing the asymptotic deviation.} Using similar steps as in Theorem \ref{thm:wz}, we have the asymptotic deviation
\begin{align}
G_{\mathbf{X},\mathbf{U}_{1},\mathbf{U}_{2},\mathbf{Y}} & =G_{\mathbf{X}}\circ P_{U_{1},U_{2}|X}\circ P_{Y|X}\nonumber \\
 & \quad+P_{X}\circ\zeta_{U_{1},U_{2}|X}(G_{\mathbf{X}})\circ P_{Y|X}\nonumber \\
 & \quad+\sqrt{P_{X,U_{1},U_{2}}}\circ G_{\mathbf{Y}|\mathbf{X},\mathbf{U}_{1},\mathbf{U}_{2}}.\label{eq:wz_GXUY-2}
\end{align}

\textbf{4) Error bound.} Using the Poisson matching lemma in a similar manner as in the proofs of Theorems \ref{thm:sc} and \ref{thm:wz},\footnote{Although now $\ell_{1},\ell_{2}$ depends on $X^{(n)}$, the same error analysis works since the Poisson matching lemma (\ref{eq:ppl}) is a pointwise bound that applies to every tuple of values of $x^{n}u_{1}^{n},u_{2}^{n},y^{n},\ell_{1},\ell_{2}$.} writing $G:=G_{\mathbf{X},\mathbf{U}_{1},\mathbf{U}_{2},\mathbf{Y}}$, we have
\begin{align}
P_{e} & \le\mathbb{P}\bigg(\left\langle G,d_{1}\right\rangle >0\;\mathrm{or}\;\left\langle G,d_{2}\right\rangle >0\label{eq:hb_error1}\\
 & \qquad\mathrm{or}\;\left\langle G,\iota_{U_{1};X}\right\rangle -\kappa(G_{\mathbf{X}})>\mathsf{W}_{1}\label{eq:hb_error2}\\
 & \qquad\mathrm{or}\;\left\langle G,\iota_{U_{2};X,U_{1}}-\iota_{U_{2};Y,U_{1}}\right\rangle +\kappa(G_{\mathbf{X}})>\mathsf{W}_{2}\bigg)\label{eq:hb_error3}\\
 & \quad+O(n^{-1/2}).
\end{align}

\textbf{5) Simplification via Gaussian vector manipulation.} Note that 
\[
\left\langle G,\iota_{U_{1};X}\right\rangle =\left\langle G_{\mathbf{X}}\circ P_{U_{1}|X}+P_{X}\circ\zeta_{U_{1},U_{2}|X}(G_{\mathbf{X}}),\,\iota_{U_{1};X}\right\rangle 
\]
is an Lipschitz function of $G_{\mathbf{X}}$. Hence, we can simply take $\kappa(G_{\mathbf{X}})=\left\langle G,\iota_{U_{1};X}\right\rangle $ and $\mathsf{W}_{1}=0$ to eliminate the error event (\ref{eq:hb_error2}). We now consider the terms in (\ref{eq:hb_error1}) and (\ref{eq:hb_error3}). Write $\iota:=\iota_{U_{1};X}+\iota_{U_{2};X,U_{1}}-\iota_{U_{2};Y,U_{1}}=\iota_{U_{1};X}+\iota_{U_{2};X|U_{1},Y}$. Then (\ref{eq:hb_error3}) becomes $\left\langle G,\iota\right\rangle >\mathsf{W}_{2}$. For $i=1,2$, let
\begin{align*}
J_{\mathbf{X}} & :=\left\langle G_{\mathbf{X}}\circ P_{U_{1},U_{2}|X}\circ P_{Y|X},\,\iota\right\rangle ,\\
D_{i,\mathbf{X}} & :=\left\langle G_{\mathbf{X}}\circ P_{U_{1},U_{2}|X}\circ P_{Y|X},\,d_{i}\right\rangle ,\\
A_{\mathbf{X}} & :=J_{\mathbf{X}}+\lambda_{1}D_{1,\mathbf{X}}+\lambda_{2}D_{2,\mathbf{X}},\\
J_{\mathbf{U}} & :=\left\langle P_{X}\circ\zeta_{U_{1},U_{2}|X}(G_{\mathbf{X}})\circ P_{Y|X},\,\iota\right\rangle ,\\
D_{i,\mathbf{U}} & :=\left\langle P_{X}\circ\zeta_{U_{1},U_{2}|X}(G_{\mathbf{X}})\circ P_{Y|X},\,d_{i}\right\rangle ,\\
J_{\mathbf{Y}} & :=\left\langle \sqrt{P_{X,U_{1},U_{2}}}\circ G_{\mathbf{Y}|\mathbf{X},\mathbf{U}_{1},\mathbf{U}_{2}},\,\iota\right\rangle ,\\
D_{i,\mathbf{Y}} & :=\left\langle \sqrt{P_{X,U_{1},U_{2}}}\circ G_{\mathbf{Y}|\mathbf{X},\mathbf{U}_{1},\mathbf{U}_{2}},\,d_{i}\right\rangle .
\end{align*}
Note that $D_{1,\mathbf{Y}}=0$ since  $G_{\mathbf{Y}|\mathbf{X},\mathbf{U}_{1},\mathbf{U}_{2}}\in\mathrm{Tan}(P_{Y|X,U_{1},U_{2}})$. By the optimality of $P_{U_{1},U_{2}|X}$, using the same arguments as Theorem \ref{thm:sc}, for any $V\in\mathrm{Tan}(P_{U_{1},U_{2}|X})$,
\[
\left\langle P_{X}\circ V\circ P_{Y|X},\,\iota+\lambda_{1}d_{1}+\lambda_{2}d_{2}\right\rangle =0,
\]
and hence $J_{\mathbf{U}}+\lambda_{1}D_{1,\mathbf{U}}+\lambda_{2}D_{2,\mathbf{U}}=0$. Letting $G\sim\mathrm{NM}(P_{U_{1},U_{2}|X})$, $[\langle P_{X}\circ G\circ P_{Y|X},d_{1}\rangle,\langle P_{X}\circ G\circ P_{Y|X},d_{2}\rangle]^{\top}$ is zero-mean Gaussian with covariance matrix (\ref{eq:hb_fullrank}), which is full rank. Hence, there exist fixed $V_{1},V_{2}\in\mathrm{Tan}(P_{U_{1},U_{2}|X})$ with $\langle P_{X}\circ V_{i}\circ P_{Y|X},d_{j}\rangle=\mathbf{1}\{i=j\}$ for $i,j=1,2$. We take
\begin{align*}
\zeta_{U_{1},U_{2}|X}(G_{\mathbf{X}}) & =\left(-\psi_{1}(\mathsf{W}_{2}-A_{\mathbf{X}})-D_{1,\mathbf{X}}\right)\cdot V_{1}\\
 & \quad+\left(-\psi_{2}(\mathsf{W}_{2}-A_{\mathbf{X}})-D_{2,\mathbf{X}}\right)\cdot V_{2},
\end{align*}
where $\psi:\mathbb{R}\to\mathbb{R}^{2}$ is a Lipschitz function. The desired result follows from taking
\begin{align*}
\psi(\alpha) & =\underset{(t_{1},t_{2})\in\mathbb{R}^{2}}{\mathrm{argmin}}\mathbb{P}\Big(J_{\mathbf{Y}}-\alpha+\lambda_{1}t_{1}+\lambda_{2}t_{2}>0\;\\
 & \qquad\qquad\mathrm{or}\;D_{1,\mathbf{Y}}-t_{1}>0\;\mathrm{or}\;D_{2,\mathbf{Y}}-t_{2}>0\Big).
\end{align*}
Refer to Appendix \ref{subsec:pf_Lipschitz} for the technical proof that $\psi$ is Lipschitz.\footnote{Although Appendix \ref{subsec:pf_Lipschitz} only proves the case for 2D Gaussian vector, the generalization to 3D is straightforward and is omitted.} The remaining steps are similar to Theorem \ref{thm:wz}, and are omitted.
\end{IEEEproof}
\medskip{}

\section{Second-Order Channel Coding with Cost Constraint}

Apart from source coding results, type deviation convergence is applicable to channel coding as well using the exact same workflow. We demonstrate this technique by recovering the second-order result on channel coding for a discrete memoryless channel with cost constraint \cite{hayashi2009information,polyanskiy2010channel,kostina2015channels}. In this setting, the encoder encodes a message $M\sim\mathrm{Unif}[\lfloor2^{n\mathsf{R}}\rfloor]$ with rate $\mathsf{R}>0$ into $X^{(n)}\in\mathcal{X}^{n}$, and sends it through a memoryless channel $P_{Y|X}^{n}$. The decoder observes the output $Y^{(n)}\in\mathcal{Y}^{n}$ and recovers $\hat{M}$. We also require that $n^{-1}\sum_{i=1}^{n}d(X_{i}^{(n)})\le\mathsf{D}$, where $d:\mathcal{X}\to\mathbb{R}$ is a cost function. The goal is to minimize the probability of error $P_{e}:=\mathbb{P}(M\neq\hat{M})$. We now recover the second-order result in \cite{hayashi2009information,kostina2015channels}.

\medskip{}

\begin{thm}
\label{thm:cc}For discrete channel coding with cost constraint, for any $0<\epsilon<1$, if $\mathsf{D}>\min_{x}d(x)$, then for every $n$, there is a scheme achieving a probability of error $P_{e}\le\epsilon$, with rate
\[
\mathsf{R}=\mathsf{C}-\sqrt{\frac{\mathsf{V}}{n}}\mathcal{Q}^{-1}(\epsilon)-O\left(\frac{\log n}{n}\right),
\]
as long as this rate is positive, where $\mathsf{C}:=\max_{P_{X}:\,\mathbb{E}[d(X)]\le\mathsf{D}}I(X;Y)$ is the capacity, $\mathsf{V}:=\mathbb{E}[\mathrm{Var}[\iota_{X;Y}(X;Y)|X]]$ (with $(X,Y)$ induced by the $P_{X}$ attaining the capacity) is the channel dispersion,\footnote{In case there are multiple $P_{X}$'s attaining the capacity, choose the one with the smallest $\mathsf{V}$ if $\epsilon\le1/2$, or the largest $\mathsf{V}$ if $\epsilon>1/2$.} and the constant in $O((\log n)/n)$ depends on $P_{Y|X}$, $d$, $\mathsf{D}$ and $\epsilon$.
\end{thm}
\medskip{}

\begin{IEEEproof}
The proof is divided into three steps.

\textbf{1) Code construction.} Consider a GCC channel $(P_{X^{(n)}})_{n}$ from $\emptyset$ (constant random variable) to $X$ with center $P_{X}$ and function $\zeta=0$. Consider the joint distribution
\[
(M,X^{(n)},Y^{(n)})\sim\mathrm{Unif}([\lfloor2^{n\mathsf{R}}\rfloor])\times(P_{X^{(n)}}\circ P_{Y|X}^{n}).
\]
Let $T_{m,x^{n}}\sim\mathrm{Exp}(1)$, i.i.d. across $m\in[\lfloor2^{n\mathsf{R}}\rfloor]$, $x^{n}\in\mathcal{X}^{n}$, which serves as a random codebook available to the encoder and the decoder. The encoder observes $M$, find 
\begin{equation}
(M,X^{(n)})=\mathrm{argmin}_{m,x^{n}}T_{m,x^{n}}/P_{M,X^{(n)}|M}(m,x^{n}|M),\label{eq:cc_enc}
\end{equation}
(note that $P_{M,X^{(n)}|M}(m,x^{n}|M)=P_{X^{(n)}}(x^{n})\cdot\mathbf{1}\{m=M\}$) and sends $X^{(n)}$ if $\left\langle \hat{P}(X^{(n)}),d\right\rangle =n^{-1}\sum_{i=1}^{n}d(X_{i}^{(n)})\le\mathsf{D}+\delta/n$, where $\delta>0$ will be specified later; or else sends an arbitrary $X^{(n)}$ satisfying this constraint. Note that although we have ``redefined'' $M$ in (\ref{eq:cc_enc}), the two $M$'s on the left-hand side and the right-hand side must match. The channel outputs $Y^{(n)}|X^{(n)}\sim P_{Y|X}^{n}$. The decoder observes $Y^{(n)}$ and computes
\[
(\hat{M},\hat{X}^{(n)})=\mathrm{argmin}_{m,x^{n}}T_{m,x^{n}}/P_{M,X^{(n)}|Y^{(n)}}(m,x^{n}|Y^{(n)}).
\]
Note that $P_{M,X^{(n)}|Y^{(n)}}(m,x^{n}|Y^{(n)})=P_{X^{(n)}|Y^{(n)}}(x^{n})P_{M}(m)$. 

\textbf{2) Computing the asymptotic deviation of $(\mathbf{X},\mathbf{Y})$.} We now utilize the type deviation convergence of the sequences. By Proposition \ref{prop:acc_product}, $\mathbf{X}$ is type deviation convergent with center $P_{X}$ and asymptotic deviation $0$, and hence
\begin{align}
\mathbb{P}\left(\left\langle \hat{P}(X^{(n)}),d\right\rangle >\mathsf{D}+\delta/n\right) & \le\mathbb{P}\left(\left\langle G_{\mathbf{X}}^{(n)},d\right\rangle >\delta/\sqrt{n}\right)\nonumber \\
 & =O(1/\sqrt{n})\label{eq:cc_cost_bd}
\end{align}
if we take $\delta$ to be the constant in Definition \ref{def:typegaussian}. By Proposition \ref{prop:gauss_memoryless}, we can assume that $(\mathbf{X},\mathbf{Y})$ is type deviation convergent with center $P_{X}P_{Y|X}$ and asymptotic deviation $G_{\mathbf{X},\mathbf{Y}}=\sqrt{P_{X}}\circ G_{\mathbf{Y}|\mathbf{X}}$, where $G_{\mathbf{Y}|\mathbf{X}}\sim\mathrm{NM}(P_{Y|X})$.

\textbf{3) Error bound.} By Proposition \ref{prop:self_info},
\begin{equation}
\iota(X^{(n)};Y^{(n)})\ge n\mathsf{C}+\sqrt{n}\left\langle \sqrt{P_{X}}\circ G_{\mathbf{Y}|\mathbf{X}},\iota_{X;Y}\right\rangle -c\log n.\label{eq:cc_density_bd}
\end{equation}
with probability $1-O(n^{-1/2}$) for some $c>0$. By Proposition \ref{prop:nm_prop}, $\left\langle \sqrt{P_{X}}\circ G_{\mathbf{Y}|\mathbf{X}},\iota_{X;Y}\right\rangle $ is Gaussian with mean $0$ and variance $\mathsf{V}=\mathbb{E}[\mathrm{Var}[\iota(X;Y)|X]]$. We have
\begin{align*}
P_{e} & \le\mathbb{P}\left(\left\langle \hat{P}(X^{(n)}),d\right\rangle >\mathsf{D}+\delta/n\;\mathrm{or}\;M\neq\hat{M}\right)\\
 & \stackrel{(a)}{\le}\mathbb{P}\left(M\neq\hat{M}\right)+O(n^{-1/2})\\
 & \stackrel{(b)}{\le}\mathbb{P}\left(\iota(M,X^{(n)};M)-\iota(M,X^{(n)};Y^{(n)})>-\frac{\log n}{2}\right)+O(n^{-1/2})\\
 & \le\mathbb{P}\left(n\mathsf{R}-\iota(X^{(n)};Y^{(n)})>-\frac{1}{2}\log n\right)+O(n^{-1/2})\\
 & \stackrel{(c)}{\le}\mathbb{P}\left(n(\mathsf{R}-\mathsf{C})-\sqrt{n}\left\langle \sqrt{P_{X}}\circ G_{\mathbf{Y}|\mathbf{X}},\iota_{X;Y}\right\rangle >-\left(c+\frac{1}{2}\right)\log n\right)+O(n^{-1/2})\\
 & \stackrel{(d)}{=}\mathbb{P}\left(\left\langle \sqrt{P_{X}}\circ G_{\mathbf{Y}|\mathbf{X}},\iota_{X;Y}\right\rangle <\sqrt{\mathsf{V}}\mathcal{Q}^{-1}\left(\epsilon-\frac{\gamma}{\sqrt{n}}\right)\right)+O(n^{-1/2})\\
 & =\epsilon,
\end{align*}
where (a) is by (\ref{eq:cc_cost_bd}), (b) is by the Poisson matching lemma on $M,(M,X^{(n)}),Y^{(n)}$, (c) is by (\ref{eq:cc_density_bd}), and (d) is by taking
\[
\mathsf{R}=\mathsf{C}-\sqrt{\frac{\mathsf{V}}{n}}\mathcal{Q}^{-1}\left(\epsilon-\frac{\gamma}{\sqrt{n}}\right)-\left(c+\frac{1}{2}\right)\frac{\log n}{n},
\]
where $\gamma$ is chosen be the constant in the $O(n^{-1/2})$ term (note that $\mathcal{Q}^{-1}(\epsilon-\gamma n^{-1/2})=\mathcal{Q}^{-1}(\epsilon)+O(n^{-1/2})$). To complete the proof, we need to strengthen $\left\langle \hat{P}(X^{(n)}),d\right\rangle \le\mathsf{D}+\delta/n$ to $\left\langle \hat{P}(X^{(n)}),d\right\rangle \le\mathsf{D}$. We can do so by concatenating $X^{(n)}$ with a constant number of symbols $x$ that minimizes $d(x)$. Since $\mathsf{D}>\min_{x}d(x)$, this can reduce the average cost by $O(1/n)$. Also, this will only incur a $O(1/n)$ penalty on $\mathsf{R}$.
\end{IEEEproof}
\medskip{}

\section{Second-Order Gelfand-Pinsker with Cost Constraint\label{sec:gp}}

\subsection{Second-Order Achievability for Gelfand-Pinsker with Cost}

We consider channel coding with noncausal state information at the encoder, also known as the Gelfand-Pinsker problem \cite{gelfand1980coding}, which was also studied by Heegard and El Gamal \cite{heegard1983capacity}. In this setting, the encoder observes a message $M\sim\mathrm{Unif}[\lfloor2^{n\mathsf{R}}\rfloor]$ with rate $\mathsf{R}>0$ and the i.i.d. channel state sequence $S^{(n)}\sim P_{S}^{n}$, and sends $X^{(n)}\in\mathcal{X}^{n}$ through a memoryless channel $P_{Y|S,X}^{n}$. The decoder observes the output $Y^{(n)}\in\mathcal{Y}^{n}$ and recovers $\hat{M}$. We also require that $n^{-1}\sum_{i=1}^{n}d(S_{i}^{(n)},X_{i}^{(n)})\le\mathsf{D}$, where $d:\mathcal{S}\times\mathcal{X}\to\mathbb{R}$ is a cost function. The goal is to minimize the probability of error $P_{e}:=\mathbb{P}(M\neq\hat{M})$. The asymptotic capacity is given by \cite{gelfand1980coding,heegard1983capacity}
\begin{equation}
\mathsf{C}(\mathsf{D}):=\max_{P_{U|S},x:\,\mathbb{E}[d(S,X)]\le\mathsf{D}}\left(I(U;Y)-I(U;S)\right),\label{eq:wz_rate-3}
\end{equation}
where the minimum is over $P_{U|S}$ and functions $x:\mathcal{S}\times\mathcal{U}\to\mathcal{X}$, subject to the constraint that $\mathbb{E}[d(S,X)]\le\mathsf{D}$ where $(S,U,X,Y)\sim P_{S}\circ P_{U|S}\circ P_{X|S,U}\circ P_{Y|S,X}$ with $P_{X|S,U}(x(s,u)|s,u)=1$.  We now show a second-order result  that improves upon \cite{watanabe2015nonasymp} and \cite{li2021unified,liu2025one} (after a straightforward generalization to include cost).

\medskip{}

\begin{thm}
\label{thm:gp}For discrete Gelfand-Pinsker coding with cost constraint, assume these two conditions are satisfied: 1) $\mathsf{D}$ is a value such that  $\lambda:=\mathrm{d}\mathsf{C}(\mathsf{D})/\mathrm{d}\mathsf{D}\ge0$ is finite at $\mathsf{D}$; and 2) letting $(P_{U|S},x)$ be a minimizer in $\mathsf{C}(\mathsf{D})$, $(S,U,X,Y)\sim P_{S}\circ P_{U|S}\circ P_{X|S,U}\circ P_{Y|S,X}$, they satisfy $\mathbb{E}[\mathrm{Var}[d(S,X)|S]]>0$. Then for every $n$, there is a scheme achieving a probability of error $P_{e}\le\epsilon$, with rate
\[
\mathsf{R}=\mathsf{C}(\mathsf{D})-\sqrt{\frac{\mathsf{V}}{n}}\mathcal{Q}^{-1}(\epsilon)-O\left(\frac{\log n}{n}\right),
\]
as long as this rate is positive, where 
\begin{align*}
\mathsf{V} & :=\mathrm{Var}[\mathbb{E}[\iota(U;S)-\iota(U;Y)+\lambda d(S,X)\,|\,S]]\\
 & \quad\quad+\mathbb{E}[\mathrm{Var}[\iota(U;Y)\,|\,S,U]],
\end{align*}
and the constant in $O((\log n)/n)$ depends on $P_{Y|S,X}$, $P_{U|S}$, $d$, $\mathsf{D}$, $\epsilon$ and the function $x$.
\end{thm}
\medskip{}

Our result recovers the second-order results in \cite{scarlett2015dispersions,li2021unified} when the cost constraint is inactive, where $\mathsf{D}$ is large enough and $\lambda=0$, giving $\mathsf{V}=\mathrm{Var}[\iota(U;S)-\iota(U;Y)]$. We note that \cite{scarlett2015dispersions} does not contain a result for general cost functions (it only covers the power constraint in dirty paper coding \cite{costa1983writing}). While the analysis in \cite{li2021unified} can be extended to incorporate a cost constraint, it requires two error events and leads to a worse second-order result. 

We now prove Theorem \ref{thm:gp}.

\medskip{}

\begin{IEEEproof}
The proof is divided into four steps.

\textbf{1) Code construction.} Consider a GCC channel $(P_{U^{(n)}|S^{(n)}})_{n}$ with center $P_{S,U}$ and a Lipschitz deviation function $\zeta_{U|S}:\mathrm{Tan}(P_{S})\to\mathrm{Tan}(P_{U|S})$ to be specified later. Consider the joint distribution
\begin{align*}
 & (M,S^{(n)},U^{(n)},Y^{(n)})\\
 & \sim\mathrm{Unif}([\lfloor2^{n\mathsf{R}}\rfloor])\times(P_{S^{(n)}}\circ P_{U^{(n)}|S^{(n)}}\circ P_{Y|S,U}^{n}),
\end{align*}
where $P_{Y|S,U}$ is the conditional distribution where $X=x(S,U)$ and $Y$ follows $P_{Y|S,X}$. Let $T_{m,u^{n}}\sim\mathrm{Exp}(1)$, i.i.d. across $m\in[\lfloor2^{n\mathsf{R}}\rfloor]$, $x^{n}\in\mathcal{X}^{n}$, which serves as a random codebook available to the encoder and the decoder. The encoder observes $M$, find 
\begin{equation}
(M,U^{(n)})=\mathrm{argmin}_{m,u^{n}}T_{m,u^{n}}/P_{M,U^{(n)}|M}(m,u^{n}|M),\label{eq:cc_enc-1}
\end{equation}
and sends $X^{(n)}$ with $X_{i}^{(n)}=x(S_{i}^{(n)},U_{i}^{(n)})$ if $\langle\hat{P}(S^{(n)},X^{(n)}),d\rangle=n^{-1}\sum_{i=1}^{n}d(S_{i}^{(n)},X_{i}^{(n)})\le\mathsf{D}+\delta/n$, where $\delta>0$ will be specified later; or else sends an arbitrary $X^{(n)}$ satisfying this constraint. The decoder observes $Y^{(n)}$ and computes
\[
(\hat{M},\hat{U}^{(n)})=\mathrm{argmin}_{m,u^{n}}T_{m,u^{n}}/P_{M,U^{(n)}|Y^{(n)}}(m,u^{n}|Y^{(n)}).
\]

\textbf{2) Computing the asymptotic deviation.} By Propositions \ref{prop:gauss_iid}, \ref{prop:acc_product} and \ref{prop:gauss_memoryless}, $(\mathbf{S},\mathbf{U},\mathbf{Y})$ can be coupled to be type deviation convergent with center $P_{S}\circ P_{U|S}\circ P_{Y|S,U}$ and asymptotic deviation 
\begin{align}
G_{\mathbf{S},\mathbf{U},\mathbf{Y}} & =(G_{\mathbf{S}}\circ P_{U|S}+P_{S}\circ\zeta_{U|S}(G_{\mathbf{S}}))\circ P_{Y|S,U}\nonumber \\
 & \quad+\sqrt{P_{S}\circ P_{U|S}}\circ G_{\mathbf{Y}|\mathbf{S},\mathbf{U}},\label{eq:gp_GYSU}
\end{align}
where $G_{\mathbf{S}}\sim\mathrm{NM}(P_{S})$, $G_{\mathbf{Y}|\mathbf{S},\mathbf{U}}\sim\mathrm{NM}(P_{Y|S,U})$.

\textbf{3) Error bound.} Let $d(s,u):=d(s,x(s,u))$. Using the Poisson matching lemma in a similar manner as in the proofs of Theorems \ref{thm:sc}, \ref{thm:wz}, \ref{thm:cc},
\begin{align}
P_{e} & \le\mathbb{P}\Big(\langle G_{\mathbf{S},\mathbf{U},\mathbf{Y}},\,d\rangle>0\;\mathrm{or}\;\nonumber \\
 & \quad\quad\langle G_{\mathbf{S},\mathbf{U},\mathbf{Y}},\iota_{U;S}-\iota_{U;Y}\rangle>\mathsf{W}\Big)+O(n^{-1/2}),\label{eq:gp_bd_On}
\end{align}
with a suitable $\delta$ (see Theorem \ref{thm:cc}), by taking
\[
\mathsf{R}=\mathsf{C}(\mathsf{D})-\frac{\mathsf{W}}{\sqrt{n}}-O\left(\frac{\log n}{n}\right),
\]
with $\mathsf{W}$ to be specified later.

\textbf{4) Simplification via Gaussian vector manipulation.} The problem is now reduced to bounding the probability that the Gaussian vector $G_{\mathbf{S},\mathbf{U},\mathbf{Y}}$ violating any of the constraints in (\ref{eq:gp_bd_On}). Similar to Theorem \ref{thm:wz}, define
\begin{align*}
J_{\mathbf{S}} & :=\langle G_{\mathbf{S}}\circ P_{U|S}\circ P_{Y|S,U},\,\iota_{U;S}-\iota_{U;Y}\rangle,\\
D_{\mathbf{S}} & :=\langle G_{\mathbf{S}}\circ P_{U|S}\circ P_{Y|S,U},\,d\rangle,\\
A_{\mathbf{S}} & :=J_{\mathbf{S}}+\lambda D_{\mathbf{S}},\\
J_{\mathbf{U}} & :=\langle P_{S}\circ\zeta_{U|S}(G_{\mathbf{S}})\circ P_{Y|S,U},\,\iota_{U;S}-\iota_{U;Y}\rangle,\\
D_{\mathbf{U}} & :=\langle P_{S}\circ\zeta_{U|S}(G_{\mathbf{S}})\circ P_{Y|S,U},\,d\rangle,\\
J_{\mathbf{Y}} & :=\langle\sqrt{P_{S}\circ P_{U|S}}\circ G_{\mathbf{Y}|\mathbf{S},\mathbf{U}},\,\iota_{U;S}-\iota_{U;Y}\rangle.
\end{align*}
By the optimality of $P_{U|S}$, using the same arguments as Theorem \ref{thm:sc}, for any $V_{U|S}\in\mathrm{Tan}(P_{U|S})$,
\begin{equation}
\left\langle P_{S}\circ V_{U|S}\circ P_{Y|S,U},\,\iota_{U;S}-\iota_{U;Y}+\lambda d\right\rangle =0.\label{eq:VUX_sum-1}
\end{equation}
Since $\zeta_{U|S}(G_{\mathbf{X}})\in\mathrm{Tan}(P_{U|S})$, we have $J_{\mathbf{U}}=-\lambda D_{\mathbf{U}}$. We first eliminate the ``$\langle G_{\mathbf{S},\mathbf{U},\mathbf{Y}},\,d\rangle>0$'' error event in (\ref{eq:gp_bd_On}). Letting $G_{U|S}\sim\mathrm{NM}(P_{U|S})$, $\langle P_{S}\circ G_{U|S}\circ P_{Y|S,U},\,d\rangle$ is zero-mean Gaussian with variance $\mathbb{E}[\mathrm{Var}[d(S,X)|S]]>0$, and hence there exists a fixed $V_{U|S}\in\mathrm{Tan}(P_{U|S})$ with $\langle P_{S}\circ V_{U|S}\circ P_{Y|S,U},\,d\rangle=1$. Take
\[
\zeta_{U|S}(G_{\mathbf{S}})=-D_{\mathbf{S}}\cdot V_{U|S}.
\]
We have $D_{\mathbf{U}}=-D_{\mathbf{S}}$ and $\langle G_{\mathbf{S},\mathbf{U},\mathbf{Y}},d\rangle=D_{\mathbf{S}}+D_{\mathbf{U}}=0$. The probability in (\ref{eq:gp_bd_On}) becomes
\begin{align*}
 & \mathbb{P}\left(J_{\mathbf{S}}+J_{\mathbf{U}}+J_{\mathbf{Y}}>\mathsf{W}\right)=\mathbb{P}\left(J_{\mathbf{S}}+\lambda D_{\mathbf{S}}+J_{\mathbf{Y}}>\mathsf{W}\right),
\end{align*}
which is at most $\epsilon$ when $\mathsf{W}=\mathcal{Q}^{-1}(\epsilon)\sqrt{\mathrm{Var}[J_{\mathbf{S}}+\lambda D_{\mathbf{S}}+J_{\mathbf{Y}}]}$, which gives the desired bound. Since $\lambda$ is finite, we have $\mathsf{D}>\mathbb{E}[\min_{x}d(S,x)]$, and hence we can concatenate $X^{(n)}$ with $O(\log n)$ symbols to strengthen $\langle\hat{P}(S^{(n)},X^{(n)}),d\rangle\le\mathsf{D}+\delta/n$ to $\langle\hat{P}(S^{(n)},X^{(n)}),d\rangle\le\mathsf{D}$ as in the last step in the proof of Theorem \ref{thm:sc}. 
\end{IEEEproof}
\medskip{}

\subsection{Comparisons with Existing Bounds}

Theorem \ref{thm:gp} improves upon the following existing achievability bounds (we assume that the same assumptions in Theorem \ref{thm:gp} hold). 
\begin{itemize}
\item Watanabe-Kuzuoka-Tan \cite{watanabe2015nonasymp}:\footnote{We slightly generalize \cite{watanabe2015nonasymp} to allow the cost to depend on the state.} Achieves $\mathsf{R}=\mathsf{C}(\mathsf{D})-\mathsf{W}/\sqrt{n}-O((\log n)/n)$ with 
\begin{equation}
P_{e}\le\min_{t,\tau\in\mathbb{R}}\mathbb{P}(\tilde{J}_{\mathbf{S}}>\mathsf{W}-\lambda t-\tau\;\mathrm{or}\;\tilde{J}_{\mathbf{Y}}>\tau\;\mathrm{or}\;\tilde{D}>t),\label{eq:gp_Pe_wkt}
\end{equation}
where $[\tilde{J}_{\mathbf{S}},\tilde{J}_{\mathbf{Y}},\tilde{D}]$ is a zero-mean Gaussian vector with covariance matrix 
\[
\mathbb{E}\big[\mathrm{Var}[[\iota(\tilde{U};S|T),-\iota(\tilde{U};Y|T),d(S,X)]^{\top}\,|\,T]\big],
\]
where $(S,\tilde{U},X,Y,T)\sim P_{S}\circ P_{T}\circ P_{\tilde{U}|S,T}\circ P_{X|S,\tilde{U},T}\circ P_{Y|S,X}$ with $I(\tilde{U};Y|T)-I(\tilde{U};S|T)=\mathsf{C}(\mathsf{D})$ and $\mathbb{E}[d(S,X)]=\mathsf{D}$. Note that the time-sharing random variable $T$ is needed in \cite{watanabe2015nonasymp}, but is unnecessary in Theorem \ref{thm:wz} since $T$ can be absorbed into $U$. We can show that Theorem \ref{thm:gp} improves upon (\ref{eq:gp_Pe_wkt}) using a similar argument as Appendix \ref{subsec:pf_compare}.
\end{itemize}
\smallskip{}

\begin{itemize}
\item Li-Anantharam \cite{li2021unified} (after a straightforward generalization to include cost; also see \cite{liu2025one}): Achieves $\mathsf{R}=\mathsf{C}(\mathsf{D})-\mathsf{W}/\sqrt{n}-O((\log n)/n)$ with 
\begin{equation}
P_{e}\le\min_{t\in\mathbb{R}}\mathbb{P}(\bar{J}_{\mathbf{X}}+\bar{J}_{\mathbf{Y}}>\mathsf{W}-\lambda t\;\mathrm{or}\;\bar{D}>t),\label{eq:gp_Pe_prev_poisson}
\end{equation}
where $[\bar{J}_{\mathbf{X}},\bar{J}_{\mathbf{Y}},\bar{D}]$ is a zero-mean Gaussian vector with covariance matrix 
\[
\mathrm{Var}\big[[\iota(U;S),-\iota(U;Y),d(S,X)]^{\top}\big],
\]
with the same variables in Theorem \ref{thm:gp}. We can show that Theorem \ref{thm:gp} improves upon (\ref{eq:gp_Pe_prev_poisson}) using a similar argument as Appendix \ref{subsec:pf_compare}. 
\end{itemize}
\smallskip{}

\section{Broadcast Channels\label{sec:bc}}

The type deviation convergence technique is applicable to a wide range of problems in network information theory. We now consider the broadcast channel with common message, where the encoder encodes the messages $M_{i}\sim\mathrm{Unif}[\lfloor2^{n\mathsf{R}_{i}}\rfloor]$ for $i=0,1,2$ into $X^{(n)}\in\mathcal{X}^{n}$, and sends it through a memoryless broardcast channel $P_{Y_{1},Y_{2}|X}^{n}$. Decoder 1 observes the output $Y_{1}^{(n)}\in\mathcal{Y}_{1}^{n}$ and recovers $\hat{M}_{01},\hat{M}_{1}$. Decoder 2 observes the output $Y_{2}^{(n)}\in\mathcal{Y}_{2}^{n}$ and recovers $\hat{M}_{02},\hat{M}_{2}$. The error probability is
\[
P_{e}:=1-\mathbb{P}\left(M_{0}=\hat{M}_{01}=\hat{M}_{02},\,M_{1}=\hat{M}_{1},\,M_{2}=\hat{M}_{2}\right).
\]
We now apply type deviation convergence on \cite[Theorem 5]{li2021unified} to derive the following achievable dispersion, which is a second-order refinement upon the generalization of Marton's inner bound \cite{marton1979broadcast} in \cite{gelfand1980capacity,liang2007broadcast,liang2011equivalence}. This recovers the currently best known bound in \cite{yassaee2013binning},\footnote{\cite{yassaee2013binning} only contains the second-order refinement of the 2-auxiliary Marton's inner bound. The 3-auxiliary bound was conveyed by M. H. Yassaee in a private communication.} which improves upon previous results \cite{verdu2012nonasymp,liu2015oneshotmutual} (with i.i.d. random codebook), \cite{yassaee2013oneshot} and \cite{li2021unified}. For the special case where $\mathsf{R}_{2}=0$ (i.e., the asymmetric broadcast channel), this result reduces to \cite{tan2013dispersions} applied on constant-composition codes.

\medskip{}

\begin{thm}
\label{thm:bc}For a discrete broadcast channel $P_{Y_{1},Y_{2}|X}$, for any $P_{U_{0},U_{1},U_{2}}$, function $x:\mathcal{U}_{0}\times\mathcal{U}_{1}\times\mathcal{U}_{2}\to\mathcal{X}$, $0<\epsilon<1$ and for any $n$, there is a scheme achieving a probability of error $P_{e}\le\epsilon$, with rates $\mathsf{R}_{0},\mathsf{R}_{1},\mathsf{R}_{2}$ if there exists $0\le\mathsf{S}_{1}\le I(U_{1};U_{2}|U_{0})$ (let $\mathsf{S}_{2}:=I(U_{1};U_{2}|U_{0})-\mathsf{S}_{1}$) and $0\le\tilde{\mathsf{R}}_{i}\le\mathsf{R}_{i}$ for $i=1,2$ such that
\begin{align*}
 & \left[\begin{array}{c}
\tilde{\mathsf{R}}_{1}+\mathsf{S}_{1}\\
\tilde{\mathsf{R}}_{2}+\mathsf{S}_{2}\\
\mathsf{R}_{0}+\mathsf{R}_{1}+\mathsf{R}_{2}-\tilde{\mathsf{R}}_{2}+\mathsf{S}_{1}\\
\mathsf{R}_{0}+\mathsf{R}_{1}+\mathsf{R}_{2}-\tilde{\mathsf{R}}_{1}+\mathsf{S}_{2}
\end{array}\right]\\
 & \in\mathbb{E}[J]-\frac{1}{\sqrt{n}}\mathcal{Q}^{-1}\left(\mathbb{E}[\mathrm{Var}[J|U_{0},U_{1},U_{2}]],\,\epsilon\right)-\frac{c\log n}{n},
\end{align*}
\[
J:=\left[\begin{array}{c}
\iota(U_{1};Y_{1}|U_{0})\\
\iota(U_{2};Y_{2}|U_{0})\\
\iota(U_{0},U_{1};Y_{1})\\
\iota(U_{0},U_{2};Y_{2})
\end{array}\right],
\]
where we assume $(U_{0},U_{1},U_{2})\sim P_{U_{0},U_{1},U_{2}}$, $X=x(U_{0},U_{1},U_{2})$, $(Y_{1},Y_{2})|X\sim P_{Y_{1},Y_{2}|X}$, $\mathcal{Q}^{-1}(\Sigma,\epsilon):=\{v\in\mathbb{R}^{k}:\,\mathbb{P}(V\le v)\ge1-\epsilon\}$ for $\Sigma\in\mathbb{R}^{k\times k}$ where $V\sim\mathrm{N}(0,\Sigma)$ and ``$\le$'' denotes entrywise comparison (as in \cite{yassaee2013oneshot}), where $c$ is a constant that depends only on $P_{Y_{1},Y_{2}|X}$, $P_{U_{0},U_{1},U_{2}}$, $\epsilon$ and the function $x$.
\end{thm}
\smallskip{}

\begin{IEEEproof}
Generate $U_{0}^{(n)},U_{1}^{(n)},U_{2}^{(n)}$ via a GCC channel $(P_{U_{0}^{(n)},U_{1}^{(n)},U_{2}^{(n)}})_{n}$ from $\emptyset$ (constant random variable) to $U_{0},U_{1},U_{2}$ with center $P_{U_{0},U_{1},U_{2}}$ and function $\zeta=0$. Using Propositions \ref{prop:gauss_memoryless} and \ref{prop:acc_product}, we have the asymptotic deviation
\begin{align*}
 & G_{\mathbf{U}_{0},\mathbf{U}_{1},\mathbf{U}_{2},\mathbf{Y}_{1},\mathbf{Y}_{2}}\\
 & =\sqrt{P_{U_{0},U_{1},U_{2}}}\circ G_{\mathbf{Y}_{1},\mathbf{Y}_{2}|\mathbf{U}_{0},\mathbf{U}_{1},\mathbf{U}_{2}}.
\end{align*}
The result follows directly from applying the achievability result in \cite[Theorem 5]{li2021unified}, and computing the relevant information density terms using Proposition \ref{prop:self_info}.
\end{IEEEproof}
\smallskip{}

\section{Acknowledgements}

This work was partially supported by two grants from the Research Grants Council of the Hong Kong Special Administrative Region, China {[}Project No.s: CUHK 24205621 (ECS), CUHK 14209823 (GRF){]}. 

\smallskip{}

\appendix{}

\subsection{Proof of Proposition \ref{prop:self_info}\label{subsec:pf_self_info}}

For $p\in\hat{P}_{n}(\mathcal{X})$, its type class is given by the preimage $\hat{P}_{n}^{-1}(p):=\{x^{n}:\,\hat{P}(x^{n})=p\}$. Consider $x^{n}\in\mathcal{X}^{n}$ with $\hat{p}:=\hat{P}(x^{n})$. Since $X^{(n)}$ is exchangeable,
\begin{align*}
\mathbb{P}(X^{(n)}=x^{n}) & =|\hat{P}_{n}^{-1}(\hat{p})|^{-1}\mathbb{P}\left(\hat{P}(X^{(n)})=\hat{p}\right)\\
 & =|\hat{P}_{n}^{-1}(\hat{p})|^{-1}\mathbb{P}\left(G_{\mathbf{X}}^{(n)}=\sqrt{n}(\hat{p}-P_{X})\right)\\
 & =|\hat{P}_{n}^{-1}(\hat{p})|^{-1}P_{G_{\mathbf{X}}^{(n)}}\left(\sqrt{n}(\hat{p}-P_{X})\right).
\end{align*}
Recall that for $\hat{p}\in\hat{P}(\mathcal{X}^{n})$, the logarithm of the size of its type class is \cite{cover2006elements}
\[
\log|\hat{P}_{n}^{-1}(\hat{p})|=nH(\hat{p})-O(|\mathcal{X}|\log n).
\]
Since $\mathbb{P}(\Vert G_{\mathbf{X}}^{(n)}-G_{\mathbf{X}}\Vert>cn^{-1/2})<cn^{-1/2}$, with probability $1-cn^{-1/2}$, $\sqrt{n}\left\langle G_{\mathbf{X}},\iota_{X}\right\rangle $ and $\sqrt{n}\left\langle G_{\mathbf{X}}^{(n)},\iota_{X}\right\rangle $ differs by at most a constant. We have
\begin{align*}
 & \iota_{X^{(n)}}(X^{(n)})-\left(nH(X)+\sqrt{n}\left\langle G_{\mathbf{X}}^{(n)},\iota_{X}\right\rangle \right)\\
 & =\log\left|\hat{P}_{n}^{-1}(\hat{P}(X^{(n)}))\right|-\log P_{G_{\mathbf{X}}^{(n)}}\left(G_{\mathbf{X}}^{(n)}\right)-nH(X)-\sqrt{n}\left\langle G_{\mathbf{X}}^{(n)},\iota_{X}\right\rangle \\
 & =\log\left|\hat{P}_{n}^{-1}\left(P_{X}+\frac{G_{\mathbf{X}}^{(n)}}{\sqrt{n}}\right)\right|-nH(X)-\sqrt{n}\left\langle G_{\mathbf{X}}^{(n)},\iota_{X}\right\rangle -\log P_{G_{\mathbf{X}}^{(n)}}\left(G_{\mathbf{X}}^{(n)}\right)\\
 & =nH\left(P_{X}+\frac{G_{\mathbf{X}}^{(n)}}{\sqrt{n}}\right)-nH(X)-\sqrt{n}\left\langle G_{\mathbf{X}}^{(n)},\iota_{X}\right\rangle -\log P_{G_{\mathbf{X}}^{(n)}}\left(G_{\mathbf{X}}^{(n)}\right)-O(|\mathcal{X}|\log n)\\
 & =n\left(H(X)+\left\langle \frac{G_{\mathbf{X}}^{(n)}}{\sqrt{n}},\iota_{X}\right\rangle -O\left(\left\Vert \frac{G_{\mathbf{X}}^{(n)}}{\sqrt{n}}\right\Vert ^{2}\right)\right)\\
 & \;\;\;\;-nH(X)-\sqrt{n}\left\langle G_{\mathbf{X}}^{(n)},\iota_{X}\right\rangle -\log P_{G_{\mathbf{X}}^{(n)}}\left(G_{\mathbf{X}}^{(n)}\right)-O(|\mathcal{X}|\log n)\\
 & =-O\left(\left\Vert G_{\mathbf{X}}^{(n)}\right\Vert ^{2}\right)-\log P_{G_{\mathbf{X}}^{(n)}}\left(G_{\mathbf{X}}^{(n)}\right)-O(|\mathcal{X}|\log n).
\end{align*}
Since there are no more than $(n+1)^{|\mathcal{X}|-1}$ possibilities for $\hat{P}(X^{(n)})$ and $G_{\mathbf{X}}^{(n)}$ \cite{cover2006elements},
\begin{align*}
\mathbb{P}\left(-\log P_{G_{\mathbf{X}}^{(n)}}\left(G_{\mathbf{X}}^{(n)}\right)>|\mathcal{X}|\log n\right) & =\mathbb{P}\left(P_{G_{\mathbf{X}}^{(n)}}\left(G_{\mathbf{X}}^{(n)}\right)<n^{-|\mathcal{X}|}\right)\\
 & \le n^{-|\mathcal{X}|}(n+1)^{|\mathcal{X}|-1}\\
 & \le2^{|\mathcal{X}|}n^{-1}.
\end{align*}
Consider a constant $c_{2}>0$ which will be specified later. By Definition \ref{def:typegaussian}, as long as $n$ is large enough such that $cn^{-1/2}\le\sqrt{\ln n}$,
\begin{align*}
 & \mathbb{P}\left(\left\Vert G_{\mathbf{X}}^{(n)}\right\Vert ^{2}>(c_{2}+1)^{2}\ln n\right)\\
 & =\mathbb{P}\left(\left\Vert G_{\mathbf{X}}^{(n)}\right\Vert >(c_{2}+1)\sqrt{\ln n}\right)\\
 & \stackrel{(a)}{\le}\mathbb{P}\left(\Vert G_{\mathbf{X}}\Vert>(c_{2}+1)\sqrt{\ln n}-cn^{-1/2}\right)+cn^{-1/2}\\
 & \le\mathbb{P}\left(\left\Vert G_{\mathbf{X}}\right\Vert >c_{2}\sqrt{\ln n}\right)+cn^{-1/2}\\
 & \stackrel{(b)}{=}O\left(e^{-(\ln n)/2}\right)+cn^{-1/2}\\
 & =O(n^{-1/2}),
\end{align*}
where (a) is because $d_{\Pi}(G_{\mathbf{X}}^{(n)},G_{\mathbf{X}})\le cn^{-1/2}$, and (b) holds for an appropriate choice of $c_{2}$ since $G_{\mathbf{X}}$ is subgaussian. The result follows.

\medskip{}

\subsection{Proof of Proposition \ref{prop:gauss_memoryless}\label{subsec:pf_gauss_memoryless}}

Generate $G_{\mathbf{Y}|\mathbf{X}}$, independent of $(\mathbf{X},G_{\mathbf{X}})$. By \cite{devroye2018total}, for $x^{n}\in\mathcal{X}^{n}$
\begin{align*}
 & d_{\mathrm{TV}}\left(\sqrt{\hat{P}(x^{n})}\circ G_{\mathbf{Y}|\mathbf{X}},\,\sqrt{P_{X}}\circ G_{\mathbf{Y}|\mathbf{X}}\right)\\
 & \le\min\{\zeta\Vert\hat{P}(x^{n})-P_{X}\Vert,1\},
\end{align*}
for some constant $\zeta>1$, where $d_{\mathrm{TV}}$ is the total variation distance. By the coupling lemma for $d_{\mathrm{TV}}$, we can define $\tilde{G}_{\mathbf{Y}|\mathbf{X}}$ which has the same distribution as $G_{\mathbf{Y}|\mathbf{X}}$ and is also independent of $(\mathbf{X},G_{\mathbf{X}})$, such that 
\begin{align}
 & \mathbb{P}\left(\sqrt{\hat{P}(X^{n})}\circ\tilde{G}_{\mathbf{Y}|\mathbf{X}}\neq\sqrt{P_{X}}\circ G_{\mathbf{Y}|\mathbf{X}}\,\Big|\,\mathbf{X},G_{\mathbf{X}}\right)\nonumber \\
 & \le\min\{\zeta\Vert\hat{P}(X^{n})-P_{X}\Vert,1\}\label{eq:G_XY_tv_bd-1}
\end{align}
almost surely. Since $\mathbf{X}$ is type deviation convergent, $\Vert G_{\mathbf{X}}^{(n)}-G_{\mathbf{X}}\Vert\le cn^{-1/2}$ with probability at least $1-cn^{-1/2}$, and hence 
\begin{align*}
 & \mathbb{E}\big[\min\{\zeta\Vert\hat{P}(X^{n})-P_{X}\Vert,1\}\big]\\
 & \le\zeta\mathbb{E}\left[\min\left\{ n^{-1/2}\left\Vert G_{\mathbf{X}}^{(n)}\right\Vert ,1\right\} \right]\\
 & \le\zeta\mathbb{E}\left[\min\left\{ n^{-1/2}\left(\left\Vert G_{\mathbf{X}}\right\Vert +cn^{-1/2}\right),1\right\} \right]+\zeta cn^{-1/2}\\
 & \le\zeta n^{-1/2}\mathbb{E}\left[\left\Vert G_{\mathbf{X}}\right\Vert \right]+\zeta cn^{-1}+\zeta cn^{-1/2}\\
 & =O(n^{-1/2}).
\end{align*}
Therefore,
\begin{align}
 & \mathbb{P}\left(\sqrt{\hat{P}(X^{n})}\circ\tilde{G}_{\mathbf{Y}|\mathbf{X}}\neq\sqrt{P_{X}}\circ G_{\mathbf{Y}|\mathbf{X}}\right)=O(n^{-1/2}).\label{eq:G_XY_tv_bd}
\end{align}
Generate $Y^{(n)}|X^{(n)}\sim P_{Y|X}^{n}$ for each $n$. Recall that $G_{\mathbf{X}}^{(n)}=\sqrt{n}(\hat{P}(X^{(n)})-P_{X})$. Let $\mathbf{n}\in\mathbb{Z}^{\mathcal{X}}$, $n_{x}:=|\{i:\,X_{i}^{(n)}=x\}|$ (note that $\mathbf{n}=n\hat{P}(X^{(n)})=\sqrt{n}G_{\mathbf{X}}^{(n)}+nP_{X}$). Since $\mathbf{X}$ is type deviation convergent, we have $\Vert G_{\mathbf{X}}^{(n)}\Vert\le c\log n$ and $n_{x}\ge nP_{X}(x)/2$ for all $x$, with probability at least $1-cn^{-1/2}$ (for some constant $c$). Consider $\tilde{Y}_{x}^{(n)}:=(Y_{i}^{(n)})_{i:X_{i}^{(n)}=x}$, i.e., the vector formed by $Y_{i}^{(n)}$ for $i\in[n]$ where $X_{i}^{(n)}=x$. Let $G_{\mathbf{Y}|\mathbf{X}}^{(n)}\in\mathbb{R}^{\mathcal{X}\times\mathcal{Y}}$, $G_{\mathbf{Y}|\mathbf{X}}^{(n)}(\cdot|x):=\sqrt{n_{x}}(\hat{P}(\tilde{Y}_{x}^{(n)})-P_{Y|X=x})$. Applying the central limit theorem in \cite{jurinskii1975smoothing}, we know that conditional on $X^{(n)}=x^{(n)}$, $G_{\mathbf{Y}|\mathbf{X}}^{(n)}(\cdot|x)$ follows $\mathrm{NM}(P_{Y|X=x})$ approximately, in the sense that there exists a constant $c_{x}$ (does not depend on $n$) such that
\begin{equation}
d_{\Pi}(G_{\mathbf{Y}|\mathbf{X}}^{(n)}(\cdot|x),\tilde{G}_{\mathbf{Y}|\mathbf{X}}(\cdot|x))<\frac{c_{x}}{\sqrt{n_{x}}},\label{eq:cvx_Gtilde}
\end{equation}
where we recall that $\tilde{G}_{\mathbf{Y}|\mathbf{X}}(\cdot|x)\sim\mathrm{NM}(P_{Y|X=x})$, independent of $(\mathbf{X},G_{\mathbf{X}})$. Recall that we have $n_{x}\ge nP_{X}(x)/2$, and hence $c_{x}/\sqrt{n_{x}}\le c_{x}/\sqrt{nP_{X}(x)/2}$, with probability at least $1-cn^{-1/2}$. By the Strassen-Dudley theorem \cite[Theorem 6.9]{billingsley2013convergence}, there exists a coupling of $G_{\mathbf{Y}|\mathbf{X}}^{(n)}$ and $\tilde{G}_{\mathbf{Y}|\mathbf{X}}$ such that 
\begin{equation}
\mathbb{P}\left(\Vert G_{\mathbf{Y}|\mathbf{X}}^{(n)}-\tilde{G}_{\mathbf{Y}|\mathbf{X}}\Vert>\frac{\tilde{c}}{\sqrt{n}}\right)<\frac{\tilde{c}}{\sqrt{n}},\label{eq:Gtx_GYX}
\end{equation}
where $\tilde{c}:=\sum_{x}(c_{x}/\sqrt{P_{X}(x)/2}+c)$. Therefore, $(Y^{(n)})_{n}$ can be couplied with $\mathbf{X}$ and $\tilde{G}_{\mathbf{Y}|\mathbf{X}}$ such that the above bound holds for all $n$. Also, recall that $\tilde{G}_{\mathbf{Y}|\mathbf{X}}$ is independent of $(\mathbf{X},G_{\mathbf{X}})$ (and hence $(\mathbf{X},G_{\mathbf{X}},G_{\mathbf{X}}^{(n)})$). We have 
\begin{align*}
G_{\mathbf{X},\mathbf{Y}}^{(n)} & =\sqrt{n}(\hat{P}(X^{(n)},Y^{(n)})-P_{X,Y})\\
 & =\left(\left(\frac{n_{x}}{\sqrt{n}}-\sqrt{n}P_{X}(x)\right)P_{Y|X=x}+\sqrt{\frac{n_{x}}{n}}G_{\mathbf{Y}|\mathbf{X}}^{(n)}(\cdot|x)\right)_{x\in\mathcal{X}}\\
 & =\sqrt{n}\left(\hat{P}(X^{(n)})-P_{X}\right)\circ P_{Y|X}+\sqrt{\hat{P}(X^{(n)})}\circ G_{\mathbf{Y}|\mathbf{X}}^{(n)}\\
 & =\sqrt{n}\left(\hat{P}(X^{(n)})-P_{X}\right)\circ P_{Y|X}+\sqrt{\hat{P}(X^{(n)})}\circ\tilde{G}_{\mathbf{Y}|\mathbf{X}}+O(n^{-1/2})\\
 & =G_{\mathbf{X}}^{(n)}\circ P_{Y|X}+\sqrt{\hat{P}(X^{(n)})}\circ\tilde{G}_{\mathbf{Y}|\mathbf{X}}+O(n^{-1/2})\\
 & =G_{\mathbf{X}}^{(n)}\circ P_{Y|X}+\sqrt{P_{X}}\circ G_{\mathbf{Y}|\mathbf{X}}+O(n^{-1/2})\\
 & =G_{\mathbf{X}}\circ P_{Y|X}+\sqrt{P_{X}}\circ G_{\mathbf{Y}|\mathbf{X}}+O(n^{-1/2}),
\end{align*}
with probability $1-O(n^{-1/2})$, due to (\ref{eq:G_XY_tv_bd}) and (\ref{eq:Gtx_GYX}), and $G_{\mathbf{X}}^{(n)}-G_{\mathbf{X}}=O(n^{-1/2})$ with probability $1-O(n^{-1/2})$ since $\mathbf{X}$ is type deviation convergent.\footnote{We say $A_{n}=B_{n}+O(n^{-1/2})$ with probability $1-O(n^{-1/2})$ if there exists a constant $c$ such that $\mathbb{P}(\Vert A_{n}-B_{n}\Vert\le cn^{-1/2})\ge1-cn^{-1/2}$ for all $n$.} This concludes the proof.

\medskip{}

\subsection{Proof of Proposition \ref{prop:acc_exists}\label{subsec:pf_acc_exists}}

For $x^{n}\in\mathcal{X}^{n}$, let $G_{x^{n}}^{(n)}:=\sqrt{n}(\hat{P}(x^{n})-P_{X})$,
\begin{align*}
\tilde{P}_{U|X}^{x^{n}}(u|x) & :=P_{U|X}(u|x)+\frac{P_{X}(x)}{\sqrt{n}\hat{P}(x^{n})(x)}\zeta(G_{x^{n}}^{(n)})(u|x).
\end{align*}
(If $P_{X}(x)=0$, take $\tilde{P}_{U|X}^{x^{n}}(u|x)=P_{U|X}(u|x)$.) We will check that $\tilde{P}_{U|X}^{x^{n}}$ is a valid conditional distribution with nonnegative entries for large enough $n$ and small enough $\Vert\hat{P}(x^{n})-P_{X}\Vert$. Since $\zeta$ takes values over $\mathrm{Tan}(P_{U|X})$, $\tilde{P}_{U|X}^{x^{n}}(u|x)=0$ whenever $P_{U|X}(u|x)=0$. Also, $\tilde{P}_{U|X}^{x^{n}}(u|x)=P_{U|X}(u|x)$ when $P_{X}(x)=0$. It is left to check the case $P_{X}(x)P_{U|X}(u|x)>0$, where we have
\begin{align*}
 & \tilde{P}_{U|X}^{x^{n}}(u|x)\\
 & =P_{U|X}(u|x)+\frac{P_{X}(x)\zeta(G_{x^{n}}^{(n)})(u|x)}{\sqrt{n}\hat{P}(x^{n})(x)}\\
 & \stackrel{(a)}{\ge}P_{U|X}(u|x)-\frac{P_{X}(x)(\Vert\zeta(0)\Vert+\tilde{c}\sqrt{n}\Vert\hat{P}(x^{n})-P_{X}\Vert)}{\sqrt{n}(P_{X}(x)-\Vert\hat{P}(x^{n})-P_{X}\Vert)}\\
 & \stackrel{(b)}{\ge}P_{U|X}(u|x)-\frac{2(\Vert\zeta(0)\Vert+\tilde{c}\sqrt{n}/c)}{\sqrt{n}}\\
 & \stackrel{(c)}{\ge}0,
\end{align*}
where (a) holds for some $\tilde{c}>0$ since $\zeta$ is Lipschitz, (b) is by restricting $\Vert\hat{P}(x^{n})-P_{X}\Vert\le1/c$ where $1/c\le(1/2)\min_{x:P_{X}(x)>0}P_{X}(x)$, and (c) is by taking $n\ge c$ and $c\ge16(\tilde{c}+\Vert\zeta(0)\Vert^{2})/\min_{u,x:P_{U|X}(u|x)>0}P_{U|X}(u|x)^{2}$. Hence, we can find $c>0$ such that as long as $n\ge c$ and $\Vert\hat{P}(x^{n})-P_{X}\Vert\le1/c$, $\tilde{P}_{U|X}^{x^{n}}$ is a valid conditional distribution with nonnegative entries. We have
\begin{align*}
 & \sqrt{n}\big(\hat{P}(x^{n})\circ\tilde{P}_{U|X}^{x^{n}}-P_{X,U}\big)\\
 & =\sqrt{n}\hat{P}(x^{n})\circ P_{U|X}+P_{X}\circ\zeta(G_{x^{n}}^{(n)})-\sqrt{n}P_{X,U}\\
 & =G_{x^{n}}^{(n)}\circ P_{U|X}+P_{X}\circ\zeta(G_{x^{n}}^{(n)}).
\end{align*}
To construct the conditional distribution $P_{U^{(n)}|X^{(n)}}$, given $X^{(n)}=x^{n}$ with $\Vert\hat{P}(x^{n})-P_{X}\Vert\le1/c$, we can let $U^{(n)}$ be uniform over sequences $u^{n}$ with conditional type $\tilde{P}_{U|X}^{x^{n}}$ (rounded off) given $x^{n}$. The round off error is $\Vert\hat{P}(x^{n},u^{n})-\hat{P}(x^{n})\circ\tilde{P}_{U|X}^{x^{n}}\Vert=O(1/n)$. Hence,
\begin{align*}
 & \big\Vert\sqrt{n}(\hat{P}(x^{n},u^{n})-P_{X,U})-G_{x^{n}}^{(n)}\circ P_{U|X}-P_{X}\circ\zeta(G_{x^{n}}^{(n)})\big\Vert\\
 & \le\big\Vert\sqrt{n}\big(\hat{P}(x^{n})\circ\tilde{P}_{U|X}^{x^{n}}-P_{X,U}\big)-G_{x^{n}}^{(n)}\circ P_{U|X}-P_{X}\circ\zeta(G_{x^{n}}^{(n)})\big\Vert+O(n^{-1/2})\\
 & =O(n^{-1/2}).
\end{align*}
This completes the proof.

\medskip{}

\smallskip{}

\subsection{Proof of $\psi(a)$ being Lipschitz in Theorem \ref{thm:wz}\label{subsec:pf_Lipschitz}}
\begin{prop}
Let $F(x,y)=\mathbb{P}(X\le x,\,Y\le y)$ be the cumulative distribution function of a Gaussian vector $(X,Y)\in\mathbb{R}^{2}$. Then there exists a Lipshitz function $\psi:\mathbb{R}\to\mathbb{R}$ such that $F(\psi(a),a-\psi(a))=\max_{t\in\mathbb{R}}F(t,a-t)$ for $a\in\mathbb{R}$.
\end{prop}
\smallskip{}

\begin{IEEEproof}
Without loss of generality, assume $(X,Y)$ is zero mean. Let the covaraince matrix be $\Sigma=\left[\begin{array}{cc}
\sigma_{X}^{2} & \sigma_{XY}\\
\sigma_{XY} & \sigma_{Y}^{2}
\end{array}\right]$. It is straightforward to check the case where $\Sigma$ is not full rank, i.e., $(X,Y)$ is supported over a line. Therefore, we assume $\Sigma$ is full rank. Let $f$ be the probability density function of $(X,Y)$, and write $f_{x}=\partial f/\partial x$, $f_{y}=\partial f/\partial y$. Let $\phi$ and $\Phi$ be the probability density function and cumulative distribution function of $\mathrm{N}(0,1)$. Let
\begin{align*}
\dot{F}_{x}(x,y) & :=\int_{-\infty}^{y}f(x,t)\mathrm{d}t,\quad\dot{F}_{y}(x,y):=\int_{-\infty}^{x}f(s,y)\mathrm{d}s,\\
\ddot{F}_{x}(x,y) & :=\int_{-\infty}^{y}f_{x}(x,t)\mathrm{d}t,\quad\ddot{F}_{y}(x,y):=\int_{-\infty}^{x}f_{y}(s,y)\mathrm{d}s.
\end{align*}

\textbf{Step 1:} Showing that $\psi'(a)$ is bounded for $a\gg0$. By direct evaluation,
\begin{align*}
\psi'(a) & =\frac{\ddot{F}_{y}(x,y)-f(x,y)}{\ddot{F}_{x}(x,y)+\ddot{F}_{y}(x,y)-2f(x,y)},
\end{align*}
where $(x,y)=(\psi(a),a-\psi(a))$. Hence, $0\le\psi'(a)\le1$ if $\ddot{F}_{x}(x,y),\ddot{F}_{y}(x,y)\le0$. We have
\[
\dot{F}_{x}(x,y)=\frac{1}{\sigma_{X}}\phi\left(\frac{x}{\sigma_{X}}\right)\Phi\left(\frac{y-\sigma_{XY}\sigma_{X}^{-2}x}{\sqrt{\sigma_{Y}^{2}-\sigma_{XY}^{2}\sigma_{X}^{-2}}}\right).
\]
If $\sigma_{XY}\ge0$, then $\dot{F}_{x}(x,y)$ is decreasing in $x$ for $x\ge0$ since both the $\phi$ and $\Phi$ terms are decreasing, and hence $\ddot{F}_{x}(x,y)\le0$ for $x\ge0$. If $\sigma_{XY}<0$, then since $\ddot{F}_{x}(x,0)/(\sigma_{X}^{-1}\phi(\sigma_{X}^{-1}x))\to-\infty$, we can find $x_{1}\ge0$ such that $\ddot{F}_{x}(x,0)<0$ for $x\ge x_{1}$, and hence, $\dot{F}_{x}(x,0)$ is decreasing in $x$ for $x\ge x_{1}$, implying that $\dot{F}_{x}(x,y)$ is decreasing in $x$ for $x\ge x_{1}$ for any fixed $y\ge0$ (since $f(x,y)$ is decreasing in $x$ for $x\ge0$ for fixed $y\ge0$). Combining both cases, we know that there exists $x_{1}\ge0$ such that $\ddot{F}_{x}(x,y)\le0$ for $x\ge x_{1}$, $y\ge0$. Applying the same arguments on $y$, there exists $x_{1},y_{1}\ge0$ such that $\ddot{F}_{x}(x,y),\ddot{F}_{y}(x,y)\le0$ for $x\ge x_{1},y\ge y_{1}$, and hence $0\le\psi'(a)\le1$ whenever $\psi(a)\ge x_{1}$, $a-\psi(a)\ge y_{1}$. Letting $a_{1}>2\max\{x_{1},y_{1}\}$ be such that $F(a_{1}/2,a_{1}/2)>\max\{\Phi(\sigma_{X}^{-1}x_{1}),\Phi(\sigma_{Y}^{-1}y_{1})\}$, we have $\psi(a)\ge x_{1}$, $a-\psi(a)\ge y_{1}$ for $a\ge a_{1}$ by the optimality of $\psi$, and hence $0\le\psi'(a)\le1$ for $a\ge a_{1}$.

\textbf{Step 2:} Showing that $\psi'(a)$ is bounded for $a\ll0$. Performing a change of coordinate $a=x+y$, $b=x-y$, $c=b-\gamma a$ (where $\gamma$ is specified later), we have
\begin{align*}
 & F(x,y)\\
 & =\int_{0}^{\infty}\frac{1}{\sigma_{A}}\phi\left(\frac{a-s}{\sigma_{A}}\right)\left(\Phi\left(\frac{c+(\gamma+1)s}{\sigma_{B|A}}\right)-\Phi\left(\frac{c+(\gamma-1)s}{\sigma_{B|A}}\right)\right)\mathrm{d}s\\
 & =\frac{\sigma_{A}}{-2a}\phi\left(\frac{a}{\sigma_{A}}\right)K(a,c),
\end{align*}
where $\left[\begin{array}{cc}
\sigma_{A}^{2} & \sigma_{AB}\\
\sigma_{AB} & \sigma_{B}^{2}
\end{array}\right]$ is the covariance matrix of $(X+Y,X-Y)$, $\gamma:=\sigma_{AB}\sigma_{A}^{-2}$, $\sigma_{B|A}^{2}:=\sigma_{B}^{2}-\sigma_{AB}^{2}\sigma_{A}^{-2}$, and
\begin{align*}
K(a,c) & :=\int_{0}^{\infty}\frac{-2a}{\sigma_{A}^{2}}\exp\left(-\frac{s(s-2a)}{\sigma_{A}^{2}}\right)\left(\Phi\left(\frac{c+(\gamma+1)s}{\sigma_{B|A}}\right)-\Phi\left(\frac{c+(\gamma-1)s}{\sigma_{B|A}}\right)\right)\mathrm{d}s\\
 & =\frac{2}{\sigma_{B|A}}\phi\left(\frac{c}{\sigma_{B|A}}\right)+O\left(\frac{\log(-a)}{-a}\right)
\end{align*}
as $a\to-\infty$ since $-2a\sigma_{A}^{-2}e^{-s(s-2a)\sigma_{A}^{-2}}$ approaches the density of the exponential distribution with rate $-2a\sigma_{A}^{-2}$. Hence,
\[
\lim_{a\to-\infty}\underset{c}{\mathrm{argmax}}K(a,c)=0.
\]
Therefore, for $a$ sufficiently negative, we can focus on $|c|\le\epsilon$ for some small $\epsilon>0$ such that $\phi''(t)<0$ for all $|t|\le\sigma_{B|A}^{-1}\epsilon/2$. For $|c|\le\epsilon$,
\begin{align*}
 & -\sigma_{B|A}^{2}\frac{\partial^{2}K(a,c)}{\partial c^{2}}\\
 & =\int_{0}^{\infty}\frac{-2a}{\sigma_{A}^{2}}\exp\left(-\frac{s(s-2a)}{\sigma_{A}^{2}}\right)\left(\phi'\left(\frac{c+(\gamma-1)s}{\sigma_{B|A}}\right)-\phi'\left(\frac{c+(\gamma+1)s}{\sigma_{B|A}}\right)\right)\mathrm{d}s\\
 & \ge\int_{0}^{\infty}\frac{-2a}{\sigma_{A}^{2}}\exp\left(-\frac{s(s-2a)}{\sigma_{A}^{2}}\right)\min_{t\in\sigma_{B|A}^{-1}(\gamma s+(s+\epsilon)[-1,1])}\left(-\phi''(t)\right)\mathrm{d}s\\
 & \to\min_{t\in\sigma_{B|A}^{-1}\epsilon[-1,1]}\left(-\phi''(t)\right)\;>\;0
\end{align*}
as $a\to-\infty$ since $\min_{t\in\sigma_{B|A}^{-1}(\gamma s+(s+\epsilon)[-1,1])}(-\phi''(t))$ is bounded and continuous in $s$. Also, for $a<0$,
\begin{align*}
 & \bigg|\sigma_{B|A}\frac{\partial^{2}K(a,c)}{\partial a\partial c}\bigg|\\
 & =\bigg|\int_{0}^{\infty}\frac{2s}{\sigma_{A}^{2}}\frac{-2a}{\sigma_{A}^{2}}\exp\left(-\frac{s(s-2a)}{\sigma_{A}^{2}}\right)\left(\phi\left(\frac{c+(\gamma+1)s}{\sigma_{B|A}}\right)-\phi\left(\frac{c+(\gamma-1)s}{\sigma_{B|A}}\right)\right)\mathrm{d}s\bigg|\\
 & \le\int_{0}^{\infty}\frac{2s}{\sigma_{A}^{2}}\frac{-2a}{\sigma_{A}^{2}}\exp\left(-\frac{s(s-2a)}{\sigma_{A}^{2}}\right)\frac{1}{\sqrt{2\pi}}\mathrm{d}s\\
 & \le\frac{1}{-a\sqrt{2\pi}}.
\end{align*}
Hence, there exists $a_{0}$ such that
\begin{align*}
\psi'(a) & =\frac{1+\gamma}{2}+\frac{1}{2}\frac{\mathrm{d}}{\mathrm{d}a}\underset{c}{\mathrm{argmax}}K(a,c)\\
 & =\frac{1+\gamma}{2}-\frac{1}{2}\frac{\partial^{2}K(a,c)/(\partial a\partial c)}{\partial^{2}K(a,c)/\partial c^{2}}
\end{align*}
is bounded for $a\le a_{0}$.

\textbf{Step 3:} Showing that $\psi'(a)$ is bounded for $a_{0}\le a\le a_{1}$. For $a\in[a_{0},a_{1}]$, we have $F(a/2,a/2)\ge F(a_{0}/2,a_{0}/2)$, and hence by the optimality of $\psi$,
\[
\psi(a)\ge\sigma_{X}\Phi^{-1}(F(a_{0}/2,a_{0}/2)),
\]
\[
a-\psi(a)\ge\sigma_{Y}\Phi^{-1}(F(a_{0}/2,a_{0}/2)).
\]
Therefore, to study the behavior of $\psi(a)$, it suffices to study $F(x,y)$ for $(x,y)$ satisfying $x\ge\sigma_{X}\Phi^{-1}(F(a_{0}/2,a_{0}/2))$, $y\ge\sigma_{Y}\Phi^{-1}(F(a_{0}/2,a_{0}/2))$ and $x+y\le a_{1}$, which is a closed bounded region. Let $\mathrm{H}(x,y):=\nabla^{2}(-\ln F(x,y))\in\mathbb{R}^{2\times2}$ be the Hessian matrix of $-\ln F(x,y)$. We will prove that $\mathrm{H}(x,y)$ is positive definite. By the Pr\'{e}kopa-Leindler inequality, for any log-concave function $g(x,y)$, the function $G(x,y)=\int_{-\infty}^{x}\int_{-\infty}^{y}g(s,t)\mathrm{d}t\mathrm{d}s$ is log-concave as well. Since $f$ is log-concave, $\mathrm{H}(x,y)$ must be positive semidefinite. By direct computation, 
\begin{align*}
\mathrm{H}_{11}(x,y) & =\big(\dot{F}_{x}(x,y)^{2}-F(x,y)\ddot{F}_{x}(x,y)\big)/F(x,y)^{2},\\
\mathrm{H}_{22}(x,y) & =\big(\dot{F}_{y}(x,y)^{2}-F(x,y)\ddot{F}_{y}(x,y)\big)/F(x,y)^{2},\\
\mathrm{H}_{12}(x,y) & =\big(\dot{F}_{x}(x,y)\dot{F}_{y}(x,y)-F(x,y)f(x,y)\big)/F(x,y)^{2}.
\end{align*}
We can see that $\mathrm{H}(x,y)$ depends only on $F(x,y)$ and the values of $f$ around the neighborhood of $\{(x,t):t\le y\}\cup\{(s,y):s\le x\}$. For a fixed $(x,y)$, if we can construct an alternative log-concave probability density function $g$ such that its cumulative distribution function $G$ satisfies that $g(x,y)=f(x,y)$ (at this particular fixed $(x,y)$), $G(x,y)=F(x,y)$, $\dot{G}_{x}(x,y)=\dot{F}_{x}(x,y)$, $\dot{G}_{y}(x,y)=\dot{F}_{y}(x,y)$, but $\ddot{G}_{x}(x,y)>\ddot{F}_{x}(x,y)$ and $\ddot{G}_{y}(x,y)>\ddot{F}_{y}(x,y)$, then $\mathrm{H}(x,y)$ would be the sum of the Hessian of $-\ln G(x,y)$ and a positive definite diagonal matrix, and hence is strictly positive definite. To do this, we can take the small perturbation $g(s,t)=f(s,t)e^{b(s-x,t-y+2)+b(t-y,s-x+2)}$, where $b(s,t)$ is a suitable bump function supported over $[-1,1]^{2}$ chosen to satisfy the requirements.\footnote{More precisely, we require $\nabla^{2}b\preceq\nabla^{2}(-\ln f)$, where $\preceq$ is the semidefinite order (note that $\nabla^{2}(-\ln f)$ is constant), $b(s,t)=0$ when $s=0$ or $t=0$, $\partial b(s,t)/\partial s\ge0$ at $s=0$ with strict inequality for some $t$, $\partial b(s,t)/\partial s\ge0$ at $s=0$ with strict inequality for some $t$, $\partial b(s,t)/\partial t\ge0$ at $t=0$ with strict inequality for some $s$, and $G(x,y)=F(x,y)$. An explicit construction would be $b(s,t)=\delta(\gamma s^{2}+s)\exp(1/((s^{2}-1)(t^{2}-1)))$ for $(s,t)\in[-1,1]^{2}$, where $\delta>0$, $\gamma\in\mathbb{R}$ are suitable small constants.} Hence, $\mathrm{H}(x,y)$ is strictly positive definite. Let
\[
D(x,y):=\frac{\mathrm{H}_{22}(x,y)-\mathrm{H}_{12}(x,y)}{\mathrm{H}_{11}(x,y)+\mathrm{H}_{22}(x,y)-2\mathrm{H}_{12}(x,y)}.
\]
We have $\psi'(a)=D(\psi(a),a-\psi(a))$. Since $\mathrm{H}(x,y)$ is positive definite, the denominator of $D(x,y)$ is positive, and hence $D(x,y)$ is a continuous function, and is bounded within a closed bounded set of $(x,y)$. Hence, $\psi'(a)$ is bounded for $a_{0}\le a\le a_{1}$. Therefore, combining all three steps, we know $\psi'(a)$ is always bounded, and $\psi$ is Lipschitz.
\end{IEEEproof}
\smallskip{}

\subsection{Proof of the Comparisons in Section \ref{subsec:wz_compare}\label{subsec:pf_compare}}

We first prove that Theorem \ref{thm:wz} improves upon (\ref{eq:wz_Pe_prev_poisson}), and hence (\ref{eq:wz_Pe_yassaee}). Let $\bar{J}:=\bar{J}_{\mathbf{Y}}+\bar{J}_{\mathbf{Y}}$. Then $[\bar{J},\bar{D}]$ has a covariance matrix $\mathrm{Var}[[\iota(U;X)-\iota(U;Y),d(X,Z)]]$. Let $[J_{\mathbf{X}},D_{\mathbf{X}}]$, $[\hat{J}_{\mathbf{U}},\hat{D}_{\mathbf{U}}]$, $[J_{\mathbf{Y}},D_{\mathbf{Y}}]$ be independent a zero mean Gaussian vectors with covariance matrices
\[
\mathrm{Var}\left[\mathbb{E}\left[\left.\left[\begin{array}{c}
\iota(U;X)-\iota(U;Y)\\
d(X,Z)
\end{array}\right]\,\right|\,X\right]\right],
\]
\[
\mathbb{E}\left[\mathrm{Var}\left[\left.\mathbb{E}\left[\left.\left[\begin{array}{c}
\iota(U;X)-\iota(U;Y)\\
d(X,Z)
\end{array}\right]\,\right|\,X,U\right]\,\right|\,X\right]\right],
\]
and
\[
\mathbb{E}\left[\mathrm{Var}\left[\left.\left[\begin{array}{c}
-\iota(U;Y)\\
d(X,Z)
\end{array}\right]\,\right|\,X,U\right]\right],
\]
respectively. Let $A_{\mathbf{X}}=J_{\mathbf{X}}+\lambda D_{\mathbf{X}}$. By the law of total variance, the covariance matrix of $[J_{\mathbf{X}},D_{\mathbf{X}}]+[\hat{J}_{\mathbf{U}},\hat{D}_{\mathbf{U}}]+[J_{\mathbf{Y}},D_{\mathbf{Y}}]$ is the same as that of $[\bar{J},\bar{D}]$, and hence we can assume $\bar{J}=J_{\mathbf{X}}+\hat{J}_{\mathbf{U}}+J_{\mathbf{Y}}$ and $\bar{D}=D_{\mathbf{X}}+\hat{D}_{\mathbf{U}}+D_{\mathbf{Y}}$. Also, by the first-order optimality of $P_{U|X}$, we have $\hat{J}_{\mathbf{U}}=-\lambda\hat{D}_{\mathbf{U}}$ (see the proof of Theorem \ref{thm:wz}). To compare (\ref{eq:wz_Pe_prev_poisson}) with Theorem \ref{thm:wz}, for any $t\in\mathbb{R}$,
\begin{align}
 & \mathbb{P}(\bar{J}>\mathsf{W}-\lambda t\;\mathrm{or}\;\bar{D}>t)\nonumber \\
 & =\mathbb{P}(J_{\mathbf{X}}+\hat{J}_{\mathbf{U}}+J_{\mathbf{Y}}>\mathsf{W}-\lambda t\;\mathrm{or}\;D_{\mathbf{X}}+\hat{D}_{\mathbf{U}}+D_{\mathbf{Y}}>t)\nonumber \\
 & =\mathbb{P}(J_{\mathbf{Y}}>\mathsf{W}-A_{\mathbf{X}}-\lambda(t-D_{\mathbf{X}}-\hat{D}_{\mathbf{U}})\;\mathrm{or}\;D_{\mathbf{Y}}>t-D_{\mathbf{X}}-\hat{D}_{\mathbf{U}})\nonumber \\
 & \ge\mathbb{E}[P_{e}^{*}(\mathsf{W}-A_{\mathbf{X}})],\label{eq:wz_compare_Pe}
\end{align}
where $P_{e}^{*}(\alpha)=\min_{t'}\mathbb{P}(J_{\mathbf{Y}}>\alpha-\lambda t'\;\mathrm{or}\;D_{\mathbf{Y}}>t')$ as in Theorem \ref{thm:wz}, by substituting $t'=t-D_{\mathbf{X}}-\hat{D}_{\mathbf{U}}$. Hence, (\ref{eq:wz_Pe_prev_poisson}) is greater than or equal to the bound in Theorem \ref{thm:wz}. Also, Corollary \ref{cor:wz_cor-1} improves upon (\ref{eq:wz_prev_poisson}) since they are the low excess distortion probability limits of Theorem \ref{thm:wz} and (\ref{eq:wz_Pe_prev_poisson}), respectively.

Next, we prove that Theorem \ref{thm:wz} improves upon (\ref{eq:wz_Pe_wkt}). First, we can assume that $Z$ is a function of $(\tilde{U},Y,T)$ since the randomness in $Z$ can be absorbed into $T$ by the functional representation lemma \cite{elgamal2011network}. Since $T$ is independent of $(X,Y)$, we have $\iota(\tilde{U};X|T)=\iota(\tilde{U},T;X)$ and $\iota(\tilde{U};Y|T)=\iota(\tilde{U},T;Y)$. Take $U=(\tilde{U},T)$. Let $\tilde{J}:=\tilde{J}_{\mathbf{X}}+\tilde{J}_{\mathbf{Y}}$. Let $K:=[\iota(U;X)-\iota(U;Y),d(X,Z)]\in\mathbb{R}^{2}$ and $\bar{K}:=\mathbb{E}[K|X,U]$. Let $[\tilde{J}_{\mathbf{U}},\tilde{D}_{\mathbf{U}}]$ be a zero mean Gaussian vector with covariance matrix
\begin{align*}
 & \mathbb{E}[\mathrm{Var}[\bar{K}|T]]-\mathrm{Var}[\mathbb{E}[\bar{K}|X]].\\
 & =\mathrm{Var}[\bar{K}]-\mathrm{Var}[\mathbb{E}[\bar{K}|T]]-\mathrm{Var}[\mathbb{E}[\bar{K}|X]]\\
 & =\mathrm{Var}\left[\bar{K}-\mathbb{E}[\bar{K}|T]-\mathbb{E}[\bar{K}|Y]\right]
\end{align*}
since $X,T$ are independent. Define $[J_{\mathbf{X}},D_{\mathbf{X}}]$, $[J_{\mathbf{Y}},D_{\mathbf{Y}}]$ as in the previous part. By the law of total variance, the covariance matrix of $[J_{\mathbf{X}},D_{\mathbf{X}}]+[\tilde{J}_{\mathbf{U}},\tilde{D}_{\mathbf{U}}]+[J_{\mathbf{Y}},D_{\mathbf{Y}}]$ is the same as that of $[\tilde{J},\tilde{D}]$, and hence we can assume $\tilde{J}=J_{\mathbf{X}}+\tilde{J}_{\mathbf{U}}+J_{\mathbf{Y}}$ and $\tilde{D}=D_{\mathbf{X}}+\tilde{D}_{\mathbf{U}}+D_{\mathbf{Y}}$. We have
\begin{align*}
 & \mathrm{Var}[J_{\mathbf{X}},D_{\mathbf{X}}]+\mathrm{Var}[\tilde{J}_{\mathbf{U}},\tilde{D}_{\mathbf{U}}]+\mathrm{Var}[J_{\mathbf{Y}},D_{\mathbf{Y}}]\\
 & =\mathrm{Var}[\tilde{J},\tilde{D}]\\
 & \preceq\mathrm{Var}[\bar{J},\bar{D}]\\
 & =\mathrm{Var}[J_{\mathbf{X}},D_{\mathbf{X}}]+\mathrm{Var}[\hat{J}_{\mathbf{U}},\hat{D}_{\mathbf{U}}]+\mathrm{Var}[J_{\mathbf{Y}},D_{\mathbf{Y}}],
\end{align*}
and hence $\mathrm{Var}[\tilde{J}_{\mathbf{U}},\tilde{D}_{\mathbf{U}}]\preceq\mathrm{Var}[\hat{J}_{\mathbf{U}},\hat{D}_{\mathbf{U}}]$, where ``$\preceq$'' denotes the positive semidefinite order. Since $\hat{J}_{\mathbf{U}}=-\lambda\hat{D}_{\mathbf{U}}$, we also have $\tilde{J}_{\mathbf{U}}=-\lambda\tilde{D}_{\mathbf{U}}$. Using the same steps as (\ref{eq:wz_compare_Pe}), for any $t,\tau\in\mathbb{R}$,
\begin{align*}
 & \mathbb{P}(\tilde{J}_{\mathbf{X}}>\mathsf{W}-\lambda t-\tau\;\mathrm{or}\;\tilde{J}_{\mathbf{Y}}>\tau\;\mathrm{or}\;\tilde{D}>t)\\
 & \ge\mathbb{P}(\tilde{J}>\mathsf{W}-\lambda t\;\mathrm{or}\;\tilde{D}>t)\\
 & \ge\mathbb{E}[P_{e}^{*}(\mathsf{W}-A_{\mathbf{X}})].
\end{align*}
Hence, (\ref{eq:wz_Pe_wkt}) is greater than or equal to the bound in Theorem \ref{thm:wz}, and Corollary \ref{cor:wz_cor-1} improves upon (\ref{eq:wz_wkt}) since they are the low excess distortion probability limits of Theorem \ref{thm:wz} and (\ref{eq:wz_Pe_wkt}), respectively.

\bibliographystyle{IEEEtran}
\bibliography{ref}

\end{document}